%% file: autoscaling-survey.tex
\documentclass[acmsmall]{acmart}

\usepackage{tabularx}

\usepackage{color}

\usepackage{multirow}
\usepackage{enumitem}
\usepackage{standalone}
\usepackage{tikz}
\usepackage{pgfplots}
\pgfplotsset{
        /pgfplots/ybar legend/.style={
        /pgfplots/legend image code/.code={%
        \draw[##1,/tikz/.cd,bar width=3pt,yshift=-0.2em,bar shift=0pt]
                plot coordinates {(0cm,0.8em)};},
},
}

\usepackage[flushleft]{threeparttable}
\usepackage{subcaption}
\usepackage{balance}
\setcopyright{usgovmixed}
\newcolumntype{P}[1]{>{\centering\arraybackslash}m{#1}}
\newcolumntype{Y}{>{\centering\arraybackslash}X}

\begin{document}
% Title portion
\title{A Survey and Taxonomy of Self-Aware and Self-Adaptive Cloud Autoscaling Systems} 
%\author{Tao Chen, Rami Bahsoon, Xin Yao}
%%\orcid{1234-5678-9012-3456}
%\affiliation{%
%  \institution{Centre of Excellence for Research in Computational Intelligence and Applications (CERCIA), School of Computer Science, University of Birmingham}
%  \streetaddress{Edgbaston}
%  \city{Birmingham}
%  \postcode{B15 2TT}
%  \country{UK}}
\author{Tao Chen}
%\orcid{1234-5678-9012-3456}
\affiliation{%
  \institution{Department of Computing and Technology, Nottingham Trent University, UK, and CERCIA, School of Computer Science, University of Birmingham, UK}
  \streetaddress{Edgbaston}
  \city{Birmingham}
  \postcode{B15 2TT}
  \country{UK}}
\author{Rami Bahsoon}  
  \affiliation{%
  \institution{CERCIA, School of Computer Science, University of Birmingham, UK}
  \streetaddress{Edgbaston}
  \city{Birmingham}
  \postcode{B15 2TT}
  \country{UK}}
  \author{Xin Yao}  
  \affiliation{%
  \institution{Department of Computer Science and Engineering, Southern University of Science and Technology, Shenzhen 518055, China, and CERCIA, School of Computer Science, University of Birmingham, UK.}
  \streetaddress{Edgbaston}
  \city{Birmingham}
  \postcode{B15 2TT}
  \country{UK}}  

\begin{abstract}
Autoscaling system can reconfigure cloud-based services and applications, through various configurations of cloud software and provisions of hardware resources, to adapt to the changing environment at runtime. Such a behavior offers the foundation for achieving elasticity in modern cloud computing paradigm. Given the dynamic and uncertain nature of the shared cloud infrastructure, cloud autoscaling system has been engineered as one of the most complex, sophisticated and intelligent artifacts created by human, aiming to achieve self-aware, self-adaptive and dependable runtime scaling. Yet, existing Self-aware and Self-adaptive Cloud Autoscaling System (SSCAS) is not mature to a state that it can be reliably exploited in the cloud. In this article, we survey the state-of-the-art research studies on SSCAS and provide a comprehensive taxonomy for this field. We present detailed analysis of the results and provide insights on open challenges, as well as the promising directions that are worth investigated in the future work of this area of research. Our survey and taxonomy contribute to the fundamentals of engineering more intelligent autoscaling systems in the cloud.

\end{abstract}

\setcopyright{rightsretained} 
\acmJournal{CSUR}
\acmYear{2018} \acmVolume{1} \acmNumber{1} \acmArticle{1} \acmMonth{1} \acmPrice{}\acmDOI{10.1145/3190507}




% We no longer use \terms command
%\terms{Design, Algorithms, Performance, Reliability}


\thanks{This work was supported by the Ministry of Science and Technology of China (Grant No. 2017YFC0804003), Science and Technology Innovation Committee Foundation of Shenzhen (Grant No. ZDSYS201703031748284), and EPSRC (Grant No. EP/J017515/01 and EP/K001523).}

%  Author's addresses: G. Zhou, Computer Science Department, College of
%  William and Mary; Y. Wu {and} J. A. Stankovic, Computer Science
%  Department, University of Virginia; T. Yan, Eaton Innovation Center;
%  T. He, Computer Science Department, University of Minnesota; C.
%  Huang, Google; T. F. Abdelzaher, (Current address) NASA Ames
%  Research Center, Moffett Field, California 94035.}

\maketitle

\renewcommand{\shortauthors}{T. Chen et al.}

\section{Introduction}

Modern IT companies, from small business to large enterprises, increasingly leverage cloud computing to improve their profits and reduce the costs. Throughout all the Software-as-a-Service (SaaS), Platform-as-a-Service (PaaS) and Infrastructure-as-a-Service (IaaS) levels, one of the pronounced benefits of the cloud is referred to as elasticity, which reflects the extent to which a system can adapt to the workload fluctuations by adjusting configurations and resource provisioning close to the demand. In certain predictable scenarios where the environmental condition has strong and stable seasonality, the configurations and resources can be approximately specified by human experts in advance. Nevertheless, for many other cases, for examples, unexpected workload changes, elasticity can be only enabled by runtime automatic scaling, or simply autoscaling: \emph{a dynamic process, often operating on a Physical Machine (PM), that adapts software configurations (e.g., threads, connections and cache, etc) and hardware resources provisioning (e.g., CPU, memory, etc) on-demand, according to the time-varying environmental conditions.} The ultimate goal of autoscaling is to continually optimize the non-functional Quality of Service (QoS) (e.g., response time and throughput) and cost objectives for all cloud-based services\footnote{Service could refer to an entire application, or any conceptual part within an application.}; thus their Service Level Agreement (SLA) and budget requirements can be better complied with. In particular, autoscaling systems help to realize elasticity by providing timely and elastic adaptation in scales, which is one of the key benefit of cloud computing that attracts a wide range of practitioners~\cite{lorido2014review}\cite{qu2016auto}. Comparing with the other cloud resources management in general, autoscaling system has been specifically designed for: (i) scaling cloud-based applications in response to dynamic changes in load, uncertainties in operations, handling multitenancy while ensuring Service Level Agreement compliance etc. In contrast, other resource management tasks, e.g., resource scheduling, often work on planned and deterministic sequence of resource demand. (ii) Adapting both the software configurations and hardware resources (and their interplays) that span over all SaaS, PaaS and IaaS levels, whereas most of the other resource management considers hardware resources and IaaS only. (iii) Taking the QoS for cloud-based applications/services at the centre of the concerned objectives (explicitly or implicitly) while the other resource management tasks often focus on resource utilization.

From the literal meaning of the word \lq autoscaling\rq, it is obvious that the process is dynamic and requires the system to adapt itself subject to the dynamic and uncertain state of the services being managed and the environment. In such a way, the cloud-based services, running on a Virtual Machine (VM) or containers, can be \lq expanded\rq~and \lq shrink\rq~according to the environmental conditions at runtime. This characteristic has made autoscaling systems well-suited to the broad category of self-aware and self-adaptive systems \cite{Chen:2015:computer} \cite{roadmap} \cite{landscape}. However, given the unique characteristics of cloud, engineering Self-aware and Self-adaptive Cloud Autoscaling System (SSCAS) poses many challenges, including efficient autoscaling architectural styles, accurate model to predict the effects of autoscaling decision\footnote{Each autoscaling decision is a specific combination of configurations and/or resource provisions that achieves certain outcomes on the targeted objectives.} on the quality attributes, appropriate granularity of runtime control and effective trade-off decision making. In particular, a SSCAS should be able to handle various dimensions of QoS attributes, software configuration and hardware resources, in the presence of \textbf{\emph{QoS interference}}~\cite{fuzzy-vm-interference} \cite{qcloud} \cite{vm-fuzzy-MIMO} \cite{2014-decision-tree-software-CP-2014} where the quality of a single cloud-based service can be influenced by the dynamic behaviors of its neighbors on a PM, i.e., the other co-located services and co-hosted VMs or containers, under the sharing infrastructure of cloud.

In this article, we provide a survey and taxonomy for the landscape of SSCAS research to better understand the state-of-the-arts and to identify the open challenges in the field. In particular, we focus on cloud autoscaling researches with respect to the well-known principles of self-awareness~\cite{epics} and self-adaptively~\cite{landscape} in computing systems, as well as the fundamental approaches and techniques that realize them. Broadly speaking, we aim to answer the following research questions: (i) What are the levels of self-awareness and self-adaptivity that have been captured in SSCAS? (ii) What are the architectural patterns used for engineering SSCAS? (iii) What are the approaches used to model the quality related to SSCAS? (iv) What is the granularity of control in SSCAS? (v) What are the approaches used for decision making in SSCAS? In a nutshell, our key findings are:

\begin{itemize}[leftmargin=.35cm]

  \item \emph{Stimulus-}, \emph{time-} and \emph{goal-awareness} are the most commonly considered self-awareness levels in current SSCAS research while \emph{self-configuring} and \emph{self-optimizing} are more attractive than the others on self-adaptivity (Please refer to Section~\ref{sec:s-aware} and~\ref{sec:s-adaptive} for their definitions, respectively).
  
  \item It was found that the general and simple feedback loop architectural pattern (and its variations) has been prominent for engineering SSCAS.
  
  \item Analytical and machine learning based modeling have been the most widely used approaches for modeling the effects of autoscaling decision on quality attributes. But surprisingly, systematic selection of model's input features and QoS interference are rarely considered for SSCAS.
  
  \item The level of service/application is the most popular granularity of control for SSCAS.
  
  \item Explicit optimization driven decision making is the most commonly used approach in SSCAS, but most of them have assumed single objective or using weighted sum aggregation of objectives. Further, we noted that QoS interference is again absent in many studies.

\end{itemize}
In addition, we obtain the following insights on the open challenges for future research of the field: 

\begin{itemize}[leftmargin=.35cm]

  \item There is a lack of considering required knowledge and its representation for SSCAS architecture with respect to the principles of self-awareness~\cite{epics}, thus urging further investigations. This can help to reason about and prevent improper design decisions, leading to better self-adaptivity.
  
  \item Despite that QoS interference has been found to be an important issue~\cite{fuzzy-vm-interference} \cite{qcloud} \cite{vm-fuzzy-MIMO} \cite{2014-decision-tree-software-CP-2014}, it is often overlooked in both the QoS modeling and decision making aspects of SSCAS. Therefore, we call for novel and effective approach to manage and mitigate QoS interference in cloud.

  \item Most studies attempt to scale hardware resources at the IaaS level only. However, a mature SSCAS should additionally consider the software configurations related to the cloud-based services and their interplay with the hardware resources, as found in recent studies~\cite{2013-JRAO-most-closest-work-2013} \cite{software-RP-two-loops}  \cite{2014-decision-tree-software-CP-2014}. 
  
  \item Instead of using fixed granularity of control (i.e., the boundary of decision making is on each service/application, VM/container, PM or cloud), future SSCAS should consider more flexible ones, e.g., dynamic and hybrid level, as discovered in recent studies~\cite{MIMO-fuzzy-hill-climbing-2013}~\cite{11icde_smartsla_full}~\cite{Chen:2014:seams}.

  \item The assumption of autoscaling bundles (e.g., the VM instances from Amazon EC2) has been made in a considerable amount of SSCAS studies. However, it is known that renting bundles cannot and does not reflect the interests of consumers and the actual demand of their cloud-based services~\cite{rule-control-elasticity-cloud}. Thus, considering arbitrary and custom combinations of configurations and resources is an inevitable trend in the cloud computing paradigm.
  
  \item Considering multi-objectivity in SSCAS is a must in order to create better diversity and possibly better trade-off quality without the needs of weights specification. In addition, how to achieve balanced trade-off over the set of non-dominated solutions is worth investigating~\cite{Chen:2015:tsc-pending}.
  
  \item More real world cases and scenarios are needed as this can be the only way to fully verify the potentials, effectiveness and impacts of SSCAS.

\end{itemize}

This article is structured as follows: Section~\ref{sec:back} introduces the background and challenges. Section~\ref{sec:related} compares our work with the other reviews. A taxonomy and findings of the survey, with respect to different aspects of a SSCAS, are presented in Section~\ref{sec:result}. Section~\ref{sec:diss} discusses on the findings and presents the open challenges we learned. Section~\ref{sec:con} presents the conclusion.

\section{Backgound}
\label{sec:back}

\subsection{Autoscaling System in Cloud} % V, H scaling
%
%In the early developments of the cloud computing paradigm, the resources are acquired and released manually, as requested by the cloud consumers. Such manual scaling can work for cloud-based services that experience stable workload, or if the workload exhibits strong seasonality. However, for unplanned and spiked loads, the manual scaling will soon become problematic as human analysis and decision can take a long time, and  the correctness of the scaling decision is difficult to be verified. As a result, automatic scaling, or simply autoscaling, is becoming a suitable solution for these scenarios.

%Depending on the given QoS attributes and the manageable cloud primitives, an autoscaling system can cover self-configuring, self-healing, self-optimising and self-protecting, or any combination of those, from the notion of self-adaptivity. A recent survey \cite{epics_survey} has established the evidences that self-awareness can improve a system that requires self-adaptivity, and therefore achieving self-awareness is a promising way to enable better autoscaling systems in cloud.

%Therefore, like many other domains \cite{epics_survey}, achieving self-awareness is a promising way to improve the self-adaptivity of autoscaling systems in cloud.

%* 78 has estalbilished the eveidance that self-awareness can improve system that requires self-adaptivity.  

\begin{figure}[t!]
\centering
\includegraphics[width=7.3cm]{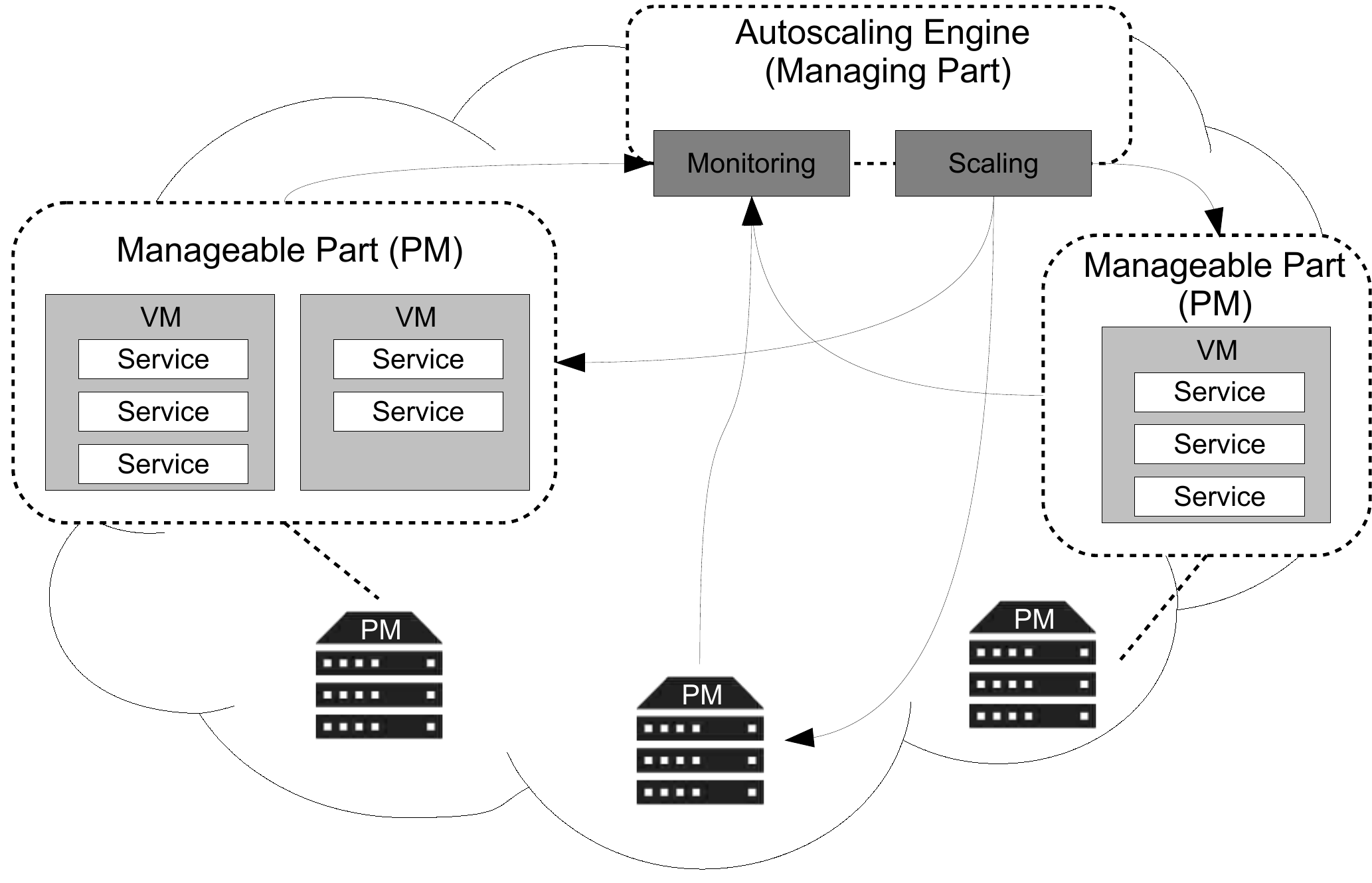}
\caption{\fontsize{8}{8}\selectfont The conceptual design of an autoscaling system. (Note that this figure represents the conceptual design of an autoscaling system in cloud. Practical deployment of the autoscaling engine can be either centralized, or decentralized where there are different engines, each of which is running on a PM.}
\label{ch1:simple_arch}
\vspace{-0.3cm}
\end{figure}

Given that it is almost impossible to access the low level details of cloud-based services (e.g., their codes and algorithms) at runtime, an autoscaling system often consist of two physical parts: a managing part containing the autoscaling engine and a manageable part encapsulating services and VMs/containers running in the cloud. The two physical parts are seamlessly and transparently connected for realizing the entire autoscaling process, known as external adaptation \cite{roadmap} \cite{landscape}.  

The external adaptation of an autoscaling system is shown in Figure \ref{ch1:simple_arch}. As we can see,  the core of an autoscaling system in the cloud is the autoscaling engine, which can consist of multiple logical aspects. A typical example of autoscaling system is a feedback loop that covers monitoring and scaling: the former gathers the service's or application's current state while the latter utilizes the information to decide an action after being analyzed and reasoned about by the autoscaling engine. Given the multi-tenant nature of cloud, cloud-based services often come with different QoS objectives, SLA requirements and budget constraints, etc. The ultimate goal of an autoscaling system is to adapt those cloud-based services, through scaling the related software configurations and hardware resources, in such a way that their objectives are continually optimized. To execute an autoscaling decision, the scaling actions could be vertical (scale up/down where changes occurs on a VM/container), horizontal (scale in/out that adds/removes other VMs/containers), or both.

%However, engineering autoscaling system faces with many challenges, including, e.g., dynamic and uncertainty caused by workload, conflicting objectives, the unforeseen QoS performance and heterogeneity of cloud-based services. Since the cloud assumes a sharing infrastructure, the autoscaling process should be aware of and be able to handle QoS interference~\cite{fuzzy-vm-interference} \cite{qcloud}. Otherwise improving the QoS performance of a service may likely to downgrade that of its neighboring services and VMs/containers, which negatively affects the overall quality of autoscaling and elasticity. 

%* put them as italic However, such a simplified form of autoscaling system tends to be limited, since it cannot effectively handle the increasing runtime complexity of the cloud environment,

%* summary the title, archtiecure, qos modeling, graunlrity and decison making
%* change the archtiecure to archtiecure pattern

%In the following sections, we provide survey and taxonomy for the most influent and recent work that is related to this paper. Particularly, we present the review and discussions based on the key logical aspects for autoscaling in the cloud, which are architectural pattern, QoS modeling, granularity of control and decision making. We then position this paper by discussing how our work differ from those existing approaches.

\subsection{Self-Adaptivity in Software Systems} % former has no general concept, the later has
\label{sec:s-adaptive}
The broad category of automatic and adaptive systems aim to deal with the dynamics that the system exhibited without human intervention; but this does not necessarily involve uncertainty, i.e., there are changes related to the system but it is easy to know when they would occur and the extent of these changes. Self-adaptivity, being a sub-category, is a particular capability of the system to handle both dynamics and uncertainty. Here, self-adaptive systems refer to the systems that are capable to adapt their behaviors according to the perception of the uncertain environment and its own state. To date, self-adaptivity in software systems remain an important and challenging research field~\cite{femosaa}\cite{2014-kalman-pair-wised-coupling-on-tiers-2014}\cite{icpechen}. According to the adaptive behaviors, self-adaptivity can be regarded as the following four properties, each of which covers a specific set of goals, as discussed by \cite{landscape}:

% the self-managing paper

\begin{itemize}[leftmargin=.35cm]

  \item \textbf{Self-configuring:} ``The capability of reconfiguring automatically and dynamically in response to change by installing, updating, integrating, and composing/decomposing software entities.''~\cite{landscape} 

  \item \textbf{Self-healing:} ``This is the capability of discovering, diagnosing, and reacting to disruptions. It can also anticipate potential problems, and accordingly take proper actions to prevent a failure.''~\cite{landscape}
    
    %Self-diagnosing refers to diagnosing errors, faults, and failures, while self-repairing focuses on recovery from them.

  \item \textbf{Self-optimizing:} ``This is also called self-tuning or self-adjusting, is the capability of managing performance and resource allocation in order to satisfy the requirements of different users, e.g., response time, throughput and utilization.''~\cite{landscape}

  \item \textbf{Self-protecting:} ``This is the capability of detecting security breaches and recovering from their effects. It has two aspects, namely defending the system against malicious attacks, and anticipating problems and taking actions to avoid them or to mitigate their effects.''~\cite{landscape}

\end{itemize}

% justify the small set of representitives set, or include a recent survey

%survey on self-adaptive projects
%There are numbers of research projects in the past decade, either for self-adaptive systems in general or for a specific application domain. In particular, there has been numbers of surveys [] on self-adaptive system in general, here, we discuss a few key developments in this area of research to serve as background information. Among the others, Rainbow [] is an architectural based framework for self-adaptive systems, it permits runtime adaptation based on explicit  knowledge about the application domain, as provided by the developers.  Rainbow covers  self-configuring, self-healing and self-optimizing.  Similarly, DEAS [] is a self-configuring framework that capable to identify objectives and reason about the optimization of those objectives for self-adaptive systems. MADAM [] is a self-configuring and self-optimizing system that assist the development of adaptive software for mobile computing domain, it allows adaptations to be reasoned at runtime. ML-IDS [] is a self-adaptive system specifically designed to detect network attack, and thus it primarily covers the self-protecting property.

\subsection{Self-Awareness in Software Systems} 
\label{sec:s-aware}

In contrast, self-awareness is about the capability of a system to acquire knowledge about its current state and the environment. Such knowledge permits better reasoning about the system's adaptive behaviors. Consequently, self-awareness is often seen as the lowest level of abstraction of self-adaptivity~\cite{landscape}, and thus it can improve the basic perceptions and self-adaptivity of a system \cite{2014_epics_handbook} \cite{7185305} \cite{epics_survey} \cite{Chen2016:book}. Inspired from the psychology domain, Becker et al.~\cite{epics} have classified self-awareness of a computing system into the following general capabilities (they have used node to represent any conceptual part of a system being managed):

\begin{itemize}[leftmargin=.35cm]

  \item \textbf{Stimulus-aware:} ``A node is stimulus-aware if it has knowledge of stimuli. The node is not
    able to distinguish between the sources of stimuli. It is a prerequisite for all other levels of self-awareness.''~\cite{epics}
%    It does not have
%    knowledge of past/future stimuli. It enables the ability in a node to
%    respond to events. It is a prerequisite for all other levels of self-awareness.

  \item \textbf{Interaction-aware:} ``A node is interaction-aware if it has knowledge that stimuli and its own
    actions form part of interactions with other nodes and the environment. It
    has knowledge via feedback loops that its actions can provoke, generate or
    cause specific reactions from the social or physical environment.''~\cite{epics}
%     It enables
%    a node to distinguish between other nodes and environments. Simple
%    interaction-awareness may just enable a node to reason about individual
%    interactions. More advanced interaction-awareness may involve the node
%    possessing knowledge of social structures such as communities or network
%    topology.

  \item \textbf{Time-aware:} ``A node is time-aware if it has knowledge of historical and/or likely future
    phenomena. Implementing time-awareness may involve the node possessing an
    explicit memory, capabilities of time series modeling and/or anticipation.''~\cite{epics}

  \item \textbf{Goal-aware:} ``A node is goal-aware if it has knowledge of current goals, objectives,
    preferences and constraints. It is important to note that there is a
    difference between a goal existing implicitly in the design of a node, and
    the node having knowledge of that goal in such a way that it can reason
    about it. The former does not describe goal-awareness; the latter does.''~\cite{epics}
%    Example implementations of such knowledge in a node include state based
%    goals (i.e. knowing what is a goal state and what is not) and utility based
%    goals (i.e. having a utility or objective function). Goal-awareness permits
%    acknowledgment of and adaptation to changes in goals. When coupled with
%    interaction-awareness or time-awareness, goal-awareness permits the ability
%    to reason about goals in relation to other nodes, or about likely future
%    goals, respectively.  

  \item \textbf{Meta-self-aware:} ``A node is meta-self-aware if it has knowledge of its own capability(ies) of
    awareness and the degree of complexity with which the capability(ies) are
    exercised. Such awareness permits a node to reason about the benefits and
    costs of maintaining a certain capability of awareness.'' ~\cite{epics}
%    It further allows the node to adapt the
%    way in which the level(s) of self-awareness are realized (e.g. by changing
%    algorithms realizing the level(s), thus changing the degree of complexity of
%    realization of the level(s)). As an example, this awareness may involve a
%    node being able to dynamically select a particular technique out of a set of
%    possibilities for realizing one or more levels, in order to meet or manage
%    trade-off between its goals or objectives. 

\end{itemize}

\section{Comparison to Related Surveys}
\label{sec:related}
Research on cloud autoscaling systems and the related topics have been reviewed in some other surveys. For example, Manvi and Shyam~\cite{manvi2014resource} present a review on resource management in the cloud, particularly at the IaaS level. They have provided a board survey on different issues related to managing cloud resources, e.g., resource adaptation, resource mapping and resource brokering etc. While resource management has some similarities to autoscaling, they lie in different levels of abstraction: the latter is more specific than the former. In other words, cloud autoscaling is one, but probably the most important part of the board cloud resource management. Another review from Mana~\cite{mann2015allocation} is explicitly concerned with VM to PM mapping problem which is also belong to the cloud resource management category, but is often regarded as a fundamentally different issue to cloud autoscaling. Ardagna et al.~\cite{ardagna2014quality} present a survey on QoS modeling and its application in the cloud. Indeed, QoS is the major concern for a cloud autoscaling system, but its management can be governed by various different approaches other than autoscaling, e.g., load balancing and admission control, which are also covered in~\cite{ardagna2014quality}. In contrast to the above, our survey has explicitly focused on automatically scaling software configuration and hardware provisioning in the cloud in order to change the capacity of cloud-based services to handle the dynamic workloads. 

Al-Dhuraibi et al.~\cite{al2017elasticity} present a review on elastic autoscaling approaches in the cloud, specifically focusing on the physical infrastructure level support, e.g., benchmarking and containerization techniques. Our survey, in contrast, is primarily concerned with the logical architecture and algorithmic level techniques for achieving different aspects of cloud autoscaling, e.g., modeling and decision making. Qu et al.~\cite{qu2016auto} also survey autoscaling approaches for a special type of cloud application, i.e., web applications, with a coarse correlation to self-adaptivity, e.g., if an approach is self-adaptive or not; while our survey is application agnostic and we present finer correlation of an approach to different levels of self-adaptivity, e.g., self-optimization. The most related survey from the literature is probably the one by Lorido-Botran et al.~\cite{lorido2014review}, in which different category of algorithmic level techniques for QoS modeling and decision making in cloud autoscaling are reviewed. However, their survey differs from ours in the following three aspects: (i) they have not provided a comprehensive taxonomy on the cloud autoscaling problem; such a taxonomy (i.e., modeling, architecture, granularity and decision making), which we will present in the next section, is important as it clearly state the open problems and challenges related to different aspects of the cloud autoscaling domain, providing better clarifications and clearer directions for researchers on this research field. (ii) In addition, Lorido-Botran et al.~\cite{lorido2014review} did not explicitly link the cloud autoscaling systems to different levels of self-awareness and self-adaptivity, which is one of the key contributions of our survey. (iii) Finally, we discuss the open problems and challenges of cloud autoscaling systems in a broader fashion.

In summary, our survey differs from the other similar reviews in the following:
\begin{itemize}[leftmargin=.35cm]
\item[--] We present a focused survey on the logical architecture and algorithmic level techniques for cloud autoscaling which are application agnostic.
\item[--] We explicitly correlate the reviewed approaches with different levels of self-awareness, self-adaptivity and the required knowledge in a fine grained manner.
\item[--] We provide a clarified taxonomy that covers different logical aspects for engineering cloud autoscaling systems, and classify every study accordingly.
\item[--] We discuss the open problems and challenges of cloud autoscaling systems in a broader fashion.
\end{itemize}

\section{Taxonomy and Survey Results for SSCAS}
\label{sec:result}
In this section, we present a taxonomy and survey results for the state-of-the-art SSCAS research obtained from our review process. 

%Discussion on the observations of existing research and the remaining open challenges for SSCAS will be presented in Section~\ref{sec:diss}.

\subsection{Review Process and Research Questions}
\label{sec:rq}
The review is intended to create a broad scope to cover the landscape of SSCAS research. Particularly, the following research questions serve as the main drivers of this review:

\begin{itemize}[leftmargin=.35cm]

\item \emph{RQ1:} What are the levels of self-awareness and self-adaptivity that have been captured in SSCAS? 
\item \emph{RQ2:} What are the architectural patterns used for engineering SSCAS? 
\item \emph{RQ3:} What are the approaches used to model the quality related to SSCAS? 
\item \emph{RQ4:} What is the granularity of control in SSCAS? 
\item \emph{RQ5:} What are the approaches used for decision making in SSCAS?

\end{itemize}

The following prominent indexing services were used during the review: IEEE Xplore, ACM Digital Library, Science Direct, ISI Web of Knowledge, and Google Scholar. The search term was "Cloud computing" AND "Autoscaling" AND ("QoS modeling" OR "Performance modeling" OR "decision making" OR "optimization" OR "Architecture" OR "Interference" OR "Resource allocation" ). After applying inclusion (e.g., considering only journal, conference, and workshop papers) and exclusion criteria (e.g., removing duplicate entries and considering only the extended version) to the initial search result, the review has ended with the total of 109 studies.

\subsection{A Taxonomy of SSCAS Research }

The overall taxonomy, concluded from the extracted studies, is given in Figure~\ref{fig:tax}. As we can see, current research on SSCAS often require sophisticated designs in different highest leveled logical aspects of the autoscaling engine, which we have classified and discussed as the following:

\begin{figure}[t!]
\centering
\includegraphics[width=7.3cm]{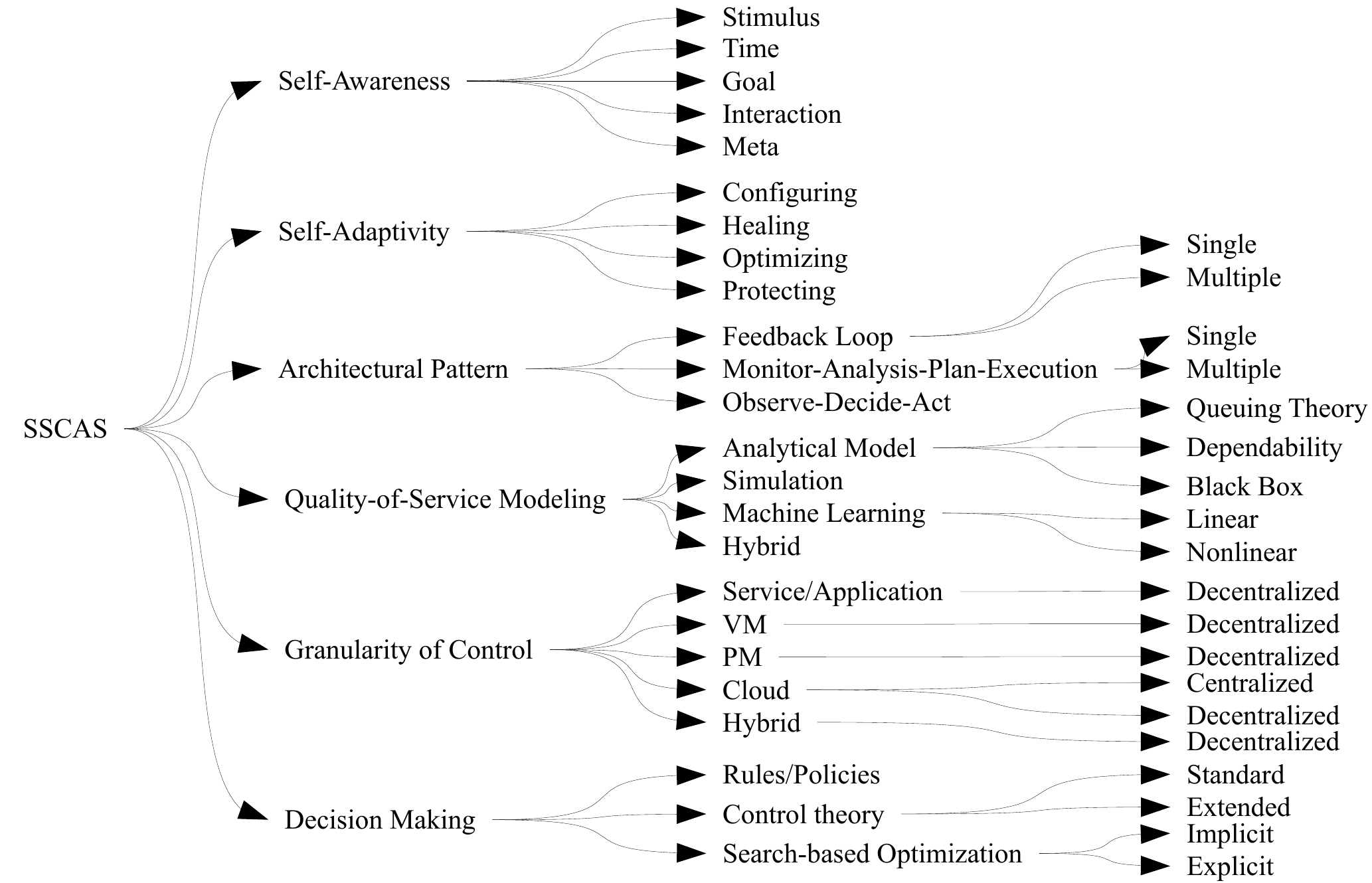}
\caption{\fontsize{8}{8}\selectfont A taxonomy of SSCAS research.}
\label{fig:tax}
\end{figure}

\begin{itemize}[leftmargin=.35cm]

\item[--] \emph{\textbf{Self-Awareness:}} This is concerned with the ability to acquire and maintain knowledge about the system's own states and the environment, as specified in Section~\ref{sec:s-adaptive}. The key challenges here are which level(s) of knowledge is required for a SSCAS, what does it means for certain level in the problem context, (e.g., what does interaction refers to?) and what is the representation for different levels of knowledge, e.g., how do we represent goals in the SSCAS?

\item[--] \emph{\textbf{Self-Adaptivity:}} This is about the ability to change the system's own behavior with specific goals in mind, as specified in Section~\ref{sec:s-aware}. Often, the required levels of self-adaptivity depending on the requirements, but they could be also related to the specific levels of knowledge that the SSCAS is able to capture, e.g., the SSCAS has to be goal-aware to achieve self-optimization.

\item[--] \emph{\textbf{Architectural Pattern:}} Autoscaling architecture is the most essential element of SSCAS. It describes the structure of the autoscaling process, the interaction between components and the modularization of the other important logical aspects in autoscaling. The challenge of architecting SSCAS is concerned with how to systematically capture different logical aspects (e.g., decision making) of SSCAS using a given architectural pattern. More importantly, how to encapsulate these aspects and the algorithms that realizes them into different components of the pattern. 
%The modeling is concerned with the model of QoS, environment conditions (e.g., workload) and demand of the control knobs (e.g., software configurations and hardware resources). T
\item[--] \emph{\textbf{QoS Modeling:}} While modeling the cost incurred by cloud-based services is straightforward, modeling the QoS is often much more complex and challenging. Here, the QoS modeling is concerned with the sensitivity of QoS with respect to the environment conditions (e.g., workload) and the control knobs (e.g., software configurations and hardware resources). The resulting model is a powerful tool to assist the autoscaling decision making process. Without loss of generality, in this article, we use \textbf{\emph{cloud primitives}} to refer to both control knobs and environmental conditions in the cloud. In particular, we further decompose the notion of primitives into two categories, termed \textit{\textbf{control primitive}} and \textit{\textbf{environmental primitive}}. Control primitives refer to the internal control knobs that can be either software or hardware. They are the fundamental features that can be controlled by the cloud providers to support QoS. Specifically, software control primitives are the key cloud configurations at the software level, e.g., the number of threads in the thread pool, the buffer size and the cache size, etc. In contrast, hardware control primitives refer to the computational resources, such as CPU, memory and bandwidth, etc. Typically, software control primitives exist on the PaaS layer while the hardware control primitives present on the IaaS layer. It is worth noting that considering software control primitives when autoscaling in the cloud is a non-trivial task, as they have been proved to be important features that can significantly influence the QoS \cite{2013-JRAO-most-closest-work-2013} \cite{software-RP-two-loops}  \cite{2014-decision-tree-software-CP-2014}. The environmental primitives, on the other hand, refer to those external stimuli that is uncontrollable but can cause dynamics and uncertainties in the cloud. These, for example, can be the workload, incoming data, node failure, etc. The above examples of primitives listed above are not exhaustive, Ghanbari et al. \cite{rule-control-elasticity-cloud} have provided a more completed and detailed list of the possible control primitives in cloud.

The challenges of QoS modeling include: (i) which primitives should be selected as model's input features; (ii) how does the QoS change in conjunction with those primitives; (iii) how to incorporate the information of QoS interference into the model; (iv) whether the model is built offline or online; (v) and whether the model is dynamic, semi-dynamic or static.

\item[--] \emph{\textbf{Granularity of control:}} Determining the granularity of control in the autoscaling engine is essential to ensure the benefits (e.g, QoS and cost objectives) for all cloud-based service. It is concerned with understanding whether certain objectives can be considered in isolation with some of the others, i.e., the boundary of decision making. This is because \textbf{\emph{objective-dependency}} (i.e., conflicted or harmonic objectives) often exist in the decisions making process, which implies that the overall quality of autoscaling can be significantly affected by the inclusion of conflicted or harmonic objectives when making decision, hence rendering it as a complex task. This is especially important for the shared cloud infrastructure where objective-dependency exists for both intra- and inter-services. That is to say,  objective-dependency is not only caused by the nature of objectives (intra-service), e.g, throughput and cost objective of a service; but also by the QoS interference (inter-services) due to the co-located services on a VM/container and co-hosted VMs/containers on a PM   \cite{2013-JRAO-most-closest-work-2013} \cite{software-RP-two-loops}  \cite{2014-decision-tree-software-CP-2014} \cite{qcloud}.  

Here, the challenges are which granularity of control to use, what is the basic entity to control (e.g., application or VM), and whether the control is in a centralized or decentralized manner.

\item[--] \emph{\textbf{Decision making:}} The final logical aspect in autoscaling logic is the dynamic decision making process that produces the optimal (or near-optimal) decision, which consists of the newly configured values of the related control primitives, for all the related objectives. In the presence of objective dependency, autoscaling decision making requires to resolve complex trade-offs, subject to the SLA and budget requirements. The trade-off decision can be then executed using either vertical (scale up/down) and/or horizontal scaling actions (scale in/out), which adapt the cloud-based services and/or VMs/containers correspondingly.

The challenges of decision making in SSCAS include: (i) how to reason about and search for the effected adaptation decisions; (ii) what are the objectives, their representations and conflicting relations, if any; (iii) and which are the control primitives to tune.

\end{itemize}
In the following, we present our detailed findings in regards to the taxonomy of SSCAS.

\subsection{The Levels of Self-Awareness and Self-Adaptivity in Cloud Autoscaling Systems}

\label{sec:sscas}

  \begin{figure*}[!t]
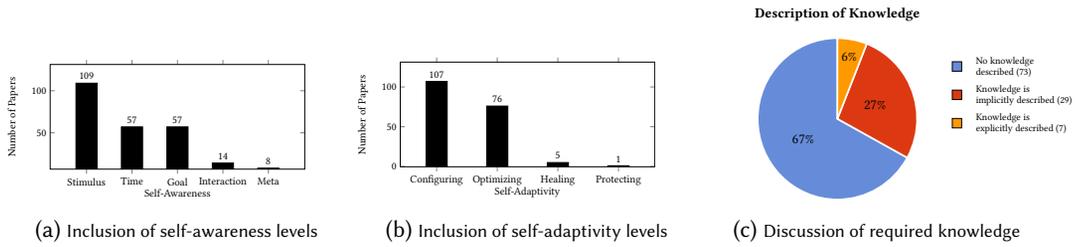

  \label{fig:sscas}
\centering
%\resizebox {\columnwidth} {3cm} {
 \begin{subfigure}[t]{0.33\textwidth}
    \includestandalone[width=4cm]{tikz/self-awareness}%     without .tex extension
    \subcaption{\fontsize{7}{8}\selectfont Inclusion of self-awareness levels}
     \label{fig:self-aw}
    \end{subfigure}
    ~
%\resizebox {\columnwidth} {3cm} {
 \begin{subfigure}[t]{0.33\textwidth}
  \includestandalone[width=4cm]{tikz/self-adaptivity}%     without .tex extension
   \subcaption{\fontsize{7}{8}\selectfont Inclusion of self-adaptivity levels}
     \label{fig:self-ad}
    \end{subfigure}
       ~  
     \begin{subfigure}[t]{0.33\textwidth}
    \includestandalone[width=5cm]{tikz/knowledge}%     without .tex extension   
   \subcaption{\fontsize{7}{8}\selectfont Discussion of required knowledge}
    \label{fig:knowledge}
    \end{subfigure}
  %}
  % or use \input{mytikz}
  \caption{\fontsize{8}{8}\selectfont Paper count on self-awareness, self-adaptivity levels and how knowledge is discussed on the SSCAS architecture.}
 
\vspace{-0.3cm}
  \end{figure*}

%
%   \begin{figure}[t!]
%\centering
%%\resizebox {\columnwidth} {3cm} {
%  \includestandalone[width=7cm]{tikz/knowledge}%     without .tex extension
%  %}
%  % or use \input{mytikz}
%  % \vspace{-0.3cm}
%  \caption{\fontsize{8}{8}\selectfont The extent to which knowledge in SSCAS architecture is discussed.}
%  \label{fig:knowledge}
%    %\vspace{-0.7cm}
%  \end{figure}

\noindent \emph{RQ1: What are the levels of self-awareness and self-adaptivity that have been captured in SSCAS? }

It is worth noting that not all the studies have explicitly declared which levels of self-awareness/self-adaptivity that they have taken into account, therefore we identified this by looking at the studies in details with respect to the definitions of self-awareness/self-adaptivity. From Figure~\ref{fig:self-aw}, we can see that \emph{stimulus-awareness}, which is the fundamental level in self-awareness, has been considered in all the studies. The \emph{time-} and \emph{goal-awareness} have attracted relativity similar amount of attention. However, \emph{interaction-} and \emph{meta-awareness} has not been widely studied in recent SSCAS research. In Figure~\ref{fig:self-ad}, we see that \emph{self-configuring} and \emph{-optimizing} have been predominately captured in the studies, whereas \emph{self-healing} and \emph{-protecting} receive little attentions. In particular, we found no study that explicit aims for \emph{self-protecting} in SSCAS. Figure~\ref{fig:knowledge} illustrates whether the required knowledge representations at the SSCAS architecture level, e.g., knowledge of goal is required in the architecture, have been discussed, implicitly discussed or explicitly discussed in the studies. We can see that the majority of the studies surveyed do not attempt to declare what knowledge is required in SSCAS architecture, leaving only 33\% of the studies have discussed the knowledge implicitly or explicitly.

%which have to a degree relevant to self-awareness and/or self-adaptivity of autoscaling system,  

\subsection{Architectural Pattern}
\label{sec:arch}

\noindent \emph{RQ2: What are the architectural patterns used for engineering SSCAS? }

We classified the predominantly applied architectural patterns for SSCAS into three categories based on their basic form; these are \emph{Feedback Loop} \cite{Brun:2009}, \emph{Observe-Decide-Act} (ODA) \cite{self-aware-ML-adaptive-control} and \emph{Monitor-Analysis-Plan-Execute} (MAPE) \cite{ibm}. 

From Figure~\ref{fig:arch}, we see that the generic feedback loop has been the predominant architecture pattern in SSCAS, following by the MAPE pattern. Particularly, as we can observe form Table~\ref{tb:arch}, single and close feedback loop are widely exploited in SSCAS. In the following, we specify some representative studies under each category in details.

\subsubsection{Feedback Loop}

Feedback loop is the most general architectural pattern for controlling self-adaptive systems, including the autoscaling systems. It is usually a closed-form loop made up of the managing system itself and the path transmitting its origin (e.g., a sensor) to its destination (e.g., an actuator). Here, we further divide the pattern as \emph{single} or \emph{multiple} loops:

\begin{itemize}[listparindent=1.5em,leftmargin=.35cm]
\item \emph{4.4.1.1 \: Single Loop:} Single loop is the simplest, yet the most commonly used pattern for SSCAS due to its flexibility.  The most common practice with single loop is to build a closed feedback control where the core is the decision making component and an optional QoS modeling component, e.g.,  Ferretti et al. \cite{05557978} , CloudOpt \cite{fine-grained-servce-LQM-MIP-power}, SmartSLA \cite{11icde_smartsla_full} and CLOUDFARM \cite{2014-linear-centralized-decision-making-2014}, etc. Some other studies have included an additional component for workload or demand prediction based on either offline profiling, e.g., Jiang et al. \cite{typical-navie-scaling-VM-number} and Fernandez et al. \cite{2014-profiling-decision-tree-scaling-2014}, or online learning, e.g., Kingfisher \cite{2011-cost-aware-v-h-scaling-2011} and PRESS \cite{signal-resource-trend-prediction}.

Open feedback exists for single loop, as presented in Cloudine \cite{2013-elasticity-primitives-2013}, where the scaling actions are partially triggered by user requests. In particular, they use a centralized \textit{Resource and Execution Manager} to handle all the scaling actions.  Apart from the general autoscaling architecture, other efforts are particularly designed upon specific cloud providers. For example, Zhang et al. \cite{MPC-price} as well as Kabir and Chiu \cite{cache-static-ANN-bi-obj-2012} propose to use a simple feedback loop for architecting autoscaling system, which is heavily tied to the properties of Amazon EC2.

  \begin{figure}[t!]
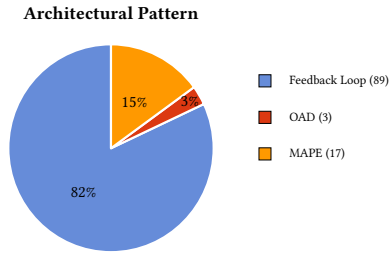

\centering
%\resizebox {\columnwidth} {3cm} {
  \includestandalone[width=6cm]{tikz/arch-pie}%     without .tex extension
  %}
  % or use \input{mytikz}
  % \vspace{-0.3cm}
  \caption{\fontsize{8}{8}\selectfont Number of papers per architectural pattern.}
  \label{fig:arch}
    %\vspace{-0.7cm}
  \end{figure}

\begin{table}[t!]

\caption{\fontsize{8}{8}\selectfont Detailed Classifications of Architectural Patterns for SSCAS.}
\label{tb:arch}
%\vspace{0.3cm}

\scriptsize
\begin{tabularx}{\textwidth}{|P{2.4cm}|P{0.8cm}|P{1.1cm}|X|} \hline
\textbf{\emph{Architectural Pattern}}&\textbf{\emph{Style}}&\textbf{\emph{Open/Close}}&\textbf{\emph{Representative Examples}}\\ \hline
\multirow{3}{*}{Feedback loop}&\multirow{2}{*}{single}&close&\cite{05557978}, \cite{fine-grained-servce-LQM-MIP-power},\cite{11icde_smartsla_full}, \cite{HPL-2008-123R1-mimo}, \cite{2014-eplison-GA-weigh-h-scaling-2014}, \cite{06119056}, \cite{2014-linear-centralized-decision-making-2014}, \cite{2014-kalman+rule-based-2014}, \cite{typical-navie-scaling-VM-number}, \cite{2014-profiling-decision-tree-scaling-2014}, \cite{2011-cost-aware-v-h-scaling-2011}, \cite{kriging-controller}, \cite{2013-self-organising-map-2013}, \cite{signal-resource-trend-prediction},  \cite{2014-fuzzy-compare-to-JRao-2014}, \cite{2013-software-CP-only-2013},\cite{acdc09}, \cite{compare-to-Rao-fuzzy-2013}, \cite{full-simulation-model}, \cite{GA-full-simulation} \cite{CloudSim}, \cite{CloudAnalysis}, \cite{DCSim}, \cite{sla-provision}, \cite{parallel-RL-vertical-QoS-2013}, \cite{MASCOTS11}, \cite{E3-R-extended}, \cite{wosp10sla}, \cite{queue-VM-group}, \cite{3-stages-game-theory}, \cite{2013-rule-based-multi-elasiticity-2013}, \cite{linear-mapping-CP-bundles-2012}, \cite{qcloud}, \cite{vm-fuzzy-MIMO}, \cite{2014-decision-tree-software-CP-2014}, \cite{dynamic-model-comparison}, \cite{change-point}, \cite{2013-single-learner-filter-wrapper-LR-2013}, \cite{kalman-AR}, \cite{kalman-clustering}, \cite{ILP-cost-only-scaling-2013}, \cite{2014-ARMA-single-agg-objective-2014}, \cite{scale-rule-based}, \cite{IWQOS11},  \cite{2013-scaling-select-one-predictor-and-error-correlation-2013}, \cite{2010-demand-pattern-match-2010}, \cite{ensemble-prediction-VM-2014-full}, \cite{seelam2015polyglot}, \cite{souza2015using}, \cite{farokhi2016hybrid}, \cite{shariffdeen2016workload}, \cite{gandhi2017model}, \cite{qu2016reliable}, \cite{sun2016roar}, \cite{da2016autoelastic}, \cite{rameshan2016augmenting}, \cite{baresi2016discrete}, \cite{zhang2016container}, \cite{li2015rest}, \cite{sun2017automated}, \cite{awada2017improving} \\ \cline{3-4}

&&open& \cite{2013-elasticity-primitives-2013}, \cite{MPC-price}, \cite{cache-static-ANN-bi-obj-2012} \\ \cline{2-4}

&multiple&close&\cite{2014-kalman-pair-wised-coupling-on-tiers-2014}, \cite{2012-app-VM-mapping-2012}, \cite{EMA-CPU-memory-PD}, \cite{MIMO-fuzzy-hill-climbing-2013}, \cite{Arc-cover-all-controller}, \cite{ICDCS2011},  \cite{PCA-model-scaling-2013}, \cite{2014-2-SVM-workload-type-2014}, \cite{06032254}, \cite{fuzzy-2-loop-control},  \cite{MASCOTS11-bu-software-CP-full} \cite{2013-JRAO-most-closest-work-2013}, \cite{TR-10-full-version}, \cite{Chen:2013:iccs}, \cite{Chen:2015:tsc-pending}, \cite{Chen:2015:computer}, \cite{multitier-resalloc-Cloud11}, \cite{profit-cent-local-search}, \cite{2014-navie-ANN-GA-2014}, \cite{JSS_Kousiouris}, \cite{Chen:2013:seams}, \cite{MOCS2014-full},  \cite{Chen:2014:ucc}, \cite{Chen:2015:tse-pending}, \cite{ma2016auto}  \\ \hline

OAD&single&close&\cite{van2015mnemos},\cite{herbein2016resource}, \cite{HuBrKo2011-SEAMS-ResAlloc} \\ \hline

\multirow{2}{*}{MAPE}&single&close&\cite{li-opt-clouds-4}, \cite{cloud_computing_2010_5_20_50060-ext}, \cite{Compsac_2010_I_Brandic},  \cite{SEASS_2010_Michael_Maurer}, \cite{tse1}, \cite{fuzzy-vm-interference}, \cite{queue-prediction}, \cite{Emeakaroha_CloudComp2010},  \cite{HPCS_IWCMC_Vincent}, \cite{self-sla-and-resource}, \cite{megahed2017stochastic}, \cite{lakew2017kpi}, \cite{aslanpour2017auto}, \cite{gill2017chopper}  \\ \cline{2-4}

&multiple&close&\cite{software-RP-two-loops}, \cite{BRGA-resource},\cite{Chen:2014:seams} \\

\hline

\end{tabularx}
\vspace{-0.3cm}
\end{table}

\item \emph{4.4.1.2 \: Multiple Loops:} It is possible to use multiple loops and controllers for autoscaling in the cloud. Here,  multiple feedback loops operate in different levels of the architecture, e.g., one operates at the cloud level while the others operate on each VM. The benefit is that multiple loops provide low coupling in the design of the loops for SSCAS. Notably, multiple loop control can be used to separate global and local controls. Among others, Kalyvianaki et al.~\cite{2014-kalman-pair-wised-coupling-on-tiers-2014} apply multiple decentralized feedback loops for autoscaling CPU in the cloud. Although it aims to exploit one loop per individual application, the controllers actually operate on each tier of an application. Chen and Bahsoon~\cite{Chen:2015:tsc-pending} also leverage multiple feedback loop to auto-scale cloud services, where each PM maintains a loop. Unlike classic feedback loop where the adaptations occur only on the manageable part of SSCAS, their adaptations also happen on the manging part.
% Different controllers do not need to interact with each others, however.

%ARUVE \cite{2012-app-VM-mapping-2012} utilizes a global controller in conjunction with the local controller to form multiple feedback loops. The local controllers are decentralized on each PM while the global controller is centralized. 
%This is also the same for CRAMP \cite{EMA-CPU-memory-PD} and \cite{MIMO-fuzzy-hill-climbing-2013} where the later has additionally relied on a centralized fuzzy controller for control the entire cloud at the global level. 

\indent Multiple loop control is also effective for isolating the logical aspects of autoscaling and management in the cloud. For instance, Wang, Xu and Zhao \cite{fuzzy-2-loop-control} propose a two layered feedback control for autoscaling in the cloud. The first layer, termed guest-to-host optimization, controls the hardware resources, e.g., CPU and memory. Subsequently, the host-to-guest optimization adapts the software configuration accordingly. 

\end{itemize}

\subsubsection{Observe-Decide-Act}

Observe-Decide-Act (ODA) loop \cite{self-aware-ML-adaptive-control} is considered as an extended pattern of the generic feedback loop. As specified in the SEEC framework \cite{self-aware-ML-adaptive-control}, ODA is unique in the sense that it decouples multiple loops to different roles (i.e., application developer, system developer, and the SEEC runtime decision infrastructure) in the development life-cycle, each role focuses on one or more steps in ODA. In such a way, ODA links the effects of human activities on the adaptive behaviors.

%A unique \emph{Decide} component separates it from the generic feedback loop, since it explicitly requires to perform reasoning about the effects of adaptation on the system's goals and objectives. 

%While an ODA loop is most commonly applied for self-adaptive systems in general, only few work (e.g., \cite{HuBrKo2011-SEAMS-ResAlloc}) has included it for the design of autoscaling system in the cloud. This is because one important aspect of ODA is to define the effects of human activities on the adaptive behaviours, which is a difficult practice for autoscaling in the cloud. 

%* why the OAD is rarely used?

%SEEC \cite{self-aware-ML-adaptive-control}, being the very first work to introduce ODA, is a general framework for self-aware and self-adaptive systems. It applies ODA for decoupling the loops to different roles (i.e., application developer, system developer, and the SEEC runtime decision infrastructure) in the development life-cycle, each role focuses  on one or more steps in ODA. \cite{5694245} adopt an ODA loop to manage FPGA-based systems, where the decision and the its translation to actions are conducted by an incorporation of the \emph{Decide} and \emph{Act} steps. Bolchini et al. \cite{6604228} have used ODA to realise the adaptation for self-adaptive systems  because of its simplicity. It observes high-level and raw data from the \emph{Observe} step, such data is then used by \emph{Decide} to know which parts of the system to reason on, and finally, actions are taken depends on the characteristic of the system being managed. 

Among others application of OAD in SSCAS, MNEMOS~\cite{van2015mnemos} has relied on OAD to realize an integrated, datacenter-wide architecture for autoscaling resources in the cloud, in which the \emph{System Monitor} acts as the observer, the \emph{Portfolio Scheduler} acts as the decider, and the \emph{VM Manager} acts as the executioner. Huber et al. \cite{HuBrKo2011-SEAMS-ResAlloc} also use ODA for self-aware autoscaling resources in the cloud. However, unlike traditional ODA loop, it has an additional \emph{Analysis} step which is used to detect the type of problems that trigger adaptation.

%\cite{herbein2016resource} have also adopted OAD pattern to self-adapt the cloud environment which an aim to allocate CPU resource, bundled by different dockers, for high performance application in the cloud. 
%* revise this sentences

%\item \emph{\textbf{Monitor-Analyze-Plan-Execute}}
\subsubsection{Monitor-Analyze-Plan-Execute}

Another pattern extended from the generic feedback loop, namely Monitor-Analyze-Plan-Execute (MAPE), is firstly proposed by IBM for architecting self-adaptive systems. In such pattern, the \emph{Decide} step in OAD is further divided into two substeps, these are \emph{Analyze} and \emph{Plan}, where the former is particularly designed to determine the causes for adaptations, e.g., SLA violation; the latter, on the other hand, is responsible for reasoning about the possible actions for adaptation. MAPE sometime can be extended by a Knowledge component (a.k.a. MAPE-K) which maintains historical data and knowledge used by the system for better adaptation. MAPE can be also realized as either \emph{single} or \emph{multiple} loops:

%* remove fusion, put this as an additional framework

%MAPE (or MAPE-K) is commonly used for self-adaptive systems. For example, FUSION \cite{2010-FUSION-2010} is a framework that designed to self-adaptive systems in general, with an aim to optimize QoS. It relies on MAPE-K where Monitor and Analyze components belong to the learning cycle, in which the QoS models are built. The Plan and Execute components are within the adaptation cycle where the optimization is performed. The additional knowledge component is used to store historical data for learning. 

\begin{itemize}[listparindent=1.5em,leftmargin=.35cm]

\item \emph{4.4.3.1 \: Single Loop:} MAPE (or MAPE-K) is also widely applied for SSCAS. For example, the architecture of the FoSII project \cite{Compsac_2010_I_Brandic} \cite{SEASS_2010_Michael_Maurer} leverages single MAPE-K to realize the self-management interface, aiming to prevent SLA violations in cloud by devising the related actions. They also use the additional \emph{Knowledge} (K) component to record cases and the related solutions, which can assist the autoscaling decision making. Chen and Bahsoon~\cite{Chen:2014:seams} have realized MAPE as a single loop where the adaptations occur on both the managing and manageable parts of the SSCAS.

%QoSMOS \cite{tse1} is designed for service-based systems rather than for cloud specific autoscaling, however, it contains many aspects similar to that of autoscaling in the cloud. To achieve continuous adaptation, it  applies MAPE with the focuses on analytical QoS modeling  and optimization of resource allocation. APPLEware \cite{fuzzy-vm-interference} is an autoscaling framework which leverages MAPE. In their architecture, the \emph{Analyze} component model the QoS while the \emph{Plan} component conducts optimization process for autoscaling.

%Li et al. \cite{li-opt-clouds-4} have treated the analysis and plan component are combined as a optimization process. Chazalet et al. \cite{cloud_computing_2010_5_20_50060-ext} also propose to use MAPE-K for managing autoscaling in the cloud. In particular, the Analyze step has been enriched with SLA driven analysis, where the concerns have been on the extent of violation and frequency of violation. 

\item \emph{4.4.3.2 \: Multiple Loops:} Realizing multiple MAPE loops for SSCAS is also possible. Zhang et al. \cite{software-RP-two-loops} introduce an architecture for autoscaling using two nested MAPE loops. The first loop is responsible for adapting the software primitives while the other loop is used to change the hardware primitives. These two loops run sequentially upon autoscaling, that is, adapting the software control primitives before changing the hardware control primitives. Similarly, BRGA \cite{BRGA-resource} utilizes MAPE to realize a framework for autoscaling in the cloud. Such solution consists of both the local and global view of the cloud-based application. 
%In particular, the  \emph{Monitor} and  \emph{Execute} phase maintain the global view whereas the  \emph{Analyze} and  \emph{Plan} phase manage the local view on each PM. The authors claim that such an approach can achieve good global quality with reasonable management overhead.
\end{itemize}

%\end{itemize}

%* the unique representation of knowledge

%Particularly in the surveyed work on cloud autoscaling, the Knowledge is usually described as a standalone component at the coarse grained. 

%\subsection{Self-Awareness}
%no explicit self-awareness architecture 

%* qos modeling may or may not be explicit for cloud autoscaling. Fundemtal research on qos modeling exists, they can be applied for autoscling or other directions.

\subsection{QoS Modeling}
\label{sec:qos}
\noindent \emph{RQ3: What are the approaches used to model the quality related to SSCAS?}

QoS modeling, or performance modeling, is a fundamental research theme in cloud computing and it can serve as useful foundations for addressing many research problems in the cloud \cite{lorido2014review}, including autoscaling. The QoS models correlate the QoS attributes to various control primitives and environmental primitives. Clearly, these models are particularly important for SSCAS, since they are powerful tools that can assist the reasoning about the effects of adaptation on objectives in the autoscaling decision making process. Note that although QoS model can provide great helps to the decision making in SSCAS, not all of the studies have considered QoS modeling as part of their solutions. In fact, some of them rely on model-free solution, e.g., control theoretic approach, which we will review in Section~\ref{sec:dm}. 

  \begin{figure}[b!]
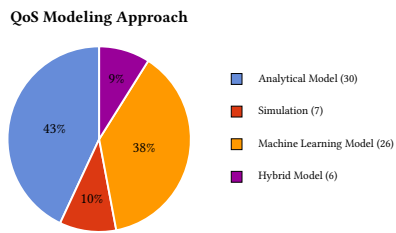

\centering
%\resizebox {\columnwidth} {3cm} {
  \includestandalone[width=6cm]{tikz/modeling-pie}%     without .tex extension
  %}
  % or use \input{mytikz}
  % \vspace{-0.3cm}
  \caption{\fontsize{8}{8}\selectfont Number of papers per QoS modeling approach.}
  \label{fig:qos}
  \vspace{-0.3cm}
  \end{figure}

Typically, QoS modeling consists of two phases: the primitives selection phase and the QoS model construction phase. More precisely, the primitives selection phase determines \emph{which} and \emph{when} the cloud primitives correlate with the QoS; while the QoS model construction phase identifies \emph{how} these primitives correlate with the QoS, i.e., their magnitudes in the correlation. The QoS models might come as three forms: (i) static models where the models' expression and their structure (e.g., the number of inputs and the coefficients) do not change over time; (ii) dynamic models which permits those changes; or (iii) semi-dynamic models in the sense that the expression (e.g., coefficients) could be dynamically updated but the input features do not. Further, those models can be built online at system runtime, or offline at the design phase of the system. In this section, we classified the studies mainly based on the modeling methods applied to the QoS model construction phase, since we found that the primitives selection is often conduced using manual and static approaches in the studies. As we can see in Figure~\ref{fig:qos}, the majority of the studies has exploited analytical model (43\%) and machine learning model (38\%) to predict QoS. In contrast, simulation and hybrid model receives much less attention. From Table~\ref{tb:qos}, we can obtain the following observations:

\begin{table}[t!]

\caption{\fontsize{8}{8}\selectfont Detailed Classifications of QoS Modeling Approaches for SSCAS. (S-QUEUE=Single Queue; M-QUEUE=Multiple Queues; LQN=Layered Queuing Network; MDP=Markov Decision Process; PCM=Palladio Component Model; PCA=Principal Component Analysis; LR=Linear Regression; ARMA=Auto-Regressive Moving-Average; KF=Kalman Filter; MIMO=Multiple-Input and Multiple-Output; KM=Kriging Model; RT=Regression Tree; ANN=Artificial Neural Network; SVM=Support Vector Machine; KNN=k-Nearest-Neighbor)}
\label{tb:qos}
%\vspace{0.3cm}
\centering
\scriptsize
\begin{tabularx}{\textwidth}{|P{1.3cm}|P{1.7cm}|P{0.7cm}|P{1.3cm}|P{1.2cm}|X|} \hline
\textbf{\emph{Modeling Approach}}&\textbf{\emph{Type}}&\textbf{\emph{Built}}&\textbf{\emph{QoS Interference}}&\textbf{\emph{State}}&\textbf{\emph{Concrete Models}}\\ \hline

\multirow{5}{*}{\parbox{1.3cm}{ \centering  Analytical model  }}& Queuing model&offline&no&static& \textbf{S-QUEUE (5):} \cite{MPC-price}, \cite{06119056}, \cite{E3-R-extended}, \cite{queue-VM-group}, \cite{typical-navie-scaling-VM-number} \newline \textbf{M-QUEUES (3):} \cite{multitier-resalloc-Cloud11}, \cite{queue-prediction}, \cite{wosp10sla} \newline \textbf{LQN (4):} \cite{3-stages-game-theory}, \cite{fine-grained-servce-LQM-MIP-power}, \cite{li-opt-clouds-4}, \cite{sla-provision}\\
\cline{2-6}
 
&\multirow{3}{*}[-0.05cm]{\parbox{1.7cm}{\centering Dependability model}}&online&no&semi& \textbf{MDP (1):} \cite{tse1} \newline \textbf{MODEL@RUNTIME (1):} \cite{2014-eplison-GA-weigh-h-scaling-2014} \\
\cline{3-6}
&&offline&no&static& \textbf{PCM (1):} \cite{HuBrKo2011-SEAMS-ResAlloc} \newline \textbf{GRAPH (1):} \cite{2013-rule-based-multi-elasiticity-2013}\\
\cline{3-6}
&&offline&no&dynamic&\textbf{PCA (1):} \cite{PCA-model-scaling-2013}\\ \cline{2-6}
&\multirow{2}{*}{Black box model}&offline&no&static& \textbf{EMPIRICAL-MODEL (9):} \cite{06032254}, \cite{Compsac_2010_I_Brandic}, \cite{2014-linear-centralized-decision-making-2014}, \cite{Emeakaroha_CloudComp2010},  \cite{HPCS_IWCMC_Vincent}, \cite{profit-cent-local-search}, \cite{cache-static-ANN-bi-obj-2012}, \cite{BRGA-resource}, \cite{shariffdeen2016workload}, \cite{megahed2017stochastic},  \cite{gill2017chopper}, \cite{awada2017improving}\\
\cline{3-6}
&&offline&yes&static& \textbf{EMPIRICAL-MODEL (1):}  \cite{ma2016auto}\\

\hline

Simulation& &offline&no&semi& \textbf{PROFILING (2):} \cite{2014-profiling-decision-tree-scaling-2014}, \cite{sun2016roar} \newline\textbf{SIMULATOR (4):} \cite{full-simulation-model},  \cite{CloudSim}, \cite{CloudAnalysis}, \cite{DCSim}, \cite{GA-full-simulation} \\
\hline

\multirow{8}{*}{\parbox{1.3cm}{\centering Machine learning model}}&\multirow{2}{*}{Linear}&online&no&semi& \textbf{LR (3):} \cite{acdc09}, \cite{software-RP-two-loops}, \cite{linear-mapping-CP-bundles-2012} \newline \textbf{ARMA (1):} \cite{HPL-2008-123R1-mimo}, \cite{MIMO-fuzzy-hill-climbing-2013} \newline \textbf{KF (1):} \cite{2014-kalman-pair-wised-coupling-on-tiers-2014} \\
\cline{3-6}
&&online&no&dynamic& \textbf{LR (1):} \cite{ISPASS07} \\
\cline{3-6}
&&online&yes&semi& \textbf{MIMO (4):} \cite{fuzzy-vm-interference}, \cite{qcloud}, \cite{vm-fuzzy-MIMO}, \cite{2014-decision-tree-software-CP-2014}\\
\cline{2-6}
&\multirow{2}{*}{Nonlinear}&online&no&semi& \textbf{KM (1):} \cite{kriging-controller} \newline \textbf{RT (1):} \cite{11icde_smartsla_full} \newline \textbf{ANN (4):} \cite{dynamic-model-comparison} \cite{2014-navie-ANN-GA-2014} \cite{JSS_Kousiouris} \cite{Chen:2013:seams} \newline \textbf{SVM (2):} \cite{2014-2-SVM-workload-type-2014} \cite{dynamic-model-comparison} \newline \textbf{CHANGE-POINT (1):}  \cite{change-point} \\
\cline{3-6}
&&online&yes&semi& \textbf{SVM (1):} \cite{rameshan2016augmenting} \\
\cline{2-6}
&\multirow{3}{*}{Ensemble}&online&no&semi& \textbf{ARMA+SVM (1):} \cite{TR-10-full-version} \newline \textbf{ANN+ANN (1):} \cite{MOCS2014-full}\\
\cline{3-6}
&&online&no&dynamic&  \textbf{KNN+LR:+RT (1):} \cite{2013-single-learner-filter-wrapper-LR-2013} \\
\cline{3-6}
&&online&yes&dynamic& \textbf{ARMA+ANN+RT (2):}  \cite{Chen:2014:ucc}, \cite{Chen:2015:tse-pending} \\
\hline
\multirow{2}{*}{\parbox{1.3cm}{\centering Hybrid model}}&&semi&no&semi& \textbf{LQN+KF (2):} \cite{2014-kalman+rule-based-2014}, \cite{kalman-AR} \newline  \textbf{LQN+KF+K-MEAN (1):} \cite{kalman-clustering} \newline \textbf{S-QUEUE+ARMA (1):} \cite{ICDCS2011} \newline \textbf{M-QUEUE+KF (1):} \cite{gandhi2017model} \\
\cline{3-6}
&&offline&yes&static& \textbf{EMPIRICAL-MODEL+PROFILING (1):} \cite{sun2017automated} \\
\hline

\end{tabularx}
%\begin{tablenotes}
%\item\textit{\scriptsize S-QUEUE=Single Queue; M-QUEUE=Multiple Queues; LQN=Layered Queuing Network; MDP=Markov Decision Process; PCM=Palladio Component Model; PCA=Principal Component Analysis; LR=Linear Regression; ARMA=Auto-Regressive Moving-Average; KF=Kalman Filter; MIMO=Multiple-Input and Multiple-Output; KM=Kriging Model; RT=Regression Tree; ANN=Artificial Neural Network; SVM=Support Vector Machine; KNN=k-Nearest-Neighbor}
%    \end{tablenotes}
\vspace{-0.3cm}
\end{table}

\begin{enumerate}[leftmargin=.55cm]

\item Despite the high importance of QoS interference, it has not received much attention when modeling the QoS (only 13\%).
\item Truly dynamic QoS modeling, i.e., changing both the input features and their coefficients, is still minority (7\%). Other studies have merely considered changing coefficients of the model while ignoring the input features' dynamics (54\% semi-dynamic), or none at all (39\% static).
\item The concrete modeling methods applied for machine learning model is more diverse than the methods in other categories.

\end{enumerate}
Additionally, from Table~\ref{tb:qos-inout}, we can observe that:

\begin{enumerate}[leftmargin=.55cm]

\item Although most studies (65\%) have only considered certain inputs/output during their experiments, they have claimed that their model is compatible with any given inputs (i.e., any cloud control primitives) and/or output (i.e., any QoS attributes).
\item The most widely considered input dimension is CPU while the most common output is response time (except the general one, i.e., QoS attribute).
\item The most explicitly modeled number of outputs is three, while the most explicitly considered number of inputs is four.

\end{enumerate}
In the following, we specify some representative studies on QoS modeling for SSCAS in details.

\begin{table}[t!]

\caption{\fontsize{8}{8}\selectfont QoS Modeling Approaches for SSCAS by Inputs and Output.}
\label{tb:qos-inout}
%\vspace{0.3cm}
\centering

\scriptsize
\begin{tabularx}{\textwidth}{|P{3cm}|X|} \hline
\textbf{\emph{Outputs}}&\textbf{\emph{Cloud Primitives}}\\ \hline

QoS attributes& \textbf{CPU (1):} \cite{HuBrKo2011-SEAMS-ResAlloc} \newline
\textbf{number of VM (1):} \cite{2014-eplison-GA-weigh-h-scaling-2014},  \cite{2014-profiling-decision-tree-scaling-2014} \newline 
 \textbf{configurations (1):} \cite{2014-2-SVM-workload-type-2014}, \cite{2014-decision-tree-software-CP-2014} \newline 
  \textbf{resources (2):} \cite{2014-navie-ANN-GA-2014}, \cite{qcloud} \newline 
     \textbf{CPU and bandwidth (1):} \cite{vm-fuzzy-MIMO} \newline 
   \textbf{CPU and memory (4):} \cite{kriging-controller}, \cite{MIMO-fuzzy-hill-climbing-2013}, \cite{fuzzy-vm-interference}, \cite{GA-full-simulation} \newline 
    \textbf{configurations and resources (6):} \cite{PCA-model-scaling-2013}, \cite{software-RP-two-loops}, \cite{tse1}, \cite{Chen:2014:ucc}, \cite{Chen:2015:tse-pending}, \cite{Chen:2013:seams}  \newline
     \textbf{configurations, CPU and memory (1):} \cite{TR-10-full-version} \newline 
      \textbf{resources and workload (8):} \cite{ISPASS07}, \cite{JSS_Kousiouris}, \cite{kalman-AR}, \cite{full-simulation-model}, \cite{CloudSim}, \cite{CloudAnalysis}, \cite{DCSim},  \cite{gill2017chopper} \newline 
  \textbf{CPU, memory and disk (2):}  \cite{2013-rule-based-multi-elasiticity-2013},  \cite{HPL-2008-123R1-mimo} \newline 
\textbf{CPU, memory and bandwidth (3):} \cite{2014-linear-centralized-decision-making-2014}, \cite{dynamic-model-comparison}, \cite{self-sla-and-resource} \newline 
\textbf{CPU, storage and bandwidth (2):} \cite{06032254}, \cite{Emeakaroha_CloudComp2010} \newline
 \textbf{configurations, resources and workload (1):} \cite{2013-single-learner-filter-wrapper-LR-2013} \newline
\textbf{CPU, bandwidth, storage and number of VM (1):} \cite{Compsac_2010_I_Brandic} \newline 
 \textbf{CPU, memory, workload and number of VM (1):} \cite{11icde_smartsla_full} \newline 
\textbf{CPU, memory, workload and bandwidth: (1)} \cite{linear-mapping-CP-bundles-2012} \newline
\textbf{CPU, memory, storage and bandwidth (1):} \cite{HPCS_IWCMC_Vincent}  \newline
\textbf{CPU, number of VMs and workload (1):} \cite{ma2016auto} \newline
 \textbf{workload and interference index (1):}  \cite{rameshan2016augmenting}\\
\hline
Response time& \textbf{CPU (1):} \cite{2014-kalman-pair-wised-coupling-on-tiers-2014} \newline \textbf{number of VM (1):} \cite{multitier-resalloc-Cloud11} \newline \textbf{workload and number of VMs (4):} \cite{3-stages-game-theory}, \cite{queue-prediction},  \cite{queue-VM-group}, \cite{sla-provision} \newline  \textbf{CPU and memory (2):} \cite{E3-R-extended}, \cite{MPC-price} \newline  \textbf{workload and CPU (2):} \cite{li-opt-clouds-4}, \cite{ICDCS2011} \newline  \textbf{thread and CPU (1):} \cite{wosp10sla} \newline \textbf{CPU, memory, workload and number of VMs (2):} \cite{fine-grained-servce-LQM-MIP-power}, \cite{2014-kalman+rule-based-2014}  \newline \textbf{CPU, workload and number of VMs (1):} \cite{gandhi2017model}    \\
\hline
Response time and workload& \textbf{number of VMs (1):}  \cite{cache-static-ANN-bi-obj-2012} \newline \textbf{CPU and memory (1):} \cite{profit-cent-local-search} \newline \textbf{workload and number of VMs (1):} \cite{typical-navie-scaling-VM-number} \newline \textbf{workload, number of VMs and VM type (1):} \cite{sun2017automated} \\
\hline
Response time and utilization & \textbf{number of VMs and VM type (1):} \cite{awada2017improving} \\
\hline
Response time and throughput & \textbf{CPU and memory (1):} \cite{06119056} \newline \textbf{CPU, memory, number of VMs and VM type (1):} \cite{sun2016roar} \\
\hline
CPU utilization &  \textbf{workload and number of VM (1):} \cite{acdc09}\\
\hline
QoS attributes and hardware demand& \textbf{configurations and resources (1):} \cite{MOCS2014-full} \newline \textbf{CPU and workload (1):} \cite{kalman-clustering} \\
\hline
QoS attributes and workload& \textbf{workload and number of PM (1):} \cite{change-point} \\
\hline
QoS attributes and overhead& \textbf{resources (1):}  \cite{BRGA-resource} \\
\hline
Cost& \textbf{workload and number of VM (2):}  \cite{shariffdeen2016workload}, \cite{megahed2017stochastic} \\

\hline

\end{tabularx}
\vspace{-0.3cm}
\end{table}

%Inputs&Output&Primitives Selection

%Modeling Approach&Inputs&Output&Primitives Selection&Online&QoS Interference&Representative Examples\\ 

%\begin{itemize}[leftmargin=.35cm]

%\item\emph{\textbf{Analytical Modeling}}

\subsubsection{Analytical Modeling}

Analytical modeling approaches rely on a closed-form structure to model the cloud-based service. These models are often built offline based on theoretical principles and assumptions.  Next, we further divide the analytical modeling approach into \emph{queuing theory}, \emph{dependability models} and \emph{black box models}.

%Given the fact that the structure and expression of the model do not change, analytical modeling usually need to make strong assumptions about the service's internal operation and/or the environment. 

\begin{itemize}[listparindent=1.5em,leftmargin=.35cm]
\item\emph{4.5.1.1 \: Queuing theory:} Queuing model and queuing network are widely applied for QoS modeling in the cloud. They model the cloud-based services as a single queue or a collection of queues that are interacting through a mixture of request arrivals and completes. Specifically, a single queue has been used to model the correlation of response time (or throughput) to CPU, number of VM and workload. For example, depending on the assumption of the distribution on arrival and service rate, the model can be built as  M/G/m queue\footnote{In queuing theory, $M$ denote Poisson distribution; $G$ denotes arbitrary distribution. A term $M/G/m$ refers to Poisson distribution of arrival rate, arbitrary distribution of service rate and there exists $m$ servers.} by Zhang et al. \cite{MPC-price}, M/G/m queue by Jiang et al. \cite{06119056},  M/M/1 queue by E$^3$-R \cite{E3-R-extended}  and JustSAT \cite{queue-VM-group}, and M/M/m queue by Jiang et al. \cite{typical-navie-scaling-VM-number}. To create more detailed modeling with respect to the internal structure of cloud-based services, multiple queues can be used to create QoS models, for example, Goudarzi and Pedram \cite{multitier-resalloc-Cloud11} apply multiple queues to model the response time for cloud-based multi-tiered applications with respect to number of VM and workload. Their work calculates average response time for the queue in the forward direction throughout the tiers. 
%Li et al. \cite{wosp10sla} apply a single queue to model the correlation between response time and CPU, workload and thread. In particular, the model contains finite capacity regions, which denote the place constraints on the maximum number of jobs circulating in a subnetwork of queues. This is because they are the simplest class of models that offer the features to describe performance scalability as a function of the software threading level and for the number of CPUs. 

%In a similar way, Bi et al. \cite{queue-prediction} use a queuing network composed of an M/M/c queue and multiple M/M/1 queues to estimate the correlation between response time and number of VMs and workload for cloud-based application. 

%Among the others, queuing theory (e.g., queuing networks and layered queuing networks) is one of the most popular techniques used in existing work for QoS modeling in the cloud. It mainly focuses on model the correlation between response time and the number of station (e..g, VM or CPU core) and the workload.

% queue theory and netowrk

%LQN

Unlike classical queuing model and queuing network, the Layer Queuing Network (LQN) additionally model the dependencies presented in a complex workflow of requests to cloud-based services and applications. For instance, Zhu et al. \cite{sla-provision} have also used LQN where the authors employ a global M/M/m queue for the entire on-demand dispatcher and then a M/G/1 queue on each tier of an application.  The former queue correlates the response time to number of VMs, while the latter queue models the relationship between response time and CPU of the VM that contains the corresponding tier.

%Chi et al. \cite{3-stages-game-theory} use LQN  (i.e., based on M/M/m queue) to model the response time of application with respect to CPU and workload.

%CloudOpt \cite{fine-grained-servce-LQM-MIP-power} relied on LQN as the aggregate QoS model for all the services contained by an application. It models only response time with respect to CPU and workload. Li et al. \cite{li-opt-clouds-4} use LQN for model services in an application. Again, it only captures response time with respect to CPU and workload. 

\item\emph{4.5.1.2 \: Dependability models:} Dependability models focus on the modeling of various states for QoS attributes. For example, in QoSMOS \cite{tse1}, the authors analytically solve the Markov Models (Discrete-Time Markov Chain and Markov Decision Process) to model the QoS for services in an application. The model correlates QoS attributes with hardware resources and workload. Huber et al. \cite{HuBrKo2011-SEAMS-ResAlloc} uses Palladio Component Model (PCM) as architecture-level QoS model since it permits to explicitly model different usage profiles and resource allocations. 

%Copil et al. \cite{2013-rule-based-multi-elasiticity-2013} uses a graph representation to model the dependency between per-service QoS and the necessary cloud primitives. Although the graph can be updated at runtime, the model is essentially analytical. 

%Kateb et al. \cite{2014-eplison-GA-weigh-h-scaling-2014} uses \emph{model@runtime} to correlate QoS attributes with the number of VM. The modeling approach is essentially based on a domain specific language, which does not only able to reason about the system at design time, but is also able to assist decision making during runtime.

\item\emph{4.5.1.3 \: Black box models:} Black-box models handle the QoS using empirical and historical domain knowledge. Among others, the CLOUDFARM framework~\cite{2014-linear-centralized-decision-making-2014} uses a empirical QoS model where the correlation between certain QoS values and the required resource is captured  (i.e., CPU). In particular, the authors assumed that the magnitudes of resources to the QoS values is known, as specified by the cloud service or application provider. Another study from Emeakaroha et al. \cite{Emeakaroha_CloudComp2010} \cite{HPCS_IWCMC_Vincent} propose an empirical model that maps the expected QoS values with CPU, memory, bandwidth and storage based on the assumptions of the system that being managed. 
%The FoSII project \cite{Compsac_2010_I_Brandic} has also applied empirical QoS models such that the correlation between hardware resources (i.e., CPU, memory and bandwidth) and QoS is hard-coded using cases, each of which contains a set of particular values of resource and their resulted utility value. 

%Their extended work \cite{HPCS_IWCMC_Vincent} is also based on a similar approach, where the authors correlate the QoS attributes to different CPU, memory, bandwidth and storage using manual and empirical mapping. The proposed mapping can be as simple as one QoS attribute to one primitives, or a complex form where multiple cloud primitives are associated with a QoS attribute.  

\end{itemize}
%Emeakaroha et al. \cite{06032254} have relied on empirical QoS models where the correlation between QoS and hardware resource  (i.e., CPU, memory, bandwidth and storage) is based on domain experience of experts.  Addis et al. \cite{profit-cent-local-search} use a simple approximation approach to assume the correlation between response time and the hardware resources, i.e., CPU and memory.Kabir and Chiu \cite{cache-static-ANN-bi-obj-2012} correlate the response time to the number of VMs and workload via an empirical utility function. In addition, they use ANN to predict future workload. The BRGA framework \cite{BRGA-resource}  has also used empirical and utility based model to correlate the QoS attributes to CPU, memory and workload.

%\item\emph{\textbf{Simulation Based Modeling}}

\subsubsection{Simulation Based Modeling}

QoS models can be also generated by various simulators; here, conducting simulations is usually a complex and expensive process  and thus they are used in an offline manner. In practice, simulation is required to be setup by the domain experts, who will often need to analyze, interpret and profile the data collected after simulation runs. Specifically, Fernandez et al.~\cite{2014-profiling-decision-tree-scaling-2014} have relied on a profiling approach that builds the QoS model for each bundle of VM offline. The process is similar to a simulation modeling approach. CDOSim~\cite{full-simulation-model} is a framework that simulates the actual application in the cloud to restrict the search-space for autoscaling and to steer the exploration towards promising decisions. CloudSim~\cite{CloudSim} is a simulation toolkit that models QoS attributes (of VM) with respect to resource allocation. It supports both single cloud and multiple clouds scenarios. As an extension of CloudSim, CloudAnalyst~\cite{CloudAnalysis} allows the simulation of QoS attributes for the application deployed on geographically-distributed datacenters. Similarly, DCSim~\cite{DCSim} simulates the overall quality of resource autoscaling for the entire cloud. 

%Another extension of CloudSim, namely EMUSim \cite{EMUSim}, focuses on simulation of QoS attributes with detailed knowledge of the applications' or services' internal structure. It is commonly used for profiling , which serve as inputs to CloudSim. GroupSim \cite{GroupSim} is explicitly designed to simulate the QoS attributes of scientific application or service running in the cloud. Given that such application or service often facilitate long running request, the simulation is designed based on events. 

%\item\emph{\textbf{Machine Learning Based Modeling}}

\subsubsection{Machine Learning Based Modeling}

\noindent The increasing complexity of cloud-based services has rendered the modeling process an extremely difficult task for human experts. To this end, recent studies have exploited the advances of machine learning algorithms and theory to create more reliable QoS models. In the following, we survey the key studies that apply machine learning approaches for QoS modeling in the cloud. In particular, we have further classified them into two categories, these are: \emph{linear} and \emph{nonlinear} modeling.

\begin{itemize}[listparindent=1.5em,leftmargin=.35cm]

\item\emph{4.5.3.1 \: Linear modeling:} Learning algorithms based on linear models for QoS modeling in the cloud can handle linear correlation between a selected set of cloud primitives (e.g, CPU, memory, number of VM, workload etc) and output (i.e., QoS attributes), and they are sometime very efficient. Simple linear models most commonly rely on linear regression, where each primitive input is associated with a time-varying weight, e.g., Lim et al. \cite{acdc09},  Zhang et al. \cite{software-RP-two-loops} and Collazo-Mojica et al. \cite{linear-mapping-CP-bundles-2012}. More advanced forms exist, e.g., Padala et al. \cite{HPL-2008-123R1-mimo} have used Auto-Regressive-Moving-Average (ARMA) model that is trained continually by Recursive Least Squares (RLS). The authors claim that the second-order linear ARMA model is easy to be estimated online and can simplify the corresponding controller design problem with adequate accuracy. 

%Notably, the authors found that the second-order ARMA model can predict the application performance with adequate accuracy. Kalivianaki et al. \cite{2014-kalman-pair-wised-coupling-on-tiers-2014} uses Kalman filter to update the QoS model. The authors claim that the Kalman filter is optimal assuming the system is represented using a linear model while the process and noise are white and Gaussian.

 %Minarolli and Freisleben \cite{MIMO-fuzzy-hill-climbing-2013} and Padala et al. \cite{HPL-2008-123R1-mimo} models the QoS using ARMA, which contains additional dimensions of inputs such as historical data point of QoS and primitives. a Recursive Least Squares (RLS) method is used to recompute the time-varying model parameters of ARMA online. 

%inteference
We found that there are limited studies, which attempt to capture the information of QoS interference in the linear QoS model and they only focus on the VM-level \cite{fuzzy-vm-interference}, \cite{qcloud}, \cite{vm-fuzzy-MIMO}, \cite{2014-decision-tree-software-CP-2014}. As an example, Q-Cloud \cite{qcloud} has explicitly considered QoS interference by using the hardware control primitives of all co-hosted VMs as inputs, rendering it in a Multi-Inputs-Multi-Output (MIMO) model, which is trained by Least Mean Square (LMS) method. 

% to capture QoS interference at VM level, APPLEware \cite{fuzzy-vm-interference} uses a fuzzy regression to model CPU, memory and QoS attributes. The inputs to the system are the resource allocation in terms of CPU and memory usage limits at various tiers of the hosted applications. The outputs of the system are the measured QoS values and average energy usage of each application.  Wang et al. \cite{vm-fuzzy-MIMO} propose a model correlates QoS attributes to hardware resources. The proposed approach learns the relationship between multiple co- hosted VMs and the multiple applications' QoS by establishing an MIMO model using a set of fuzzy rules.  The model is also designed to handle QoS interference by incorporating the resources of all co-hosted VM as inputs in the models.

\item\emph{4.5.3.2 \: Nonlinear modeling:} Learning algorithms based on nonlinear models for QoS modeling in the cloud is able to capture complex and nonlinear correlation, in addition to the linear one. However, it can also produce relatively large overhead than the linear modeling. Here, existing studies often aim to model the correlation between hardware control primitives (e.g., CPU, memory and bandwidth) and QoS. The nonlinear modeling can be relied on kriging model \cite{kriging-controller}, Regression Tree (RT)\cite{11icde_smartsla_full}, Artificial Neural Network (ANN) \cite{dynamic-model-comparison} \cite{2014-navie-ANN-GA-2014} \cite{JSS_Kousiouris} \cite{Chen:2013:seams}, Support Vector Machine (SVM) \cite{2014-2-SVM-workload-type-2014} \cite{dynamic-model-comparison}, change-point detection  \cite{change-point}. For example, SmartSLA \cite{11icde_smartsla_full} employs Regression Tree (RT) and boosting to model the QoS. RT partitions the parameter space in a top-down fashion, and organizes the regions into a tree style. The tree is then trained by M5P where the leaves are regression models. The study from Kunda et al. \cite{dynamic-model-comparison} presents sub-modeling based on ANN and SVM for correlating QoS with hardware control primitives in the cloud. Instead of building a single model for a QoS attribute, they train \emph{n} sub-models, whereby \emph{n} is determined by performing k-mean clustering based on the similarity between data values of QoS, creating more accurate and finer grained models.

\item\emph{4.5.3.3 \: Ensemble modeling:} Examples exist for cases where multiple linear and/or nonlinear machine learning algorithms are explored together. Among others, Chen, Bahsoon and Yao~\cite{Chen:2014:ucc} \cite{Chen:2015:tse-pending} exploit a bucket of learning algorithms (both linear and non-linear models). The model accuracies are tracked continually at runtime, considering QoS interference. The best model for a given input values, according to both local and global errors, will be used to make prediction.

%Zhu and Agrawal \cite{TR-10-full-version} use a variant of ARMA and SVM to model the correlation of QoS attributes to software and hardware control primitives. In particular, the ARMA variant, which is trained by SVM, is used to link QoS and software control primitives. Subsequently, another dedicated SVM is used to model the relationship between software control primitives and hardware control primitives (i.e., CPU and memory in the work). 
%Another work from Kousiouris et al. \cite{MOCS2014-full} correlates QoS attributes with various primitives using ANN. Additionally, it applies a time-series ANN to predict the workload in conjunction with the QoS models. This aims to provide more accurate information when making prediction online.

\item\emph{4.5.3.4 \: Comparison of different learning algorithms:} Given the various types of machine learning algorithms, it can be difficult to determine which one(s) are the appropriate algorithms for QoS modeling in the cloud, with respect to both accuracy and overhead. There are researches that have conducted empirical comparisons of different possible learning algorithms for QoS modeling in the cloud \cite{2013-offline-profiling-2013}  \cite{determine-CP-compare-models} \cite{tp38}. 

%For example, Lloyd et al. \cite{determine-CP-compare-models} conduct an extensive experiment over different machine learning algorithms in cloud. They select the primitives based on manual analysis. The results show that the relevant and useful primitives could be different depending on the characteristics of services and application; and that different machine learning algorithms achieve a variety of accuracy depending on the scenarios. 

\end{itemize}

%Therefore, we can conclude that majority of the approaches that apply machine learning for QoS modeling are semi-dynamic.

%\item\emph{\textbf{Hybrid Modeling}}

\subsubsection{Hybrid Modeling}

We discovered that linear machine learning algorithms are also commonly used with analytical approaches to form QoS models. Specifically, Grandhi et al.  \cite{2014-kalman+rule-based-2014}  and  Zheng et al. \cite{kalman-AR} have proposed hybrid model:  to model the multi-tiered application, they have relied on a modified LQN containing some time-varying coefficients. The authors then employ the Kalman filter as an online parameter estimator to continually estimate those coefficients. The approach proposed by Xiong et al. \cite{ICDCS2011} has relied on a combined model, where a M/G/1 queue is used to model the correlation of workload to response time; while ARMA is used to model the relationship between response time and CPU.

% Ghanbari et al. \cite{kalman-clustering} have also followed a similar approach, but through the use of k-mean clustering, they additionally cluster the model into multiple sub-models based on different types of workload.

%\item\emph{\textbf{Dynamic Primitives Selection}}

\subsubsection{Dynamic Primitives Selection}
%% dynamic PS
We noticed that the majority of the aforementioned studies regard the primitives selection as a manual and offline process, most commonly, they have relied on empirical knowledge and heavy human analysis to select the important primitives as the input features of QoS models. Although not many, there are some studies that explicitly consider dynamic process in primitives selection, which tends to be more accurate and can be easily applied \cite{PCA-model-scaling-2013} \cite{ISPASS07} \cite{2013-single-learner-filter-wrapper-LR-2013} \cite{Chen:2015:tse-pending}. As an example, vPerfGuard \cite{2013-single-learner-filter-wrapper-LR-2013} is a framework that correlates QoS attributes with respect to software control primitives, hardware control primitives and environmental primitives. The authors achieve primitive selection based on both filter (relevance based correlation coefficient) and wrapper (i.e., hill-climbing comparison for different learning algorithms). Chen and Bahsoon~\cite{Chen:2015:tse-pending} dynamically select primitives that maximize both information relevance (between a primitive and quality) and minimize redundancy (between already selected primitives). While explicitly modeling the effects of QoS interference, the authors propose a fully self-adaptive approach that selects primitives that improve prediction accuracy given a learning algorithm.

\subsubsection{Workload and Demand Modeling}

We found that some existing studies (e.g., \cite{2011-cost-aware-v-h-scaling-2011}, \cite {ILP-cost-only-scaling-2013}, \cite{JSS_Kousiouris}) attempts to model the workload and demand for assisting autoscaling decision making. In those cases, the modeling is reduced to a single dimension, where the core is to model the trend of the workload or demand using its historical data. However, unlike QoS modeling which is often multi-dimensional, the single dimension in workload or demand models do not offer the ability to reason about the effects of autoscaling decisions and the possible trade-offs.

\subsection{Granularity of Control}
\label{sec:gra}

\noindent \emph{RQ4: What is the granularity of control in SSCAS?}

The ultimate goal of autoscaling is to optimize the QoS and cost objectives, which are referred to as benefit, for all cloud-based services. To this end, the granularity of control in autoscaling plays an integral role, since it determines the boundary of decision making: which and how many objectives should be considered in a decision making process of autoscaling. In the following, we classify existing SSCAS studies depending on what level of granularity they operate at.

     \begin{figure}[b!]
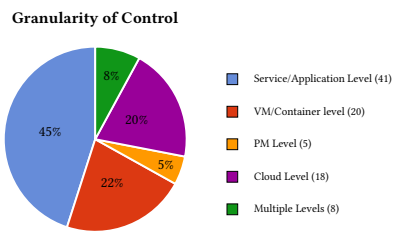

\centering
%\resizebox {\columnwidth} {3cm} {
  \includestandalone[width=6cm]{tikz/granularity-pie}%     without .tex extension
  %}
  % or use \input{mytikz}
  % \vspace{-0.3cm}
  \caption{\fontsize{8}{8}\selectfont Number of papers per granularity of control.}
  \label{fig:gra}
\vspace{-0.3cm}
  \end{figure}

As we can see from Figure~\ref{fig:gra}, the predominant granularity of control is at the service/application level where the boundary of decisions making is grouped by each service/application. Notably, controlling at the cloud level tends to be the second most popular, leaving the other levels being minority. Generally, the finer granularity of control implies that it is harder to achieve globally-optimal benefit but likely to generate smaller overhead. On the other hand, globally-optimal benefit can be easier reached with large overhead if the granularity of control is coarser. According to Table~\ref{tb:gra}, we can obtained the following observations:

\begin{table}[t!]

\caption{\fontsize{8}{8}\selectfont The Granularity of Control in SSCAS.}
\label{tb:gra}
%\vspace{0.3cm}
\centering
\scriptsize
\begin{tabularx}{\textwidth}{|P{1.6cm}|P{1.3cm}|P{1.6cm}|X|} \hline
\textbf{\emph{Granularity}}&\textbf{\emph{Entity}}&\textbf{\emph{Style}}&\textbf{\emph{Representative Examples}}\\ \hline
\multirow{2}{*}{\parbox{1.6cm}{\centering Service \& Application level}}&Application&Decentralized& \cite{2011-cost-aware-v-h-scaling-2011}, \cite{queue-VM-group}, \cite{2014-profiling-decision-tree-scaling-2014}, \cite{linear-mapping-CP-bundles-2012}, \cite{TR-10-full-version},  \cite{wosp10sla}, \cite{06119056}, \cite{cloud_computing_2010_5_20_50060-ext}, \cite{signal-resource-trend-prediction}, \cite{PCA-model-scaling-2013}, \cite{acdc09}, \cite{ILP-cost-only-scaling-2013}, \cite{typical-navie-scaling-VM-number}, \cite{multitier-resalloc-Cloud11}, \cite{2014-decision-tree-software-CP-2014}, \cite{2014-aco-app-to-VM-consolidation-2014}, \cite{scale-rule-based}, \cite{2014-kalman+rule-based-2014}, \cite{ICDCS2011}, \cite{IWQOS11}, \cite{compare-to-Rao-fuzzy-2013}, \cite{HPL-2008-123R1-mimo}, \cite{souza2015using}, \cite{farokhi2016hybrid},  \cite{shariffdeen2016workload}, \cite{megahed2017stochastic}, \cite{lakew2017kpi}, \cite{aslanpour2017auto}  \\
\cline{2-4}
&Service&Decentralized&  \cite{2013-rule-based-multi-elasiticity-2013}, \cite{2014-ARMA-single-agg-objective-2014}, \cite{Compsac_2010_I_Brandic}, \cite{SEASS_2010_Michael_Maurer}, \cite{kriging-controller}, \cite{tse1}, \cite{2014-eplison-GA-weigh-h-scaling-2014}, \cite{E3-R-extended}, \cite{GA-full-simulation},  \cite{2013-scaling-select-one-predictor-and-error-correlation-2013}, \cite{2010-demand-pattern-match-2010}, \cite{cache-static-ANN-bi-obj-2012}, \cite{HuBrKo2011-SEAMS-ResAlloc} \\
\hline
\multirow{2}{*}{\parbox{1.6cm}{\centering VM \& Container level}}& Application&Decentralized& \cite{fuzzy-2-loop-control}, \cite{parallel-RL-vertical-QoS-2013}, \cite{2013-software-CP-only-2013}, \cite{software-RP-two-loops}, \cite{2014-2-SVM-workload-type-2014}, \cite{vm-fuzzy-MIMO}, \cite{qcloud}, \cite{gandhi2017model},    \\
\cline{2-4}
&Application&Centralized& \cite{gill2017chopper},  \cite{sun2017automated}, \cite{awada2017improving}  \\
\cline{2-4}
&VM&Decentralized& \cite{2014-fuzzy-compare-to-JRao-2014}, \cite{2014-kalman-pair-wised-coupling-on-tiers-2014}, \cite{ensemble-prediction-VM-2014-full} \\
\cline{2-4}
&VM&Centralized& \cite{qu2016reliable}, \cite{da2016autoelastic},  \cite{rameshan2016augmenting} \\
\cline{2-4}
&Container&Decentralized& \cite{baresi2016discrete}, \cite{li2015rest}  \\
\hline
&Container&Centralized&  \cite{zhang2016container} \\
\hline

PM level&Application&Decentralized&\cite{MASCOTS11}, \cite{MASCOTS11-bu-software-CP-full}, \cite{2013-JRAO-most-closest-work-2013}, \cite{2014-navie-ANN-GA-2014}, \cite{fuzzy-vm-interference}  \\
\hline
\multirow{2}{*}{Cloud level}&Application&Centralized&\cite{3-stages-game-theory}, \cite{2012-app-VM-mapping-2012}, \cite{2013-elasticity-primitives-2013}, \cite{profit-cent-local-search}, \cite{2014-linear-centralized-decision-making-2014}, \cite{li-opt-clouds-4}, \cite{sla-provision}, \cite{06032254}, \cite{fine-grained-servce-LQM-MIP-power}, \cite{MPC-price}, \cite{EMA-CPU-memory-PD}, \cite{BRGA-resource}, \cite{self-sla-and-resource}, \cite{van2015mnemos}, \cite{herbein2016resource}, \cite{2013-self-organising-map-2013}   \\
\cline{2-4}
&Application&Decentralized&\cite{05557978},\cite{Arc-cover-all-controller}  \\
\hline
\multirow{2}{*}{Hybrid levels}&Application&Decentralized&\cite{MIMO-fuzzy-hill-climbing-2013}, \cite{11icde_smartsla_full}, \cite{ma2016auto}  \\
\cline{2-4}
&Application&Centralized&\cite{seelam2015polyglot}    \\
\cline{2-4}
&Service&Decentralized&\cite{Chen:2013:iccs}, \cite{Chen:2014:seams}, \cite{Chen:2015:tsc-pending}, \cite{Chen:2015:computer}    \\

\hline

\end{tabularx}
\vspace{-0.3cm}
\end{table}

\begin{enumerate}[leftmargin=.55cm]

\item Most of the studies (72\%) see each application as the basic entity regardless to the granularity of control in SSCAS.
\item Decentralization (74\%) is the most popular approach for all granularity of control, except the cloud level where  centralized (or partially centralized) control is predominately exploited.

\end{enumerate}
In the following, we specify each granularity of control for SSCAS in details.

%On one hand, control at the cloud level can easily guarantee the global benefit for all cloud-based services, but it tends to result in large computation overhead as there can be a large number of objectives to be considered in the decision making process. On the other hand, the number of objective can be reduce if we only consider local benefit, e.g., at the service level, and the decision making process for different groups of local benefit runs independently and simultaneously; in such a way, the overhead can be reduced. However, the downside is that multiple local optimums may not necessarly imply global optimal benefit in the cloud. 

%\begin{itemize}[leftmargin=.35cm]

%\item\emph{\textbf{Controlling at Service/Application Level}}

\subsubsection{Controlling at Service and Application Level}

Service/Application level is the finest level of control in a SSCAS. It is worth noting that by service, we refer to any conceptual part of the system being managed.  As a result, control granularity at the service/application level may refer to independently controlling/scaling an application, a tier of an application or a cloud-based service.

% applicaiton
We found that most of the studies have focused on controlling each cloud-based application. These approaches have relied on controlling the QoS and/or cost for each individual application in isolation, and therefore, they sometimes regard an application as a service.  Examples of such include:  Lim et al. \cite{acdc09} control the application and its required VM, in which case  an application is regarded as a service. Sedaghat et al. \cite{ILP-cost-only-scaling-2013} regard application as a service, and considered the required number of VMs and the fixed VM bundles for such service. 

There are studies that explicitly controls cloud-based service in general. Among others, Copli et al. \cite{2013-rule-based-multi-elasiticity-2013} control the QoS, cost and their elasticity for each service deployed in the cloud. Yang et al. \cite{2014-ARMA-single-agg-objective-2014} control the cost of individual cloud-based services. The FoSII project \cite{Compsac_2010_I_Brandic} controls individual cloud-based service, their QoS and cost. Gambi et al.~\cite{kriging-controller} control at the service level, where the controller decides on the optimal autoscaling decision for cloud-based service in isolation. 

%QoSMOS \cite{tse1} explicitly focuses on each individual service, and adapting it in isolation. Kateb et al. \cite{2014-eplison-GA-weigh-h-scaling-2014} consider many services are encapsulated in the application, and the authors focus on each service in isolation. The E$^3$-R framework \cite{E3-R-extended} and Frey et al. \cite{GA-full-simulation} control each service, including their composition and autoscaling.

%\item\emph{\textbf{Controlling at Virtual Machine Level}}

\subsubsection{Controlling at Virtual Machine and Container Level}

VM hypervisor and container are two fundamental infrastructure that underpin cloud computing. In particular, VM and container differ in the sense that the former requires a full Operating System to be installed on a VM while the latter does not. This fact allows the container to set naively with the host PM, providing much faster creation and removal time of VM image. However, such benefit comes in the expenses of weaker security guarantees and potentially greater chances of interference, given that the container instances have less isolation. Despite such a difference, the two infrastructures are conceptually similar as they both aim to provide certain level of isolation on top of the hosting PM, and thus they can be regarded as the same granularity of control.

VM level means that the control and decision making operate at each VM in the SSCAS. In particular, certain studies assumes a one-to-one mapping between application (or a tier) and VM and thus they can be categorized as either service level or VM level granularity. To better separate them from the pure service/application level granularity of control, these studies are regarded as VM level granularity. Specifically, FC2Q \cite{2014-fuzzy-compare-to-JRao-2014} regards application tier and VM interchangeably, therefore controlling each tier of an application is equivalent to control each individual VM. Similarly, Kalyvianaki et al. \cite{2014-kalman-pair-wised-coupling-on-tiers-2014} control a tier of an application that resides on a VM, and the authors only focus on CPU allocation of a VM.

%Wang, Xu and Zhao \cite{fuzzy-2-loop-control} control the cloud in a per-VM basis, and each VM is adapted in isolation. VScale \cite{parallel-RL-vertical-QoS-2013} focuses on vertical scaling only and hence the control granularity is per VM. Zhang et al. \cite{software-RP-two-loops} assume only one application per VM, and control the deployed VM in isolation correspondingly. 

%Guo et al. \cite{2013-software-CP-only-2013} control the application deployed on the VM and the authors assume one application per VM. Similarly, Matrix \cite{2014-2-SVM-workload-type-2014} has used VM level control as there is only one application per VM.

%\item\emph{\textbf{Controlling at Physical Machine Level}}

\subsubsection{Controlling at Physical Machine Level}

Autoscaling decision making on each PM independently is referred to as PM level control in the SSCAS. The primary intention of PM level control is to manage the QoS interference caused by co-hosted VMs. Among the others, Xu et al. \cite{MASCOTS11} \cite{MASCOTS11-bu-software-CP-full} control the VMs collectively at the PM level, thus the autoscaling promotes better management of QoS interference at the VM level. 

%The extended study from Xu et al. \cite{MASCOTS11-bu-software-CP-full} consider QoS interference at the VM level, therefore the granularity of control is based on each PM. 

%Similarly, Bu et al. \cite{2013-JRAO-most-closest-work-2013} considers VM level QoS interference and thus the control is for each PM. Minarolli and Freisleben \cite{2014-navie-ANN-GA-2014} consider all the co-hosted VM in conjunction with each others and thus its control granularity is at the PM level.  Lama et al. \cite{fuzzy-vm-interference} control at the PM level in order to handle QoS interference.

%\item\emph{\textbf{Controlling at Cloud Level}}
\subsubsection{Controlling at Cloud Level}

The most coarse level of control granularity is at the cloud level for SSCAS. The majority of the studies achieves autoscaling at the cloud level by using a centralized and global controller, with an aim to manage utility (\cite{3-stages-game-theory}, \cite{2012-app-VM-mapping-2012}, \cite{6008793}, \cite{2013-elasticity-primitives-2013}, \cite{profit-cent-local-search}, \cite{2014-linear-centralized-decision-making-2014}), profits (\cite{li-opt-clouds-4}, \cite{sla-provision}) and availability (\cite{06032254}). Among others, Ferretti et al. \cite{05557978} control the QoS for all cloud-based services in a global manner. However, the actual deployment can be either centralized or decentralized. Similarly, CRAMP \cite{EMA-CPU-memory-PD} uses a centralized and global controller, it controls the entire cloud for cost and QoS. CloudOpt \cite{fine-grained-servce-LQM-MIP-power} also controls the entire cloud using centralized control, as the considered optimization involves all the PM in the cloud. 

%Zhang et al. \cite{MPC-price} control the cost of the entire cloud with respect to how the VM instances of Amazon can be utilized.  BRGA \cite{BRGA-resource} maintains global view of the entire cloud, and thus it belongs to cloud level control. The FOSII project \cite{self-sla-and-resource} also controls the entire cloud in a centralized manner, where the goal is to manage the entire cloud at infrastructure level. 

Some of the studies have relied on a decentralized manner where a consensus protocol is employed for controlling at the cloud granularity. For example, Wuhib et al. \cite{Arc-cover-all-controller} aim to control the entire cloud, and thus the QoS and the overall power consumption of cloud can be collectively managed. Further, they have relied on decentralized deployment, which can reduce the overhead of cloud-level control.

%\item\emph{\textbf{Controlling at Hybrid Levels}}

\subsubsection{Controlling at Hybrid Level}

We found that it is also possible for SSCAS to operate at multiple and hybrid levels, with an aim to better manage the overhead and global benefit. For example, Minarolli and Freisleben \cite{MIMO-fuzzy-hill-climbing-2013} combine both PM level and cloud level control, where the PM level is decentralized and the objective is to optimize the utility locally.  Similarly, SmartSLA \cite{11icde_smartsla_full} aims to control the resource allocation for all the cloud-based services, therefore it utilizes a global, cloud-level control in addition to the decentralized local control on each VM. Different to the others, Chen and Bahsoon~\cite{Chen:2014:seams} exploit a dynamic schema where the multiple simultaneously presented granularity are changed at runtime, according to the objective-dependency.

%\end{itemize}

%In summary, the finer granularity of control implies that it is harder to achieve globally-optimal benefit but likely to generate smaller overhead. On the other hand, globally-optimal benefit can be easier reached with large overhead if the granularity of control is coarser. The majority of the approaches surveyed operate at static and fixed granularity of control, even for some of the hybrid ones. As a result, given the time-varying QoS sensitivity and interference in cloud, they can be inflexible for any runtime changes on the effects of control granularity to the global benefit.

\subsection{Trade-off Decision Making}
\label{sec:dm}
\noindent \emph{RQ5: What are the approaches used for decision making in SSCAS?}

The final important logical aspect in cloud autoscaling is the challenging decision making process, with the goal to optimize QoS and cost objectives. It is even harder to handle the trade-off between possibly conflicting objectives. Such decision making process is essentially a combinatorial optimization problem where the output is the optimal (or near-optimal) decision containing the newly configured values for all related control primitives. In the following, we survey the key studies on the decision making for SSCAS. In particular, we classify them into three categories, these are \emph{rule based control}, \emph{control theoretic approach} and \emph{search-based optimization}.

\begin{figure}[t!]
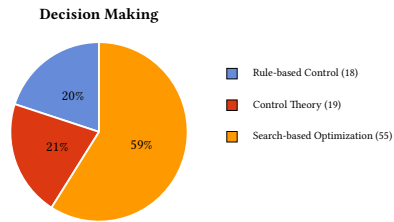

\centering
%\resizebox {\columnwidth} {3cm} {
  \includestandalone[width=6cm]{tikz/decision-pie}%     without .tex extension
  %}
  % or use \input{mytikz}
  % \vspace{-0.3cm}
  \caption{\fontsize{8}{8}\selectfont Number of papers per decision making approach.}
  \label{fig:dm}
    %\vspace{-0.7cm}
  \end{figure}

As we can see from Figure~\ref{fig:dm}, while rule-based control and control theoretic approach share similar popularity in the studies, search-based optimization receives much more attentions than the other two. From Table~\ref{tb:dm}, we can observe that:

\begin{enumerate}[leftmargin=.55cm]

\item Most of the studies (63\%) in SSCAS do not attempt to consider the trade-off between objectives during the decision making, or they handle such trade-off in the way that different objectives are aggregated using weighted sum (29\%), which essentially converting the multiple objectives into a single one.
\item Explicit consideration of QoS interference (16\%) is still rare during the decision making.

\end{enumerate}

From Table~\ref{tb:dm-inout1} and \ref{tb:dm-inout2} we can see that:

\begin{enumerate}[leftmargin=.55cm]

\item Most of the studies (78\%) claim that they can work on any given objectives, thus the QoS attributes and cost being the most popular objectives to be improved during the decision making process of SSCAS.

\item Hardware resources, particularly CPU and memory, are the most commonly considered dimension of control primitives to be tuned in SSCAS. However, there are little studies (9\%) that consider the interplay between software and hardware control primitives.

\item Considerable amount of studies (34\%) has assumed bundles on the autoscaling decision, which will reduce the search space but might negatively constrain the quality of decision making. 

\end{enumerate}
Observations from Table~\ref{tb:dm-vh} shows us that:

\begin{enumerate}[leftmargin=.55cm]

\item The majority of the studies (66\%) has considered both vertical and horizontal scaling.
\item Horizontal scaling receives much more attentions than the vertical one.

\end{enumerate}
In the following, we specify some of the decision making approaches for SSCAS in details.

\begin{table}[t!]

\caption{\fontsize{8}{8}\selectfont The Decision Making Approaches for SSCAS. (PDC=Proportional-Derivative Control;KC=Kalman Control;FC=Fuzzy Control;PIC=Proportional-Integral Control;PIDC=Proportional-Integral-Derivative Control;LA=Lagrange Algorithm;ANN=Artificial Neural Network;MPC=Model Predictive Control;MA=Moving Average;QP=Quadratic Programming;MIMO=Multiple Input, Multiple Output;RL=Reinforcement Learning;DP=Dynamic Programming;ES=Exhaustive Search;ILP=Integer Linear Programming;MIP=Mixed Integer Programming;ACO=Ant Colony Optimization;DT=Decision Tree;NFM=Network Flow Model;FDS=Force-Directed Search;BS=Binary Search;LS=Local Search;GS=Grid Search;TS=Tabu Search;GA=Genetic Algorithm;PSO=Particle Swan Optimization;SMS-MOEA=S-metric Selection Multi-Objective Evolutionary Algorihtm;NSGA-II=Non-dominated Sorting Genetic Algorithm-II;MOACO=Multi-Objective Ant Colony Optimization;SAA=Sample Average Approximation}
\label{tb:dm}
%\vspace{0.3cm}
\centering
\scriptsize
\begin{tabularx}{\textwidth}{|P{2cm}|P{0.8cm}|P{1.5cm}|P{1.2cm}|Xp{2.5cm}|} \hline
\textbf{\emph{Decision Making Approach}}&\textbf{\emph{Form}}&\textbf{\emph{Trade-off}}&\textbf{\emph{QoS Interference}}&\textbf{\emph{Concrete Methods}}&\\ \hline

\multirow{2}{*}{\parbox{1.3cm}{\centering Rule-based control}}&&None&No&\multicolumn{2}{l|}{\parbox{6.5cm}{\textbf{RULES (17):}  \cite{2013-elasticity-primitives-2013}, \cite{2013-rule-based-multi-elasiticity-2013}, \cite{05557978}, \cite{06032254}, \cite{Arc-cover-all-controller}, \cite{self-sla-and-resource}, \cite{scale-rule-based}, \cite{cloud_computing_2010_5_20_50060-ext}, \cite{van2015mnemos}, \newline \cite{herbein2016resource}, \cite{seelam2015polyglot}, \cite{souza2015using}, \cite{aslanpour2017auto}, \cite{qu2016reliable}, \cite{gill2017chopper}, \cite{da2016autoelastic}, \cite{li2015rest}   }} \\
\cline{3-6}

&&None&Yes&\multicolumn{2}{l|}{\parbox{6.5cm}{\textbf{RULES (1):} \cite{rameshan2016augmenting}  }} \\
\hline

\multirow{6}{*}{\parbox{1.3cm}{\centering Control theory}}&\multirow{2}{*}{Standard}&None&No&\textbf{PDC (3):} \cite{2012-app-VM-mapping-2012}, \cite{acdc09}, \cite{EMA-CPU-memory-PD} \newline \textbf{KC (2):} \cite{2014-kalman-pair-wised-coupling-on-tiers-2014}, \cite{2014-kalman+rule-based-2014} \newline \textbf{FC (3):} \cite{2014-fuzzy-compare-to-JRao-2014}, \cite{compare-to-Rao-fuzzy-2013}, \cite{fuzzy-2-loop-control}&\\
\cline{3-6}
&&Weighted sum&No&\textbf{FC (1):} \cite{IWQOS11}&\\
\cline{2-6}
&\multirow{4}{*}{Extended}&None&No&\textbf{PIC+LA (1):} \cite{ICDCS2011} \newline \textbf{MPC+MA (1):} \cite{farokhi2016hybrid} \newline \textbf{PIC+ILP (1):}  \cite{baresi2016discrete} & \textbf{ANN+FC (1):} \cite{2013-software-CP-only-2013} \newline \textbf{MIMO(1):}  \cite{lakew2017kpi} \\
\cline{3-6}
&&Weighted sum&No&\textbf{FC+QPS (1):} \cite{MIMO-fuzzy-hill-climbing-2013} & \textbf{PIDC+RL+ES (1):} \cite{TR-10-full-version} \\
\cline{3-6}
&&Weighted sum&Yes& \textbf{MPC+QP (1):} \cite{fuzzy-vm-interference}&\\
\cline{3-6}
&&None&Yes& \textbf{FUZZY-MIMO (1):} \cite{vm-fuzzy-MIMO} & \textbf{MIMO (1):} \cite{qcloud} \\
\hline

\multirow{8}{*}{\parbox{1.3cm}{\centering Search-based optimization}}&\multirow{2}{*}{Implicit}&None&Yes&\textbf{RL (3):}\cite{MASCOTS11-bu-software-CP-full}, \cite{MASCOTS11}, \cite{2013-JRAO-most-closest-work-2013}& \\
\cline{3-6}
&&None&No&\multicolumn{2}{l|}{\parbox{6.5cm}{\textbf{RL (1):} \cite{parallel-RL-vertical-QoS-2013} \newline \textbf{DEMAND (5):} \cite{2013-scaling-select-one-predictor-and-error-correlation-2013}, \cite{signal-resource-trend-prediction}, \cite{2010-demand-pattern-match-2010}, \cite{2013-self-organising-map-2013}, \cite{ensemble-prediction-VM-2014-full}}}\\
\cline{2-6}
&\multirow{6}{*}{Explicit}&Single&No&\textbf{HEURISTIC (2):} \cite{2014-ARMA-single-agg-objective-2014}, \cite{ma2016auto} \newline \textbf{DP (1):} \cite{MPC-price} \newline \textbf{ES (7):} \cite{PCA-model-scaling-2013}, \cite{queue-VM-group}, \cite{linear-mapping-CP-bundles-2012}, \cite{HuBrKo2011-SEAMS-ResAlloc}, \cite{shariffdeen2016workload}, \cite{megahed2017stochastic},  \cite{zhang2016container}  \newline \textbf{ILP (2):} \cite{2011-cost-aware-v-h-scaling-2011}, \cite{ILP-cost-only-scaling-2013} & \textbf{MIP (1):} \cite{fine-grained-servce-LQM-MIP-power} \newline \textbf{ACO (1):} \cite{2014-aco-app-to-VM-consolidation-2014} \newline \textbf{LA (1):} \cite{2014-2-SVM-workload-type-2014}\\
\cline{3-6}
&&Single&Yes& \textbf{HEURISTIC+DT (1):} \cite{2014-decision-tree-software-CP-2014}&\\
\cline{3-6}
&&Weighted sum&Yes&\textbf{ES (2):} \cite{3-stages-game-theory}, \cite{gandhi2017model}  \textbf{RS (1):} \cite{Chen:2014:seams} & \textbf{HEURISTIC (1):} \cite{sun2017automated}\\
\cline{3-6}
&&Weighted sum&No&\textbf{ES (6):} \cite{tse1}, \cite{kriging-controller}, \cite{typical-navie-scaling-VM-number}, \cite{Compsac_2010_I_Brandic}, \cite{SEASS_2010_Michael_Maurer}, \cite{awada2017improving} \newline \textbf{NFM (1):} \cite{li-opt-clouds-4} \newline \textbf{FDS (1):} \cite{multitier-resalloc-Cloud11} \newline  \textbf{BS (1):} \cite{cache-static-ANN-bi-obj-2012} \newline \textbf{DP (1):} \cite{2014-linear-centralized-decision-making-2014} & \textbf{LS (1):} \cite{profit-cent-local-search} \newline \textbf{GS (1):} \cite{11icde_smartsla_full} \newline \textbf{DT (1):} \cite{2014-profiling-decision-tree-scaling-2014} \newline \textbf{QP (1):} \cite{HPL-2008-123R1-mimo}\\
\cline{3-6}

&&Weighted sum&No&\textbf{TS (1):} \cite{sla-provision} \newline \textbf{GA (3):} \cite{software-RP-two-loops}, \cite{2014-navie-ANN-GA-2014}, \cite{BRGA-resource} \newline \textbf{PSO (1):}  \cite{software-RP-two-loops}& \\
\cline{3-6}
&&Pareto&No&\textbf{SMS-MOEA (1):} \cite{wosp10sla} \newline \textbf{NSGA-II (3):} \cite{2014-eplison-GA-weigh-h-scaling-2014}, \cite{E3-R-extended}, \cite{GA-full-simulation}& \\
\cline{3-6}
&&Pareto&Yes&\textbf{MOACO (3):} \cite{Chen:2013:iccs}, \cite{Chen:2015:tsc-pending}, \cite{Chen:2015:computer} &\\
\hline

\end{tabularx}
\vspace{-0.3cm}
\end{table}

%Objective&Bundles&Control Primitives

%\begin{itemize}[leftmargin=.35cm]

%\item\emph{\textbf{Rule Based Control}}

\subsubsection{Rule Based Control}

Rule-based control is the most classic approaches for making decision in SSCAS. Commonly,
one or more conditions are manually specified and mapped to a decision, e.g., increase CPU and memory by \emph{x} if the throughput is lower than \emph{y}. Therefore, the possible trade-off is often implicitly handled by the conditions and actions mapping. Specifically, Cloudline \cite{2013-elasticity-primitives-2013} allows programmable elasticity rules to drive autoscaling decisions. It is also possible to modify these rules at runtime as required by the users. Copil et al. \cite{2013-rule-based-multi-elasiticity-2013} handle the decision making process by specifying different condition-and-actions mapping for autoscaling in the cloud. In addition, the rules can be defined at different levels, e.g., PaaS and IaaS. Similarly, Ferretti et al. \cite{05557978} allow to setup mapping between QoS expectation and actions using XML like notations.

\begin{table}[t!]

\caption{\fontsize{8}{8}\selectfont Decision Making Approaches for SSCAS by Control Primitives and Objectives.}
\label{tb:dm-inout1} 
%\vspace{0.3cm}
\centering
\scriptsize
\begin{tabularx}{\textwidth}{|P{3cm}|P{1.2cm}|X|} \hline
\textbf{\emph{Objective}}&\textbf{\emph{Bundles}}&\textbf{\emph{Control Primitives}}\\ \hline

\multirow{2}{*}{Cost}&Yes&\textbf{number of VMs (3):} \cite{2014-aco-app-to-VM-consolidation-2014},  \cite{shariffdeen2016workload}, \cite{qu2016reliable} \newline \textbf{CPU and memory (2):}  \cite{2011-cost-aware-v-h-scaling-2011}, \cite{MPC-price} \newline \textbf{CPU, memory and number of VM (1):} \cite{ILP-cost-only-scaling-2013}, \cite{fine-grained-servce-LQM-MIP-power}\\
\cline{2-3}
&No&\textbf{number of VMs (1):} \cite{megahed2017stochastic} \newline \textbf{configurations (1):} \cite{2014-2-SVM-workload-type-2014} \newline \textbf{CPU (1):} \cite{ICDCS2011}  \\
\hline
\multirow{2}{*}{Response time and cost}&Yes&\textbf{number of VMs (3):} \cite{3-stages-game-theory}, \cite{typical-navie-scaling-VM-number}, \cite{cache-static-ANN-bi-obj-2012} \newline \textbf{resources (1):}  \cite{2014-ARMA-single-agg-objective-2014} \\
\cline{2-3}
&No&\textbf{number of VM (1):} \cite{sla-provision} \newline \textbf{CPU, memory and bandwidth (1):} \cite{scale-rule-based} \newline \textbf{CPU, memory and number of VM (1):} \cite{2014-kalman+rule-based-2014}  \\
\hline
\multirow{2}{*}{QoS attributes and cost}&Yes&\textbf{number of VMs (1):} \cite{2012-app-VM-mapping-2012} \newline \textbf{configurations and resources (1):} \cite{PCA-model-scaling-2013} \newline  \textbf{CPU and memory (1):} \cite{2014-profiling-decision-tree-scaling-2014} \newline \textbf{CPU, memory and number of VMs (2):} \cite{2013-elasticity-primitives-2013}, \cite{EMA-CPU-memory-PD} \newline \textbf{CPU, memory and disk (1):} \cite{2013-elasticity-primitives-2013} \newline \textbf{CPU, memory and bandwidth (2):} \cite{2014-linear-centralized-decision-making-2014}, \cite{linear-mapping-CP-bundles-2012}  \\
\cline{2-3}
&No&\textbf{CPU (4):} \cite{2014-fuzzy-compare-to-JRao-2014}, \cite{li-opt-clouds-4}, \cite{IWQOS11}, \cite{compare-to-Rao-fuzzy-2013} \newline  \textbf{number of VM (2):} \cite{multitier-resalloc-Cloud11}, \cite{2014-eplison-GA-weigh-h-scaling-2014} \newline \textbf{resources (2):} \cite{tse1}, \cite{2014-navie-ANN-GA-2014} \newline  \textbf{CPU and memory (6):} \cite{kriging-controller}, \cite{MIMO-fuzzy-hill-climbing-2013}, \cite{parallel-RL-vertical-QoS-2013}, \cite{MASCOTS11}, \cite{fuzzy-vm-interference}, \cite{GA-full-simulation} \newline \textbf{configurations and resources (4):} \cite{Chen:2013:iccs}, \cite{Chen:2015:tsc-pending}, \cite{Chen:2015:computer}, \cite{Chen:2014:seams} \newline \textbf{ CPU, memory and disk (1):} \cite{HPL-2008-123R1-mimo} \newline \textbf{CPU, memory and configurations (1):} \cite{TR-10-full-version} \newline \textbf{CPU, storage and bandwidth (1):} \cite{06032254} \newline  \textbf{CPU, memory and thread (2):} \cite{fuzzy-2-loop-control}, \cite{MASCOTS11-bu-software-CP-full} \newline \textbf{CPU, bandwidth, storage and number of VM (1):} \cite{Compsac_2010_I_Brandic} \newline  \textbf{CPU, thread, session, buffer and memory (1):} \cite{2013-JRAO-most-closest-work-2013}.   \\
\hline
QoS attributes&No& \textbf{configurations (1):} \cite{2013-software-CP-only-2013} \newline \textbf{configurations and resources (1):} \cite{software-RP-two-loops} \newline  \textbf{CPU, memory and bandwidth (1):} \cite{05557978} \newline \textbf{thread, CPU and memory (1):} \cite{cloud_computing_2010_5_20_50060-ext} \newline  \textbf{CPU and memory (1):} \cite{lakew2017kpi} \newline  \textbf{resources (1):}  \cite{gill2017chopper}  \newline  \textbf{number of VMs (1):} \cite{rameshan2016augmenting} \\
\hline
\multirow{2}{*}{Response time}&No&\textbf{CPU (1):} \cite{2014-kalman-pair-wised-coupling-on-tiers-2014} \newline \textbf{memory (1):} \cite{farokhi2016hybrid}  \newline \textbf{CPU and bandwidth (2):} \cite{vm-fuzzy-MIMO}, \cite{qcloud} \newline \textbf{CPU and thread (1):}  \cite{wosp10sla} \newline \textbf{CPU and number of VMs (1):} \cite{gandhi2017model} \newline  \textbf{CPU, memory and number of VMs (1):} \cite{li2015rest}  \\
\cline{2-3}
&Yes& \textbf{CPU, memory and number of VMs (1):} \cite{baresi2016discrete} \\

\hline

\end{tabularx}
\vspace{-0.3cm}
\end{table}

\begin{table}[t!]

\caption{\fontsize{8}{8}\selectfont Decision Making Approaches for SSCAS by Control Primitives and Objectives (continued).}
\label{tb:dm-inout2} 
%\vspace{0.3cm}
\centering
\scriptsize
\begin{tabularx}{\textwidth}{|P{5.8cm}|P{1cm}|X|} \hline
\textbf{\emph{Objective}}&\textbf{\emph{Bundles}}&\textbf{\emph{Control Primitives}}\\ \hline
CPU utilization&No&\textbf{CPU and number of VMs (1):} \cite{acdc09}\\
\hline
CPU utilization&Yes&\textbf{number of VMs (1):}\cite{da2016autoelastic} \newline \textbf{CPU and number of VMs (1):} \cite{ma2016auto}\\
\hline
Throughput&Yes&\textbf{number of VMs (1):} \cite{seelam2015polyglot}\\
\hline
General utilization&No& \textbf{configurations (1):} \cite{2014-decision-tree-software-CP-2014} \newline \textbf{number of VMs (1):}  \cite{2013-scaling-select-one-predictor-and-error-correlation-2013} \newline \textbf{resources (1):} \cite{signal-resource-trend-prediction} \newline \textbf{CPU (2):} \cite{2010-demand-pattern-match-2010}, \cite{herbein2016resource} \newline \textbf{CPU and memory (1):} \cite{van2015mnemos}  \\
\hline
General utilization&Yes&\textbf{number of VMs (1):} \cite{ensemble-prediction-VM-2014-full}  \newline \textbf{CPU, memory, number of VMs and bandwidth (1):} \cite{zhang2016container}\\
\hline
QoS attributes and power&Yes&\textbf{CPU, memory and number of VM (1):} \cite{Arc-cover-all-controller} \\
\hline
Response time and utilization&Yes&\textbf{number of VM (1):} \cite{aslanpour2017auto} \newline \textbf{number of VM and VM type (1):} \cite{awada2017improving} \\
\hline
 Response time, cost and availability&No&\textbf{CPU and memory (1):} \cite{profit-cent-local-search} \\
\hline
VM consumption&Yes&\textbf{number of VM (2):} \cite{queue-prediction}, \cite{queue-VM-group}\\
\hline
SLA penalty&No&\textbf{CPU, memory and number of VM (1):} \cite{11icde_smartsla_full}\\
\hline
Response time, throughput, CPU utilization and cost&No&\textbf{CPU and memory (1):} \cite{06119056} \\
\hline
Response time, throughput, CPU utilization&Yes&\textbf{number of VMs and VM type (1):} \cite{sun2017automated}\\
\hline
SLA and power&No&\textbf{resources (1):} \cite{2013-self-organising-map-2013} \\
\hline
SLA and power&No&\textbf{CPU (1):} \cite{souza2015using} \\
\hline
Response time, throughput and cost&No&\textbf{CPU and memory (1):} \cite{E3-R-extended}  \\
\hline
Benefits and overhead&No&\textbf{CPU and memory (1):} \cite{BRGA-resource}\\
\hline
QoS attributes, utilization and cost&No&\textbf{CPU (1):} \cite{HuBrKo2011-SEAMS-ResAlloc} \\
\hline
QoS attributes, utilization, number of actions&No&\textbf{CPU, memory and bandwidth (1):} \cite{self-sla-and-resource}\\

\hline

\end{tabularx}
\vspace{-0.3cm}
\end{table}

\begin{table}[t!]

\caption{\fontsize{8}{8}\selectfont Decision Making Approaches for SSCAS by Scaling Actions.}
\label{tb:dm-vh} 
%\vspace{0.3cm}
\centering
\scriptsize
\begin{tabularx}{\textwidth}{|P{1.3cm}|X|} \hline
\textbf{\emph{Scaling}}&\textbf{\emph{Decision Making Approaches}} \\ \hline

Vertical& \textbf{RULES (1):} \cite{souza2015using} \newline \textbf{CONTROL THEORY (4):} \cite{2014-kalman-pair-wised-coupling-on-tiers-2014}, \cite{2013-software-CP-only-2013}, \cite{farokhi2016hybrid}, \cite{lakew2017kpi} \newline \textbf{SEARCH-BASED OPTIMIZATION (6):} \cite{parallel-RL-vertical-QoS-2013},  \cite{tse1}, \cite{wosp10sla}, \cite{06119056}, \cite{2010-demand-pattern-match-2010}, \cite{2014-decision-tree-software-CP-2014}\\
\hline
Horizontal& \textbf{RULES (6):} \cite{herbein2016resource}, \cite{Arc-cover-all-controller}, \cite{seelam2015polyglot}, \cite{aslanpour2017auto}, \cite{da2016autoelastic}, \cite{rameshan2016augmenting} \newline  \textbf{CONTROL THEORY (1):} \cite{2012-app-VM-mapping-2012} \newline  \textbf{SEARCH-BASED OPTIMIZATION (13):} \cite{3-stages-game-theory}, \cite{MPC-price}, \cite{multitier-resalloc-Cloud11}, \cite{queue-VM-group}, \cite{sla-provision}, \cite{typical-navie-scaling-VM-number}, \cite{cache-static-ANN-bi-obj-2012}, \cite{2014-aco-app-to-VM-consolidation-2014}, \cite{2013-scaling-select-one-predictor-and-error-correlation-2013}, \cite{ensemble-prediction-VM-2014-full}, \cite{shariffdeen2016workload}, \cite{megahed2017stochastic}, \cite{sun2017automated}\\
\hline
Both& \textbf{RULES (12):} \cite{2013-elasticity-primitives-2013}, \cite{2013-rule-based-multi-elasiticity-2013}, \cite{05557978}, \cite{06032254}, \cite{Compsac_2010_I_Brandic}, \cite{scale-rule-based}, \cite{self-sla-and-resource}, \cite{cloud_computing_2010_5_20_50060-ext}, \cite{van2015mnemos}, \cite{qu2016reliable}, \cite{gill2017chopper}, \cite{li2015rest} \newline \textbf{CONTROL THEORY (14):}\cite{2014-fuzzy-compare-to-JRao-2014}, \cite{acdc09}, \cite{EMA-CPU-memory-PD}, \cite{fuzzy-2-loop-control}, \cite{MIMO-fuzzy-hill-climbing-2013}, \cite{2014-kalman+rule-based-2014}, \cite{ICDCS2011}, \cite{IWQOS11}, \cite{TR-10-full-version}, \cite{fuzzy-vm-interference}, \cite{compare-to-Rao-fuzzy-2013}, \cite{vm-fuzzy-MIMO}, \cite{qcloud}, \cite{baresi2016discrete} \newline \textbf{SEARCH-BASED OPTIMIZATION (35):} \cite{MASCOTS11-bu-software-CP-full}, \cite{MASCOTS11}, \cite{2011-cost-aware-v-h-scaling-2011}, \cite{2014-ARMA-single-agg-objective-2014}, \cite{2014-linear-centralized-decision-making-2014}, \cite{fine-grained-servce-LQM-MIP-power}, \cite{kriging-controller}, \cite{li-opt-clouds-4}, \cite{PCA-model-scaling-2013}, \cite{profit-cent-local-search}, \cite{software-RP-two-loops}, \cite{2014-2-SVM-workload-type-2014}, \cite{2014-navie-ANN-GA-2014}, \cite{2014-profiling-decision-tree-scaling-2014}, \cite{HPL-2008-123R1-mimo}, \cite{linear-mapping-CP-bundles-2012}, \cite{2014-eplison-GA-weigh-h-scaling-2014}, \cite{BRGA-resource}, \cite{E3-R-extended}, \cite{GA-full-simulation}, \cite{HuBrKo2011-SEAMS-ResAlloc}, \cite{2013-JRAO-most-closest-work-2013}, \cite{signal-resource-trend-prediction}, \cite{2013-self-organising-map-2013}, \cite{ILP-cost-only-scaling-2013}, \cite{Chen:2014:seams},\cite{SEASS_2010_Michael_Maurer}, \cite{11icde_smartsla_full}, \cite{Chen:2013:iccs}, \cite{Chen:2015:tsc-pending}, \cite{Chen:2015:computer}, \cite{gandhi2017model}, \cite{ma2016auto}, \cite{zhang2016container}, \cite{awada2017improving}\\

\hline

\end{tabularx}
\vspace{-0.3cm}
\end{table}

%\item\emph{\textbf{Control Theoretic Approach}}

\subsubsection{Control Theoretic Approach}

%% V-H
%2014-fuzzy-compare-to-JRao-2014
%acdc09
%EMA-CPU-memory-PD
%fuzzy-2-loop-control
%MIMO-fuzzy-hill-climbing-2013
%2014-kalman+rule-based-2014
%ICDCS2011
%IWQOS11
%TR-10-full-version
%fuzzy-vm-interference
%compare-to-Rao-fuzzy-2013

%% H
%2012-app-VM-mapping-2012

%% V
%2014-kalman-pair-wised-coupling-on-tiers-2014
%2013-software-CP-only-2013

Advanced control theory is another widely investigated approach for autoscaling decision making in SSCAS because of its low latency and dynamic nature. Studies in this category could be either \emph{standard}, i.e., they rely solely on the classic control theory; or \emph{extended} where additional methods are considered.
% However, it is difficult to explicitly reason about the effects of possible trade-off decisions in a control theoretic approach. 

% pure control theory
Among the others, standard controllers (e.g., Proportional-Derivative control \cite{2012-app-VM-mapping-2012}  \cite{acdc09} \cite{EMA-CPU-memory-PD}, Kalman control \cite{2014-kalman-pair-wised-coupling-on-tiers-2014} \cite{2014-kalman+rule-based-2014}  and Fuzzy control  \cite{2014-fuzzy-compare-to-JRao-2014} \cite{compare-to-Rao-fuzzy-2013} \cite{fuzzy-2-loop-control} ) are commonly designed as a sole approach to make autoscaling decisions in the cloud. Specifically, ARUVE \cite{2012-app-VM-mapping-2012} and CRAMP \cite{EMA-CPU-memory-PD} utilizes a Proportional-Derivative (PD) controller, where the proportional and derivative factors are not sensitive to a concrete QoS model while supporting proactive autoscaling of cloud services and applications in a shared hosting environment. Anglano et al. \cite{2014-fuzzy-compare-to-JRao-2014} and Albano et al. \cite{compare-to-Rao-fuzzy-2013}  apply fuzzy control that is updated by fuzzy rules at runtime. The aim is to optimize both QoS, cost and energy by autoscaling hardware resources. Although the authors claim they can cope with any hardware resources, only CPU tuning is explored. They have also ignored the QoS interference. 

%Kalyvianaki et al. \cite{2014-kalman-pair-wised-coupling-on-tiers-2014} and Grandhi et al. \cite{2014-kalman+rule-based-2014} use MIMO model and Kalman controller for making autoscaling decisions, and the authors aim at response time by autoscaling CPU on a VM.

 %Similarly, Wang, Xu and Zhao \cite{fuzzy-2-loop-control} has also relied on fuzzy control and they additionally conduct offline profiling for the software control primitives. Grandhi et al. \cite{2014-kalman+rule-based-2014} has additionally considered the removal and migration of VMs in the deicions. Lim et al. \cite{acdc09} use a typical proportional thresholding control for autoscaling decision making. Here, the control policy modifies an integral control by using a dynamic target range, instead of a single target value. 

% control theory and weighted-sum and others, for the same or different aspect
We have also found that control theoretic approaches can be sometime used with other algorithms to better facilitate the autoscaling decision making, forming extended controller: \cite{ICDCS2011}, \cite{MIMO-fuzzy-hill-climbing-2013}, \cite{2013-software-CP-only-2013},  \cite{IWQOS11},  \cite{TR-10-full-version}, \cite{fuzzy-vm-interference}. Particularly, the gains in the controllers can be further tuned by optimization and/or machine learning algorithms, and this is especially useful for Model Predictive Control (MPC). Among others, Zhu and Agrawal \cite{TR-10-full-version} utilize a Proportional-Integral-Derivative (PID) and reinforcement learning controller for decision making with respect to adapting software control primitives. Such result is then tuned in conjunction with the hardware control primitives using exhaustive search. The QoS attributes and cost are formulated as weighted-sum relation. The autoscaling decision making process in APPLEware \cite{fuzzy-vm-interference} have relied on Model Predictive Control (MPC), with the aim to optimize a cost function that represents the local objectives and resource constraints at a point in time. The state of an application, together with the other autoscaling decisions from the neighboring VMs, are collectively considered in a quadratic programming solver. 

\subsubsection{Search-Based Optimization}

A large amount of existing studies of SSCAS relies on search-based optimization, in which the decisions and trade-offs are extensively reasoned in a finite, but possibly large search space. Depending on the algorithms, search-based optimization for autoscaling decision making in the cloud can be either \emph{explicit} or \emph{implicit}\textemdash the former performs optimization as guided by explicit system models; while this process is not required for the later. 

%
%% V-H
%MASCOTS11-bu-software-CP-full
%MASCOTS11
%2011-cost-aware-v-h-scaling-2011
%2014-ARMA-single-agg-objective-2014, cost only, heuristic
%2014-linear-centralized-decision-making-2014, weighted-sum and dynamic programming
%fine-grained-servce-LQM-MIP-power, cost only, MIP
%kriging-controller, weighted-sum and exhaustive
%li-opt-clouds-4, weighted-sum and NFM
%PCA-model-scaling-2013, throughput only,  exhaustive
%profit-cent-local-search, weighted-sum and general local search
%software-RP-two-loops, weighted-sum and meta-herustics
%11icde_smartsla, SLA penalty only, grid search
%2014-2-SVM-workload-type-2014, cost only, lagrange algorithm
%2014-navie-ANN-GA-2014,  weighted-sum and GA 
%2014-profiling-decision-tree-scaling-2014,  weighted-sum and decision tree search
%HPL-2008-123R1-mimo, weighted-sum and quadratic programming
%linear-mapping-CP-bundles-2012, single QoS attribute and exhaustive
%2014-eplison-GA-weigh-h-scaling-2014, pareto GA and eplison
%BRGA-resource, weighted-sum and GA
%E3-R-extended, pareto and GA
%GA-full-simulation, pareto and GA
%HuBrKo2011-SEAMS-ResAlloc, exhuastive

%%% H
%3-stages-game-theory
%MPC-price, profit only LQP 
%multitier-resalloc-Cloud11, weighted-sum and force-directed search
%queue-VM-group, VM number only, exhaustive
%sla-provision, weighted-sum and tabu search
%typical-navie-scaling-VM-number, weighted-sum and exhaustive
%cache-static-ANN-bi-obj-2012,  weighted-sum and linear search
%2014-aco-app-to-VM-consolidation-2014, number of VM, aco

%%% V
%parallel-RL-vertical-QoS-2013
%tse1
%wosp10sla, pareto, SMS-EMOA
%06119056, weighted-sum and GA

\begin{itemize}[listparindent=1.5em,leftmargin=.35cm]

\item \emph{4.7.3.1 \: Implicit search:} The implicit and search-based optimization approaches for autoscaling decision making do not use QoS models. Similar to the control theoretic approaches, the implicit search is also limited in reasoning about the possible trade-offs. For example, the study from Xu et al.  \cite{MASCOTS11} , \cite{MASCOTS11-bu-software-CP-full} applies a model-free Reinforcement Learning (RL) approach for adapting thread, CPU and memory for QoS and cost. The approach is however implicit, providing that there is neither explicit system models nor explicit optimization. The authors have considered QoS interference during autoscaling. Similarly, VScale \cite{parallel-RL-vertical-QoS-2013} utilizes RL for making autoscaling decisions, which are then achieved by vertical scaling. The RL is realized by using parallel learning, that is to say, the authors intend to speed up agent's learning process of approximated model by learning in parallel, without visiting every state-action pair in a given environment.

%Therefore, the agent does not have to visit every state-action pair in a given environment. 

The approaches that rely on demand prediction (e.g., the Autoflex \cite{2013-scaling-select-one-predictor-and-error-correlation-2013}, PRESS \cite{signal-resource-trend-prediction}, \cite {ILP-cost-only-scaling-2013}, \cite{2010-demand-pattern-match-2010}, \cite{2013-self-organising-map-2013} and \cite{ensemble-prediction-VM-2014-full}) are also regarded as implicit search. This is because the autoscaling decision is directly predicted by the demand models, without the needs of reasoning and optimization.

\item \emph{4.7.3.2 \: Explicit search:} In search-based optimization category, the explicit approaches for autoscaling decision making rely on the explicit QoS models to evaluate and guide the search process. Depending on the different formulations of the decision making problem for autoscaling in the cloud, explicit search can reason about the effects of decisions and the possible trade-offs in details. According to our survey, we found three most commonly used formulations, these are single objective optimization, weighted-sum optimization, and pareto-based optimization.

% single
We discovered that it is not uncommon to optimize only a single objective (e.g., cost or profit) for SSCAS, providing that the requirements of the other objectives are satisfied (i.e., they are often regarded as constraints). For example, Kingfisher \cite{2011-cost-aware-v-h-scaling-2011} and Sedaghat et al. \cite{ILP-cost-only-scaling-2013} use Integer Linear Programming (ILP) to optimize the cost for scaling the CPU and memory for VMs of an application while regarding the demand for satisfying QoS as constraint. 
To apply explicit search-based optimization for SSCAS, the most widely solution for handling the multi-objectivity is to aggregate all related objectives into a weighted (usually weighted-sum) formulation, which converts the decision making process into a single objective optimization problem. The search-based algorithm include:  exhaustive search \cite{3-stages-game-theory}  \cite{tse1}  \cite{kriging-controller} \cite{typical-navie-scaling-VM-number}, auxiliary network flow model  \cite{li-opt-clouds-4}, force-directed search \cite{multitier-resalloc-Cloud11}, binary search \cite{cache-static-ANN-bi-obj-2012}. For example, the FoSII project \cite{Compsac_2010_I_Brandic}  \cite{SEASS_2010_Michael_Maurer} regards the autoscaling decision making as case based reasoning process, where the decision is made by looking for the similar historical cases using exhaustive search. The solution of the most similar case is reused o solve the current one.  

We noted that some studies have relied on more advanced and nonlinear search algorithms, ranging from relatively simple ones: dynamic programming \cite{2014-linear-centralized-decision-making-2014} and local-search strategy  \cite{profit-cent-local-search}, to more complex forms: grid search  \cite{11icde_smartsla_full}, decision tree search \cite{2014-profiling-decision-tree-scaling-2014} \cite{2014-decision-tree-software-CP-2014} and quadratic programming \cite{HPL-2008-123R1-mimo}. For example, CLOUDFARM \cite{2014-linear-centralized-decision-making-2014} addresses the decision making based on a weighted-sum utility function of all cloud-based application and services. The decision making process is formulated as a knapsack problem, which can be resolved by dynamic programming.  
%In SmartSLA \cite{11icde_smartsla_full},  the decision making for autoscaling is aimed for optimizing the aggregated SLA penalty using a grid search algorithm. 

%To optimize weighted-sum utilisation, Amiya et al. \cite{2014-decision-tree-software-CP-2014} have explicitly aimed to mitigate QoS interference in the cloud using heuristic based decision tree search, however, they only intend to autoscale software control primitives.

%Addis et al. \cite{profit-cent-local-search} formulate the decision making process of autoscaling as a weighted-sum of response time and cost. The optimization is resolved by local-search strategy algorithm consists of two parts: a method for finding an initial decision and a local-search method for improving that decision. The  decision making in the work of Fernandez et al. \cite{2014-profiling-decision-tree-scaling-2014} has also usd weighed-sum formulation of QoS and cost. The formulation is then evaluated using a decision tree search, the decision which has the highest score from the formulation is used for autoscaling. Padala et al. \cite{HPL-2008-123R1-mimo} have also used weighted-sum equation containing the QoS and cost. The formulation is different depends on cases, e.g., CPU contention or disk contention. The optimization is then resolved using quadratic programming. 

We have also found that metaheuristic algorithms are popular for autoscaling decision making in SSCAS, because they can often efficiently address NP-hard problems with approximated results under no assumptions of the problem. The most common algorithms include: Tabu Search \cite{sla-provision},  Genetic Algorithm (GA) \cite{software-RP-two-loops}  \cite{2014-navie-ANN-GA-2014}  \cite{BRGA-resource} , Particle Swarm optimization (PSO) \cite{software-RP-two-loops}. As an example, Zhu et al. \cite{sla-provision} formulate the autoscaling decision making as optimize a weighted-sum formulation of response time and cost. To optimize the objectives, the authors apply a hybrid Tabu Search, which relied on iterative gradient descent.

%in every iteration, the current matrix is disturbed and a new decision is generated as initial solution of gradient descent. After reaching a particular fixed point, the variation of profit is calculated and the best configuration is returned.  

%Zhang et al. \cite{software-RP-two-loops} applies a weighted-sum of different QoS attributes for autoscaling decision making. The authors claim that the optimization problem can be easily resolved using meta-heuristics algorithm, e.g., GA or Piratical Swarm Optimization(PSO).   Minarolli and Freisleben \cite{2014-navie-ANN-GA-2014} formulate the decision making of autoscaling as a weighted-sum relation, which contains the QoS and cost of all co-hosted VMs. The optimization is resolved using GA. In the work from Jiang et al\cite{06119056}, the decision making is a weighted-sum formulation of response time, throughput, CPU utilization and cost. The optimization is resolved by GA. The BRGA framework \cite{BRGA-resource} has also relied on weighted-sum representation, which contains QoS attributes and cost, to express the decision making process. The optimization is resolve by using GA.

% pareto
Finally, pareto relation can explicitly handle multi-objectivity for autoscaling in the cloud without the need to specify weights on the objectives \cite{2014-eplison-GA-weigh-h-scaling-2014},  \cite{wosp10sla}, \cite{E3-R-extended}, \cite{GA-full-simulation}, \cite{Chen:2015:tsc-pending}. For example, in E$^3$-R \cite{E3-R-extended}, the decision making problem is formulated using Pareto relation, where it is resolved by using Non-dominated Sorting Genetic Algorithm-II (NSGA-II). Further, the approach applies objective reduction technique with an aim to remove the objectives, which are not significantly conflicted with the others, from the decision making process. Chen and Bahsoon~\cite{Chen:2015:tsc-pending} exploit Multi-Objective Ant Colony Optimization (MOACO) for trade-off decision making when autoscaling cloud-based services. The authors consider the trade-off between naturally conflicting objectives and between competing services (i.e., QoS interference). Further, a compromise-dominance mechanism is proposed to find well-compromised trade-off decision. 
%Kateb et al. \cite{2014-eplison-GA-weigh-h-scaling-2014}  formulate the decision making process as Pareto-based multi-objective optimization, which is solved by Multi-objective Evolutionary Algorithm (MOEA), e.g., Non-dominated Sorting Genetic Algorithm-II (NSGA-II)
%

%At each generation, NSGA-II identifies non-dominating solutions, within which crowding distance is used to calculate the distance between an individual and its neighbors in order to preserve diversity. In particular, each generation of the search is evaluated using epsilon dominance which is a relaxed form of the commonly used Pareto-dominance metric. The decisions, which have better epsilon dominance and smaller crowding distance, will be selected to the next generation. 

%In this way, the author aims to reduce the overhead while not affecting the quality of decisions. 

\end{itemize}
%Similarly, Li et al. \cite{wosp10sla} have also leveraged on pareto front for autoscaling decision. This is achieved by using S-Metric-Selection Evolutionary Multi-objective Optimization Algorithm (SMS-EMOA) as the search-based optimization algorithm.  Frey et al. \cite{GA-full-simulation} have also applied pareto analysis on the autoscaling decision making process, where GA is used for optimizing the QoS and cost. But the search space is restricted by bundles of VMs.

%\end{itemize}

\subsection{Experimental Evaluation on SSCAS}

Another important step in SSCAS research is to quantitatively evaluate the proposed solution, which is often achieved via experimental analysis. To this end, setting the experiments are of high importance to researchers in this field. Table~\ref{tb:exp} summarizes the common infrastructure, benchmarks and workload trace from the considered studies. As we can see that there are studies chose to use simulator for controlled experiments and ease of complexity. However, simulators may not fully capture the realistic environment. As a result, custom private cloud and public cloud are also exploited for both controlled and open experiments. Notably, custom private cloud can be much more flexible on choosing the underlying software, e.g., one may utilize the hypervisor or container directly or choose to use higher level software such as OpenStack~\cite{open}. It is worth noting that custom private and public cloud can share similar underlying software and tools, but they may require expertise on different levels of abstraction. For example, one may choose to deploy SSCAS on a custom private cloud that makes use of Xen~\cite{xen}\textemdash the same hypervisor that underpins Amazon EC2~\cite{iaas} which contains additional high-level interfaces and restrictions. Benchmarks are not required for simulator, but it is crucial for both private and public cloud infrastructure. A wide range of benchmarks have been exploited to evaluate SSCAS, from simple web hosting to complex multi-tier software. Finally, the workload traces can be either synthetic in which a fixed pattern is generated by the workload generator (e.g., JMeter~\cite{jmeter}); or real where recorded traces form different real domains are used to stress the benchmark and SSCAS.

\subsection{Implementation of Scaling}

The actual implementation of scaling depends on the underlying scenarios, e.g., the type of hypervisor/container, the cloud-based applications and the actual cloud control primitives among the others. Existing virtulization techniques have provided readily available commands and tools to support autoscaling at runtime. For example, if the underlying hypervisor was Xen~\cite{xen}, then resource such as CPU and memory of a VM, as well as create/destroy VMs can be scaled dynamically using the $xm$ command. Regarding the actual scaling methods, vertical scaling will have trivial effects on the states of the cloud-based applications, thus they can be directly applied using the command support by hypervisor/container. For horizontal scaling, making new replicas or removing old one needs consistency guarantee on stateful applications/services, which can be ensured by the underlying hypervisor through various readily available protocols. For example in Xen, horizontal scaling can be achieved via \emph{primary-backup replication} or \emph{asynchronous checkpointing}, \emph{etc}.

\begin{table}[t!]

\caption{\fontsize{8}{8}\selectfont Infrastructure, Benchmarks and Workload Traces for Experimental Evaluation on SSCAS.}
\label{tb:exp}
%\vspace{0.3cm}

\scriptsize
\begin{tabularx}{\textwidth}{|p{2cm}|p{2cm}|X|} \hline
\multicolumn{2}{|c|}{\textbf{\emph{Experiment Setup}}}&\textbf{\emph{Software and Tools}}\\ \hline 

\multirow{3}{*}{Infrastructure}&Simulator&CloudSim~\cite{CloudSim}, CDOSim~\cite{full-simulation-model}, CloudAnalyst~\cite{CloudAnalysis}, DCSim~\cite{DCSim}\\ \cline{2-3}
&Private Cloud&Xen~\cite{xen}, VMWare ESXi~\cite{vmware}, KVM~\cite{kvm}, Docker~\cite{docker}, OpenStack~\cite{open} , Eucalyptus~\cite{euc} \\ \cline{2-3}
&Public Cloud&Amazon EC2~\cite{iaas}, RackSpace~\cite{rack}, Azure~\cite{azure}, Google Compute Engine~\cite{google} \\   \hline
\multicolumn{2}{|c|}{Benchmarks}&RUBiS~\cite{rubis}, RUBBoS~\cite{rubbos}, TPC-W~\cite{tpcw}, WikiBench~\cite{wikibench} \\ \hline
 \multicolumn{2}{|c|}{Workload Traces}&Synthetic trace, FIFA98~\cite{f98}, Wikipedia~\cite{wikipedia}, ClarkNet~\cite{clark}\\
\hline

\end{tabularx}
\vspace{-0.3cm}
\end{table}

\section{Reflections and Open Challenges for SSCAS Research}
\label{sec:diss}

In this section, we reflect on the finding of our survey and taxonomy; state the open challenges as well as discuss industrial situation and pricing strategy for SSCAS.

%In this section, we reflect on the key findings obtained form the survey and taxonomy. To this end, we firstly discuss the observations from the survey with respect to different logical aspects of SSCAS research and provide inferences on the causes of those phenomenas. Next, we present a deeper insights into the open problems and challenges for future research on SSCAS.

\subsection{Discussion and Comparison on Existing SSCAS Research}
We now discuss the most noticeable observations by reviewing the existing SSCAS research. We carefully position our discussions in light of the different logical aspects of SSCAS.

\subsubsection{The Levels of Self-Awareness and Self-Adaptivity in Cloud Autoscaling Systems}  

\emph{Stimulus-awareness} has been considered in all the 109 studies as it is the most fundamental levels in self-awareness principles, because it is the basic requirement for a software system to be able to adapt. \emph{Time-} (52\%) and \emph{goal-awareness} (52\%) receives same attention due to the fact that the objective models often contain historical information and they can be used to reason about goals. In contrast, handling \emph{interaction-} (13\%) and \emph{meta-self-awareness} (7\%) are less popular as the former requires to handle QoS interference while the latter often come with extra complexity.

While \emph{self-optimizing} and \emph{self-configuring} have been the major themes for SSCAS, we found very little studies that considered \emph{self-healing} (5\%) and only one work targets for \emph{self-protecting}. This is obvious as the fundamental idea of autoscaling is not for security related purposes but for the performance related quality, which is often much more appealing for cloud consumers.

Also, 67\% of the studies has ignored the importance of specifying the required knowledge at the architecture level, entailing the risk of limited awareness~\cite{epics}. 
%and thus negatively affecting the adaptation of SSCAS~\cite{epics}.

\subsubsection{Architectural Pattern}

The generic feedback (82\%), particularly the single and close loop, has been the predominant architectural pattern for SSCAS. MAPE (15\%) is ranked the second and OAD (3\%) being the much less popular one, as OAD often assume the involvement of human decisions maker which is difficult in the case of SSCAS.

The reason could be due to the fact that the feedback loops are flexible and simple to be realized, providing the basic components to achieve self-adaptivity. However, such design can limit the consideration of required knowledge for the autoscaling system to perform adaptations, or the consideration is rather simple and coarse-grained. In contrast, such an issues has been relaxed by OAD and MAPE as they are more stricted by predefining components to capture different aspects of a SSCAS. We see that MAPE is clearly more popular than OAD because the former can be good for separation of concepts (e.g., \emph{Analyze},  \emph{Plan} and \emph{Knowledge}) and for expressing the sequential interactions between those concepts while the latter fails to capture runtime aspects of the SSCAS, as it is mainly designed for decoupling loops of different human activities. Nevertheless, these architectural patterns lack of fine-grained representation of the required knowledge. Thus, it is not immediately intuitive what level(s) of the knowledge is required by each logical aspect of a SSCAS.

\subsubsection{QoS Modeling}

Both analytical model and machine learning model, included in 43\% and 38\% of the studies respectively, are widely exploited in QoS modeling for SSCAS while the simulation (10\%) and hybrid (9\%) approach are clearly less popular. This could be because analytical model is good for runtime efficiency, simplicity, interoperability, and they could be very effective if all of their assumptions are satisfied. However, analytical approaches generally require in-depth knowledge about the likely behaviors of the system being modeled, i.e., some knowledge about the system's internal structure or environmental conditions. Such an issues is resolved by using machine learning model which are often assumptions free, and more importantly, they are able to continually evolve themselves at runtime in order to cope with dynamics and uncertainty. Nevertheless, depending on the learning algorithm, the resulting overhead can be high (e.g., the nonlinear ones) and the accuracy is sensitive to the situation (e.g., fluctuation of the data trend).

In contrast, simulation exhibits static nature and it is restricted by a wide set of assumptions, including e.g., the distribution of workload and the effects of QoS interference, etc. needs complex human intervention and assumptions. However, it is believe that simulations model could be the most accurate way to model QoS when the all assumptions are satisfied~\cite{full-simulation-model}. Hybrid model, as in 6 of the studies surveyed, could potentially combine the strengths from different models. 

%However, hybrid model often create extra overhead which needs to be carefully considered.

We noted that only 13\% of the studies intend to address QoS interferences when modeling the QoS in SSCAS. This might be because considering QoS interference will significantly increase the dimensionality in the model, which in turn, rendering the problem much more complex. Such a complexity makes human analysis very difficult, if not impossible. As a result, for those studies that do consider QoS interference, machine learning algorithms are often exploited.

There are plenty of studies (61\%) that consider dynamic (or semi-dynamic) structure of a QoS model (i.e., those that denoted as both \emph{dynamic} and \emph{semi} in Table~\ref{tb:qos}), however, the dynamic related to the input features have been rarely researched simultaneously, i.e, only 7\% of the studies (for those that denoted as \emph{dynamic} in Table~\ref{tb:qos}). Indeed, changing the inputs of a model could be useful only when the dimensionality of a model is high; that is to say, changing the input features might not cause significant difference if the considered total number of inputs is around, e.g., less than five. The majority of the \emph{dynamic} (or \emph{semi-dynamic}) modeling are machine learning based, leaving the \emph{static} ones are largely analytical or simulation based. This is obvious, as the nature of those modeling approaches determine the extents to which they can be changed when they are built.

%The machine learning modeling exploited in the studies exhibits high diversity due to the no-free-lunch theorem, which states that there is no single algorithms that can outperforms the others across all the scenarios.

A considerable amount of studies (65\%) have claimed that their QoS models can work on any given inputs and/or output, as shown in Table~\ref{tb:qos-inout}. This happens mostly for machine learning and simulation approaches. However, during experimental analysis, the highest number of QoS that being modeled and the number of inputs were both four~\cite{Chen:2014:ucc}~\cite{Chen:2015:tse-pending}. The most commonly considered output is response time and the inputs are hardware resources, particularly CPU, memory and number of VM. This complies with the current trend in the cloud computing market.

\subsubsection{Granularity of Control}

In SSCAS research, the service/application level of granularity is the most popular one, which yields 45\% of the studies. This is because focusing on the finest granularity of control can achieve the maximum level of scalability, which particularly fits the cloud. However, fine granularity of control is achieved in the expenses of the globally optimal quality of the cloud, since no interactions between service/application are considered. In contrast, focusing on cloud-level is another extreme, which trades scalability for global optimality. Considering multiple levels could be a solution to reach a better trade-off as discussed in a small amount of studies (8\%), but how and when to select the levels to consider imposes additional challenges.

Notably, most of the studies (72\%) has relied on the control of each application in cloud regardless to the actual granularity of control. The reason being could be due to the fact that cloud-based application (or a collection of services) is the most crucial unit for consumers to experience the benefit of cloud, which is a common interest for both cloud consumers and providers.  

%Decentralization, adopted by 74\% of the studies, is the most popular for all the granularity due to the high scalability in the expense of global optimality. The only exception is the cloud level as the it often aims for global optimal result and thus centralized decision making is more preferable. 

\subsubsection{Decision Making}

Generally, rule based control is a highly intuitive approach for autoscaling decision making, and it also has negligible overhead. However, the static nature of the rules requires to assume all the possible conditions and the effects of those decisions that are mapped to the conditions, which is highly depending on the assumptions. To resolve such an issue, control theoretic approach appear to be an effective solutions as it is also efficient while require very little assumptions. However, the major drawback of control theoretic approaches is that they require to make many actuations on the physical system, in order to collect the 'errors' for stabilizing itself. This means that amateur decisions are very likely to be made. In addition both approaches lack of handling multi-objectivity and the trade-off; they often fail to cope with the problem where there is a large number of autoscaling decisions, which is common for cloud. In contrast, search-based optimization, especially the explicit search, makes loose assumption about the number of autoscaling decisions and is able to find optimality (or near-optimality) under highly dynamic and uncertain environment. Therefore as we have shown, search-based optimization, either implicit or explicit, is the most popular approach for making decisions in SSCAS. 

We noted that there is a considerable amount studies (63\%) that do not attempt to explicitly consider trade-off during decision making of SSCAS, as they assumed single objective or rely completely on human preferences. The reason might be attributed to the fact that such formalization is simple and straightforward, which can work well when there is a strong preference on an objective. The rest studies handle multi-objectivity via either weighted sum or pareto relation.

QoS interference is again absent in many studies (84\%) due to the fact that considering it will unavoidably increase the dimensionality (i.e., objectives) during the decisions making, leading to a more complex problem. This would make the problem unsolvable by many existing approaches.

We found that most of the studies (78\%) have claimed that their decision making approach could handle any given objectives, thus they have considered arbitrary QoS attributes and cost as the objectives in SSCAS since these are the most critical indicator for cloud-based services and applications. The highest number of objectives that were considered during the experiments are five~\cite{Chen:2015:tsc-pending} though. CPU, memory and number of VM are the most popular control primitives in decision making because they are the most straightforward dimension to be scaled. However, only as little as 9\% of the studies consider the interplay between software and hardware control primitives. To apply solution that can not handle a large search space of the decision making, one often reduce the search space by introducing fixed bundles, which is an assumption made by a considerable amount of studies, i.e., 34\%. However, such reduction has the risk that some good solution can be ruled out during the process. Finally, while both vertical and horizontal scaling have been considered in the majority of the studies (66\%), focusing solely on horizontal scaling is more popular than the vertical one as the former is more widely supported by major cloud vendors.

\subsection{Open Problems and Challenges for SSCAS research}

%According to the observations drawn from the survey and taxonomy, there are a few key points that have not been explicitly treated or have not been resolved properly. Those key points, when explicitly taken into account, could rise new problems and challenges for future SSCAS research. 
Drawing on the survey and taxonomy, in the following, we specify the open problems and challenges for future SSCAS research and make suggestions for potential research directions where appropriate.

\begin{itemize}[leftmargin=.35cm]

\item \emph{Explicit Knowledge Representations are Required in SSCAS Architecture:} As we can see from Section~\ref{sec:sscas}, only 33\% of the studies intend to discuss the required knowledge at the architecture level. This means that, in the remaining 67\% work, it is more difficult to capture more complex and advanced levels of knowledge, as evident by the fact that most work does not go beyond the basic \emph{stimulus-awareness}. Indeed, studies~\cite{epics}~\cite{2014_epics_handbook} \cite{7185305} \cite{epics_survey} \cite{Chen2016:book} have found that, for self-aware and self-adaptive software systems in general, the absence of explicit consideration for the fine-grained representation of the knowledge in the architecture can results in, e.g., improper inclusion of unnecessary knowledge and/or missing important knowledge that can improve adaptation quality when developing autoscaling systems. 67\% studies which do not discuss knowledge at the architecture level implies that such an issue is often overlooked and it is remain unresolved in the SSCAS context, urging the need of further investigations.

The challenge here lies in the fact of how can one systematically distinguish different levels of knowledge and how they can be architected into SSCAS in a principal way. We argue that the required levels of knowledge and their representations can be declared in light with the formal principle of self-awareness. In particular, a potential way is to follow the handbook~\cite{2014_epics_handbook} for mapping different levels of knowledge into a concrete SSCAS architecture.

%\item \emph{Open Looped Architectural Pattern for SSCAS is Worth Investigated:} From Section we note that only limited studies have considered open architectural pattern where the information from human can directly influence the adaptation loop. This might seem to be difficult for SSCAS as human intervention is challenge at runtime. However, we argue that open does not necessarily mean that the architectural pattern extracts information from human under the awareness of them, it can actually achieve such without human notification.

\item \emph{Multiple Loops can Create More Benefit for SSCAS Architecture:} From Section~\ref{sec:arch} we noted that the majority of the studies has considered single loop, which could cause the problem of high coupling in the design of SSCAS. The multiple loops, on the other hand, helps to achieve better separation between different aspects of SSCAS, leading to fine-grained and localized adaptation. The challenge here is how many and at what levels of abstraction one should place the loops within the SSCAS Architecture. We suggest that designing multiple looped SSCAS architecture with respect to what levels of knowledge the system required could be a neat solution~\cite{Chen:2015:computer}.

\item \emph{QoS Interference Should be Explicitly Handled in QoS Modeling and Decision Making Process:} Our survey results (see Section~\ref{sec:qos} and ~\ref{sec:dm}) indicate that only less than 16\% of the studies took QoS interference into account. Missing QoS interference in the model and decision making could lead to incorrect or misleaded autoscaling decisions, as the cloud-based services would be unavoidably affected by the dynamic behaviors of its neighbors. However, incorporating QoS interference rises the challenge of dimensionality which causes the model and decision making much more complex. Therefore, this challenge calls for novel approach to reduce the dimensionality, or mitigate its negative effects, during the model and decision making in SSCAS~\cite{Chen:2015:tse-pending} \cite{Chen:2015:tsc-pending}.

\item \emph{The Interplay Between Software Configuration and Hardware Resources is Important:} Most existing studies of SSCAS focus on hardware resources as IaaS level only. However, as shown in~\cite{2013-JRAO-most-closest-work-2013} \cite{software-RP-two-loops} \cite{2014-decision-tree-software-CP-2014} \cite{Chen:2015:tse-pending}, various software configurations at PaaS level could interplay with each other and the hardware resources, which in turn, affect the QoS of cloud-based services. The challenge is how to create a holistic approach that combines both PaaS and IaaS level in SSCAS.

\item \emph{Dynamic Feature Selection is Required for QoS Modeling in SSCAS:} Section~\ref{sec:qos} indicates that the majority of the existing studies have ignored primitive selection in the QoS modeling or have been relying on manual approach, because the assumed dimensions of inputs is rather limited. However, when both QoS interference and software configurations are considered, selecting the most significant features in the model becomes a crucial task since there could be an explosion of the primitives space~\cite{Chen:2015:tse-pending}~\cite{2013-single-learner-filter-wrapper-LR-2013}. Challenge here lies in how to evaluate the effectiveness of feature combination on the model accuracy while generating reasonable overhead. Given the arbitrary types of feature, which calls for generic and efficient feature selection design for SSCAS.

\item \emph{More Flexible Granularity of Control in SSCAS is Needed:} Single, static and fixed granularity of control is predominately exploited in existing SSCAS research. To better handle dynamic and uncertainty, it could be more beneficial to introduce multiple granularity and/or dynamically adjust the granularity of control at runtime~\cite{11icde_smartsla_full} \cite{Chen:2014:seams}, as the granularity of control implies a trade-off between the global optimality of SSCAS and the imposed overhead. The challenge is how to explicitly capture the objective-dependency when designing granularity of control.

%thus creating the ability to reason about the effects of granularity on global optimality and the imposed overhead.

\item \emph{The Assumptions on the Bundles of Resources Needs to be Relaxed:} From Section~\ref{sec:dm} we noted that while most of the studies have not constrained the possible autoscaling decisions with respect to the fixed bundles, there are still certain amount of studies that heavily rely on the fixed types of bundles, e.g., a search space of 57 VM instance types on Amazon EC2. However, renting bundles cannot and does not reflect the interests of consumers and the actual demand of their cloud-based services. We argue that future cloud autoscaling would inevitably needs to take arbitrary combination of software configurations and resources into account, as what has already been supported in Google Compute Engine~\cite{google}. As a result, autoscaling decision making imposes a challenging problem that faces with an explosion of decision space (e.g, millions of alternatives), calling for novel and efficient approach to achieve optimal or near-optimal quality.

%\item \emph{Vertical Scaling Needs More Investigations:} 

\item \emph{The Trade-off Between Conflicting Objectives Should be Explicitly Handled:}  As shown in Section~\ref{sec:dm}, the approaches have mostly ignored trade-off. There is also certain amount of explicit search-based optimization studies has assumed only single objective. However, this can restrict the applicability of SSCAS as the decision making would fail in identifying good trade-off points or strongly bias to the single objective. Alternatively, there is also considerably large amount of studies exploit weighted sum objective aggregation, which embed the trade-off in a single representation. However, it is well-known that the relative weights are difficult to be tailored and a single aggregation could restrict the search, causing limitation when searching for good decisions spread over the search space. Further, achieving balanced trade-off have only being explored in very limited studies, e.g.,~\cite{Chen:2015:tsc-pending}. The challenge here is how to search for decisions that contain good convergence and diversity, and eventually selecting the one that has the most balanced trade-off for scaling. We advocate that stochastic optimization approach, particularly nature inspired algorithms, can be promising in addressing such a challenge.

\item \emph{More Real World Case Studies of SSCAS are Needed:} We found that real world cases and scenarios of SSCAS, especially those with large scale and practical application, are absent in many studies. Indeed, those studies impose many challenges beyond the perspective of research, but they can be the only way to fully verify the potentials, effectiveness and impacts of SSCAS.

\end{itemize}

\subsection{Current Industrial Situation of SSCAS}

Industrial cloud providers (e.g., Amazon~\cite{iaas} and RightScale~\cite{rightscale}) have been relying on model-free, simple rule and policy based autoscaling approaches for decades. These approaches leave the difficult problem of how to specify rules to cloud consumers, which may work well in the beginning when the demand and complexity of cloud-based applications are simple and straightforward. However, recently the level of complexity (e.g., in terms of the number of cloud control primitives) of cloud is changing to a state that makes human analysis very difficult, especially under conflicting objectives and a large number of alternative autoscaling decisions~\cite{2014-eplison-GA-weigh-h-scaling-2014},  \cite{wosp10sla}, \cite{E3-R-extended}, \cite{Chen:2015:tsc-pending}. Specifically, as discussed in Section 5.1.5, those approaches suffer two significant pitfalls. (i) They requires understanding of the application and domain knowledge to determine the mapping between conditions and actions, which can significantly affect the quality of scaling~\cite{al2013impact}, \cite{software-RP-two-loops}. (ii) They cannot adapt to dynamically changing workload or state of the applications~\cite{2014-navie-ANN-GA-2014}, \cite{Chen:2015:tsc-pending}. 

As a result, engineering advanced SSCAS is an inevitable trend in this area; the reason why current big cloud providers have not yet widely implemented them could be due to the fact that SSCAS itself is not mature to the state which it can be reliably adopted. However, as our survey reports, researchers and practitioners have been working on overcoming these challenges for almost a decade. There has been some attempts to apply SSCAS commercially, for example, Microsoft Azure~\cite{azure} has recently benefited from Aneka~\cite{vecchiola2009aneka}, a research effort supporting high level framework, which contains a more advanced and complicated SSCAS that relies on search-based optimization as part of its subsystems. Aneka's work is an evidence of how pending industrial challenges had informed research; the results are now incorporated in Azure. Nevertheless, we envision more progress on enhanced, scalable and cost-effective effective solutions for both the cloud providers and consumers.

\subsection{Discussion on the Pricing for Cloud Autoscaling}

Indeed, more advanced autoscaling approaches in SSCAS (e.g., machine learning and search-based optimization) may impose additional computational resources. Furthermore, advances in autoscaling cannot be done in isolation of pricing (dynamic metering and pricing in particular), as both the cost and revenue are acknowledged among the drivers for the industrial need of more advanced solutions. However, upfront investment in additional recourse can be arguably paid off, in situations where scale and dynamic demand is effectively enabled. This can be observed through greater pay off through better utilization and SLA guarantee, which in turn, improve the overall reputation and thus attracting more consumers. As analyzed in numerous existing work~\cite{lorido2014review}\cite{qu2016auto}\cite{GA-full-simulation}\cite{Chen:2015:tse-pending}\cite{Chen:2015:tsc-pending}, the additional resources spent are actually marginal compared with the savings obtained through more accurate, effective autoscaling. In case the cloud provider wishes to charge the consumers for the services provided by SSCAS, there could be two ways to achieve this: (i) charging the computation utilized by the SSCAS through existing pricing schema, e.g., the \emph{reserved} or \emph{spot instance} from Amazon EC2. Here, the SSCAS is an optional service which would be priced as normal instance for the consumers' application/services. (ii) The charge of SSCAS is combined with the normal price per time unit in existing pricing schema, e.g., instead of charging \$1/hour of an instance, it can be priced as \$1.3/hour where the extra \$0.3/hour is for the SSCAS. Here, the SSCAS is a default and mandatory service to the cloud consumers.

\section{Conclusion}
\label{sec:con}
In this article, we survey the state-of-the-art research on SSCAS and provide a taxonomy based on our findings. Specifically, we review the literature with respect to the research questions presented in Section~\ref{sec:rq}. According to our survey, the key findings are:

\begin{itemize}[leftmargin=.35cm]

\item[--] \emph{Stimulus-}, \emph{time-} and \emph{goal-awareness} are the most widely considered levels of knowledge in SSCAS. \emph{Self-configuring} and \emph{Self-optimizing} are the most popular self-adaptivity notions in SSCAS.
\item[--] Feedback loop is the most commonly exploited architectural pattern for engineering SSCAS.
\item[--] Analytical model and machine learning based model are prominent for QoS modeling in SSCAS.
\item[--] Controlling at the level of service/application is the mostly applied granularity.
\item[--] Search-based optimization is the most common approach for making autoscaling decisions.

\end{itemize}
Apart from those observations, we also gain many insights on the open problems and challenges for future SSCAS research. The most noticeable ones are:

\begin{itemize}[leftmargin=.35cm]
\item[--] Explicit knowledge representations are required in SSCAS architecture.
\item[--] Multiple loops can create non-trivial Benefits for SSCAS architecture. 
\item[--] QoS Interference should be explicitly handled in QoS modeling and decision making process.
\item[--] The interplay between software configurations and hardware resources is non-trivial.
\item[--] Dynamic Feature selection is required for QoS modeling in SSCAS.
\item[--] More flexible granularity of control in SSCAS is needed: 
\item[--] The assumptions on the bundles of resources needs to be relaxed.
\item[--] The trade-off  between conflicting objectives should be explicitly handled.
\item[--] More real world cases and scenarios of SSCAS are needed.
\end{itemize}

We hope that our survey and taxonomy will motivate further research for more intelligent cloud autoscaling system and its interactions with the other problems in the cloud computing paradigm.
%we described the background and definitions for cloud autoscaling, self-aware and self-adaptive systems in general. Subsequently, we outlined the major requirements for the key logical aspects of autoscaling in the cloud, and discuss the key state-of-the-art developments proposed for each of the logical aspects, from which a taxonomy is provided. We hope that our systematic survey and taxonomy will motivate further research for more intelligent cloud autoscaling and its interaction with the other problems in the cloud.

\balance

% Bibliography
\bibliographystyle{acm}
\def\bibfont{\fontsize{8}{9.5}\selectfont} 
\bibliography{references}

\end{document}

%% file: tikz/self-awareness.tex
\begin{tikzpicture}
\begin{axis}[
    ybar=0.1cm,
    bar width=0.7cm,
    ymax=130,
 enlarge x limits=0.2,
 enlarge y limits=0.01,
    legend style={at={(0.15,0.95)},
      anchor=north,legend columns=3},
    ylabel={Number of Papers},
     xlabel={Self-Awareness},
    symbolic x coords={Stimulus,Time,Goal,Interaction,Meta},
    xtick=data,
    nodes near coords,
    nodes near coords align={vertical},
    width=10cm,
    height=5cm
    ]
\addplot[fill=black!100!white] coordinates {(Stimulus,109) (Time,57) (Goal,57) (Interaction,14) (Meta,8)};
%\addplot[fill=black!60!white] coordinates {(Training,0.53) (Prediction,0.33)};
%\addplot[fill=black!40!white] coordinates {(Training,0.12) (Prediction,0.12)};
%\addplot[fill=black!20!white] coordinates {(Training,0.37) (Prediction,0.54)};
%\addplot[fill=white!100!white] coordinates {(Training,60.68) (Prediction,0.16)};
%
%\legend{HT,NB,SGD,$k$NN,MLP}
\end{axis}
\end{tikzpicture}

%% file: tikz/self-adaptivity.tex
\begin{tikzpicture}
\begin{axis}[
    ybar=0.1cm,
    bar width=0.7cm,
    ymax=130,
 enlarge x limits=0.2,
 enlarge y limits=0.01,
    legend style={at={(0.15,0.95)},
      anchor=north,legend columns=3},
    ylabel={Number of Papers},
     xlabel={Self-Adaptivity},
    symbolic x coords={Configuring,Optimizing,Healing,Protecting},
    xtick=data,
    nodes near coords,
    nodes near coords align={vertical},
    width=10cm,
    height=5cm
    ]
\addplot[fill=black!100!white] coordinates {(Configuring,107) (Optimizing,76) (Healing,5) (Protecting,1)};
%\addplot[fill=black!60!white] coordinates {(Training,0.53) (Prediction,0.33)};
%\addplot[fill=black!40!white] coordinates {(Training,0.12) (Prediction,0.12)};
%\addplot[fill=black!20!white] coordinates {(Training,0.37) (Prediction,0.54)};
%\addplot[fill=white!100!white] coordinates {(Training,60.68) (Prediction,0.16)};
%
%\legend{HT,NB,SGD,$k$NN,MLP}
\end{axis}
\end{tikzpicture}

%% file: tikz/knowledge.tex
\definecolor{rosso}{RGB}{220,57,18}
\definecolor{giallo}{RGB}{255,153,0}
\definecolor{blu}{RGB}{102,140,217}
\definecolor{verde}{RGB}{16,150,24}
\definecolor{viola}{RGB}{153,0,153}

\makeatletter

\tikzstyle{chart}=[
    legend label/.style={font={\scriptsize},anchor=west,align=left},
    legend box/.style={rectangle, draw, minimum size=5pt},
    axis/.style={black,semithick,->},
    axis label/.style={anchor=east,font={\tiny}},
]

\tikzstyle{bar chart}=[
    chart,
    bar width/.code={
        \pgfmathparse{##1/2}
        \global\let\bar@w\pgfmathresult
    },
    bar/.style={very thick, draw=white},
    bar label/.style={font={\bf\small},anchor=north},
    bar value/.style={font={\footnotesize}},
    bar width=.75,
]

\tikzstyle{pie chart}=[
    chart,
    slice/.style={line cap=round, line join=round, very thick,draw=white,font={\large}},
    pie title/.style={font={\bf}},
    slice type/.style 2 args={
        ##1/.style={fill=##2},
        values of ##1/.style={}
    }
]

\pgfdeclarelayer{background}
\pgfdeclarelayer{foreground}
\pgfsetlayers{background,main,foreground}

\newcommand{\pie}[3][]{
    \begin{scope}[#1]
    \pgfmathsetmacro{\curA}{90}
    \pgfmathsetmacro{\r}{1}
    \def\c{(0,0)}
    \node[pie title] at (90:1.3) {#2};
    \foreach \v/\s in{#3}{
        \pgfmathsetmacro{\deltaA}{\v/100*360}
        \pgfmathsetmacro{\nextA}{\curA + \deltaA}
        \pgfmathsetmacro{\midA}{(\curA+\nextA)/2}

        \path[slice,\s] \c
            -- +(\curA:\r)
            arc (\curA:\nextA:\r)
            -- cycle;
        \pgfmathsetmacro{\d}{max((\deltaA * -(.5/50) + 1) , .5)}

        \begin{pgfonlayer}{foreground}
        \path \c -- node[pos=\d,pie values,values of \s]{$\v\%$} +(\midA:\r);
        \end{pgfonlayer}

        \global\let\curA\nextA
    }
    \end{scope}
}

\newcommand{\legend}[2][]{
    \begin{scope}[#1]
    \path
        \foreach \n/\s in {#2}
            {
                  ++(0,-10pt) node[\s,legend box] {} +(5pt,0) node[legend label] {\n}
            }
    ;
    \end{scope}
}

\begin{tikzpicture}
[
    pie chart,
    slice type={no}{blu},
    slice type={implicitly}{rosso},
    slice type={explicitly}{giallo},
   % slice type={sedia}{viola},
    %slice type={caffe}{verde},
    pie values/.style={font={\normalsize}},
    scale=2
]

    \pie{Description of Knowledge}{67/no,27/implicitly,6/explicitly}

    \legend[shift={(1.5cm,1cm)}]{{No knowledge \\ described (73)}/no, {Knowledge is \\implicitly described (29)}/implicitly, {Knowledge is \\explicitly described (7)}/explicitly}

\end{tikzpicture}

%% file: tikz/arch-pie.tex
\definecolor{rosso}{RGB}{220,57,18}
\definecolor{giallo}{RGB}{255,153,0}
\definecolor{blu}{RGB}{102,140,217}
\definecolor{verde}{RGB}{16,150,24}
\definecolor{viola}{RGB}{153,0,153}

\makeatletter

\tikzstyle{chart}=[
    legend label/.style={font={\scriptsize},anchor=west,align=left},
    legend box/.style={rectangle, draw, minimum size=5pt},
    axis/.style={black,semithick,->},
    axis label/.style={anchor=east,font={\tiny}},
]

\tikzstyle{bar chart}=[
    chart,
    bar width/.code={
        \pgfmathparse{##1/2}
        \global\let\bar@w\pgfmathresult
    },
    bar/.style={very thick, draw=white},
    bar label/.style={font={\bf\small},anchor=north},
    bar value/.style={font={\footnotesize}},
    bar width=.75,
]

\tikzstyle{pie chart}=[
    chart,
    slice/.style={line cap=round, line join=round, very thick,draw=white},
    pie title/.style={font={\bf}},
    slice type/.style 2 args={
        ##1/.style={fill=##2},
        values of ##1/.style={}
    }
]

\pgfdeclarelayer{background}
\pgfdeclarelayer{foreground}
\pgfsetlayers{background,main,foreground}

\newcommand{\pie}[3][]{
    \begin{scope}[#1]
    \pgfmathsetmacro{\curA}{90}
    \pgfmathsetmacro{\r}{1}
    \def\c{(0,0)}
    \node[pie title] at (90:1.3) {#2};
    \foreach \v/\s in{#3}{
        \pgfmathsetmacro{\deltaA}{\v/100*360}
        \pgfmathsetmacro{\nextA}{\curA + \deltaA}
        \pgfmathsetmacro{\midA}{(\curA+\nextA)/2}

        \path[slice,\s] \c
            -- +(\curA:\r)
            arc (\curA:\nextA:\r)
            -- cycle;
        \pgfmathsetmacro{\d}{max((\deltaA * -(.5/50) + 1) , .5)}

        \begin{pgfonlayer}{foreground}
        \path \c -- node[pos=\d,pie values,values of \s]{$\v\%$} +(\midA:\r);
        \end{pgfonlayer}

        \global\let\curA\nextA
    }
    \end{scope}
}

\newcommand{\legend}[2][]{
    \begin{scope}[#1]
    \path
        \foreach \n/\s in {#2}
            {
                  ++(0,-10pt) node[\s,legend box] {} +(5pt,0) node[legend label] {\n}
            }
    ;
    \end{scope}
}

\begin{tikzpicture}
[
    pie chart,
    slice type={fb}{blu},
    slice type={oad}{rosso},
    slice type={mape}{giallo},
    %slice type={sedia}{viola},
    %slice type={caffe}{verde},
    pie values/.style={font={\small}},
    scale=2
]

    \pie{Architectural Pattern}{82/fb,3/oad,15/mape}

    \legend[shift={(1.5cm,1cm)}]{{Feedback Loop (89)}/fb, {OAD (3)}/oad, {MAPE (17)}/mape}

\end{tikzpicture}

%% file: tikz/modeling-pie.tex
\definecolor{rosso}{RGB}{220,57,18}
\definecolor{giallo}{RGB}{255,153,0}
\definecolor{blu}{RGB}{102,140,217}
\definecolor{verde}{RGB}{16,150,24}
\definecolor{viola}{RGB}{153,0,153}

\makeatletter

\tikzstyle{chart}=[
    legend label/.style={font={\scriptsize},anchor=west,align=left},
    legend box/.style={rectangle, draw, minimum size=5pt},
    axis/.style={black,semithick,->},
    axis label/.style={anchor=east,font={\tiny}},
]

\tikzstyle{bar chart}=[
    chart,
    bar width/.code={
        \pgfmathparse{##1/2}
        \global\let\bar@w\pgfmathresult
    },
    bar/.style={very thick, draw=white},
    bar label/.style={font={\bf\small},anchor=north},
    bar value/.style={font={\footnotesize}},
    bar width=.75,
]

\tikzstyle{pie chart}=[
    chart,
    slice/.style={line cap=round, line join=round, very thick,draw=white},
    pie title/.style={font={\bf}},
    slice type/.style 2 args={
        ##1/.style={fill=##2},
        values of ##1/.style={}
    }
]

\pgfdeclarelayer{background}
\pgfdeclarelayer{foreground}
\pgfsetlayers{background,main,foreground}

\newcommand{\pie}[3][]{
    \begin{scope}[#1]
    \pgfmathsetmacro{\curA}{90}
    \pgfmathsetmacro{\r}{1}
    \def\c{(0,0)}
    \node[pie title] at (90:1.3) {#2};
    \foreach \v/\s in{#3}{
        \pgfmathsetmacro{\deltaA}{\v/100*360}
        \pgfmathsetmacro{\nextA}{\curA + \deltaA}
        \pgfmathsetmacro{\midA}{(\curA+\nextA)/2}

        \path[slice,\s] \c
            -- +(\curA:\r)
            arc (\curA:\nextA:\r)
            -- cycle;
        \pgfmathsetmacro{\d}{max((\deltaA * -(.5/50) + 1) , .5)}

        \begin{pgfonlayer}{foreground}
        \path \c -- node[pos=\d,pie values,values of \s]{$\v\%$} +(\midA:\r);
        \end{pgfonlayer}

        \global\let\curA\nextA
    }
    \end{scope}
}

\newcommand{\legend}[2][]{
    \begin{scope}[#1]
    \path
        \foreach \n/\s in {#2}
            {
                  ++(0,-10pt) node[\s,legend box] {} +(5pt,0) node[legend label] {\n}
            }
    ;
    \end{scope}
}

\begin{tikzpicture}
[
    pie chart,
    slice type={am}{blu},
    slice type={sim}{rosso},
    slice type={ml}{giallo},
    slice type={hm}{viola},
    %slice type={caffe}{verde}, 60
    pie values/.style={font={\small}},
    scale=2
]

    \pie{QoS Modeling Approach}{43/am,10/sim,38/ml,9/hm}

    \legend[shift={(1.5cm,1cm)}]{{Analytical Model (30)}/am, {Simulation (7)}/sim, {Machine Learning Model (26)}/ml,{Hybrid Model (6)}/hm}

\end{tikzpicture}

%% file: tikz/granularity-pie.tex
\definecolor{rosso}{RGB}{220,57,18}
\definecolor{giallo}{RGB}{255,153,0}
\definecolor{blu}{RGB}{102,140,217}
\definecolor{verde}{RGB}{16,150,24}
\definecolor{viola}{RGB}{153,0,153}

\makeatletter

\tikzstyle{chart}=[
    legend label/.style={font={\scriptsize},anchor=west,align=left},
    legend box/.style={rectangle, draw, minimum size=5pt},
    axis/.style={black,semithick,->},
    axis label/.style={anchor=east,font={\tiny}},
]

\tikzstyle{bar chart}=[
    chart,
    bar width/.code={
        \pgfmathparse{##1/2}
        \global\let\bar@w\pgfmathresult
    },
    bar/.style={very thick, draw=white},
    bar label/.style={font={\bf\small},anchor=north},
    bar value/.style={font={\footnotesize}},
    bar width=.75,
]

\tikzstyle{pie chart}=[
    chart,
    slice/.style={line cap=round, line join=round, very thick,draw=white},
    pie title/.style={font={\bf}},
    slice type/.style 2 args={
        ##1/.style={fill=##2},
        values of ##1/.style={}
    }
]

\pgfdeclarelayer{background}
\pgfdeclarelayer{foreground}
\pgfsetlayers{background,main,foreground}

\newcommand{\pie}[3][]{
    \begin{scope}[#1]
    \pgfmathsetmacro{\curA}{90}
    \pgfmathsetmacro{\r}{1}
    \def\c{(0,0)}
    \node[pie title] at (90:1.3) {#2};
    \foreach \v/\s in{#3}{
        \pgfmathsetmacro{\deltaA}{\v/100*360}
        \pgfmathsetmacro{\nextA}{\curA + \deltaA}
        \pgfmathsetmacro{\midA}{(\curA+\nextA)/2}

        \path[slice,\s] \c
            -- +(\curA:\r)
            arc (\curA:\nextA:\r)
            -- cycle;
        \pgfmathsetmacro{\d}{max((\deltaA * -(.5/50) + 1) , .5)}

        \begin{pgfonlayer}{foreground}
        \path \c -- node[pos=\d,pie values,values of \s]{$\v\%$} +(\midA:\r);
        \end{pgfonlayer}

        \global\let\curA\nextA
    }
    \end{scope}
}

\newcommand{\legend}[2][]{
    \begin{scope}[#1]
    \path
        \foreach \n/\s in {#2}
            {
                  ++(0,-10pt) node[\s,legend box] {} +(5pt,0) node[legend label] {\n}
            }
    ;
    \end{scope}
}

\begin{tikzpicture}
[
    pie chart,
    slice type={s}{blu},
    slice type={v}{rosso},
    slice type={p}{giallo},
    slice type={c}{viola},
    slice type={m}{verde},
    pie values/.style={font={\small}},
    scale=2
]

    \pie{Granularity of Control}{45/s,22/v,5/p,20/c,8/m}

    \legend[shift={(1.5cm,1cm)}]{{Service/Application Level (41)}/s, {VM/Container level (20)}/v, {PM Level (5)}/p,{Cloud Level (18)}/c,{Multiple Levels (8)}/m}

\end{tikzpicture}

%% file: tikz/decision-pie.tex
\definecolor{rosso}{RGB}{220,57,18}
\definecolor{giallo}{RGB}{255,153,0}
\definecolor{blu}{RGB}{102,140,217}
\definecolor{verde}{RGB}{16,150,24}
\definecolor{viola}{RGB}{153,0,153}

\makeatletter

\tikzstyle{chart}=[
    legend label/.style={font={\scriptsize},anchor=west,align=left},
    legend box/.style={rectangle, draw, minimum size=5pt},
    axis/.style={black,semithick,->},
    axis label/.style={anchor=east,font={\tiny}},
]

\tikzstyle{bar chart}=[
    chart,
    bar width/.code={
        \pgfmathparse{##1/2}
        \global\let\bar@w\pgfmathresult
    },
    bar/.style={very thick, draw=white},
    bar label/.style={font={\bf\small},anchor=north},
    bar value/.style={font={\footnotesize}},
    bar width=.75,
]

\tikzstyle{pie chart}=[
    chart,
    slice/.style={line cap=round, line join=round, very thick,draw=white},
    pie title/.style={font={\bf}},
    slice type/.style 2 args={
        ##1/.style={fill=##2},
        values of ##1/.style={}
    }
]

\pgfdeclarelayer{background}
\pgfdeclarelayer{foreground}
\pgfsetlayers{background,main,foreground}

\newcommand{\pie}[3][]{
    \begin{scope}[#1]
    \pgfmathsetmacro{\curA}{90}
    \pgfmathsetmacro{\r}{1}
    \def\c{(0,0)}
    \node[pie title] at (90:1.3) {#2};
    \foreach \v/\s in{#3}{
        \pgfmathsetmacro{\deltaA}{\v/100*360}
        \pgfmathsetmacro{\nextA}{\curA + \deltaA}
        \pgfmathsetmacro{\midA}{(\curA+\nextA)/2}

        \path[slice,\s] \c
            -- +(\curA:\r)
            arc (\curA:\nextA:\r)
            -- cycle;
        \pgfmathsetmacro{\d}{max((\deltaA * -(.5/50) + 1) , .5)}

        \begin{pgfonlayer}{foreground}
        \path \c -- node[pos=\d,pie values,values of \s]{$\v\%$} +(\midA:\r);
        \end{pgfonlayer}

        \global\let\curA\nextA
    }
    \end{scope}
}

\newcommand{\legend}[2][]{
    \begin{scope}[#1]
    \path
        \foreach \n/\s in {#2}
            {
                  ++(0,-10pt) node[\s,legend box] {} +(5pt,0) node[legend label] {\n}
            }
    ;
    \end{scope}
}

\begin{tikzpicture}
[
    pie chart,
    slice type={r}{blu},
    slice type={c}{rosso},
    slice type={s}{giallo},
    pie values/.style={font={\small}},
    scale=2
]

    \pie{Decision Making}{20/r,21/c,59/s}

    \legend[shift={(1.5cm,1cm)}]{{Rule-based Control (18)}/r, {Control Theory (19)}/c, {Search-based Optimization (55)}/s}

\end{tikzpicture}

%% file: autoscaling-survey.bbl
\begin{thebibliography}{100}

\bibitem{iaas}
Amazon elastic compute cloud.
\newblock http://aws.amazon.com/ec2/.

\bibitem{jmeter}
Apache jmeter.
\newblock http://jmeter.apache.org/.

\bibitem{clark}
Clarknet http trace.
\newblock http://ita.ee.lbl.gov/html/contrib/ClarkNet-HTTP.html.

\bibitem{docker}
Docker: The cloud container.
\newblock https://www.docker.com/.

\bibitem{euc}
Eucalyptus cloud.
\newblock http://www.eucalyptus.com/.

\bibitem{google}
Google compute engine.
\newblock http://cloud.google.com/products/compute- engine.html/.

\bibitem{kvm}
Kernel based virtual machine.
\newblock http://www.linux-kvm. org/.

\bibitem{azure}
Microsoft windows azure.
\newblock https://www.windowsazure.com/en-us/.

\bibitem{open}
Openstack framework.
\newblock https://www.openstack.org/.

\bibitem{rack}
Rackspace cloud.
\newblock https://www.rackspace.com/.

\bibitem{rubis}
Rice university bidding systems.
\newblock http://rubis.ow2.org/.

\bibitem{rubbos}
Rubbos: Bulletin board benchmark.
\newblock http://jmob.ow2. org/rubbos.html/.

\bibitem{tpcw}
Tpc-w.
\newblock http://www.tpc.org/tpcw/default.asp.

\bibitem{rightscale}
Understanding the voting process.
\newblock https://support.rightscale.com/12-Guides/ RightScale 101/System
  Architecture/RightScale Alert System/Alerts based on Voting Tags/
  Understanding the Voting Process/.

\bibitem{vmware}
Vmware vsphere esx and esxi.
\newblock https://www.vmware.com/products/esxi-and-esx.html.

\bibitem{wikibench}
Wikibench: A web hosting benchmark.
\newblock http://www. wikibench.eu.

\bibitem{wikipedia}
Wikipedia access traces.
\newblock http://www.wikibench.eu/?pageid=60.

\bibitem{f98}
World cup 98 trace.
\newblock http://ita.ee.lbl.gov/html/contrib/WorldCup.html.

\bibitem{xen}
Xen: a virtual machine monitor.
\newblock http://xen.xensource.com/.

\bibitem{BRGA-resource}
{\sc Abdul-Rahman, O., Munetomo, M., and Akama, K.}
\newblock Toward a genetic algorithm based flexible approach for the management
  of virtualized application environments in cloud platforms.
\newblock In {\em Computer Communications and Networks (ICCCN), 2012 21st
  International Conference on\/} (July 2012), pp.~1--9.

\bibitem{profit-cent-local-search}
{\sc Addis, B., Ardagna, D., Panicucci, B., and Zhang, L.}
\newblock Autonomic management of cloud service centers with availability
  guarantees.
\newblock In {\em Cloud Computing, 2010 IEEE 3rd International Conference on\/}
  (2010), pp.~220--227.

\bibitem{al2017elasticity}
{\sc Al-Dhuraibi, Y., Paraiso, F., Djarallah, N., and Merle, P.}
\newblock Elasticity in cloud computing: State of the art and research
  challenges.
\newblock {\em IEEE Transactions on Services Computing\/} (2017).

\bibitem{al2013impact}
{\sc Al-Haidari, F., Sqalli, M., and Salah, K.}
\newblock Impact of cpu utilization thresholds and scaling size on autoscaling
  cloud resources.
\newblock In {\em Cloud Computing Technology and Science, IEEE 5th
  International Conference on\/} (2013), pp.~256--261.

\bibitem{compare-to-Rao-fuzzy-2013}
{\sc Albano, L., Anglano, C., Canonico, M., and Guazzone, M.}
\newblock Fuzzy-q amp;e: Achieving qos guarantees and energy savings for cloud
  applications with fuzzy control.
\newblock In {\em Cloud and Green Computing (CGC), 2013 Third International
  Conference on\/} (Sept 2013), pp.~159--166.

\bibitem{2013-scaling-select-one-predictor-and-error-correlation-2013}
{\sc Almeida~Morais, F., Vilar~Brasileiro, F., Vigolvino~Lopes, R.,
  Araujo~Santos, R., Satterfield, W., and Rosa, L.}
\newblock Autoflex: Service agnostic auto-scaling framework for iaas deployment
  models.
\newblock In {\em Cluster, Cloud and Grid Computing (CCGrid), 2013 13th
  IEEE/ACM International Symposium on\/} (May 2013), pp.~42--49.

\bibitem{2014-fuzzy-compare-to-JRao-2014}
{\sc Anglano, C., Canonico, M., and Guazzone, M.}
\newblock Fc2q: exploiting fuzzy control in server consolidation for cloud
  applications with sla constraints.
\newblock {\em Concurrency and Computation: Practice and Experience 27}, 17
  (2015), 4491--4514.

\bibitem{ardagna2014quality}
{\sc Ardagna, D., Casale, G., Ciavotta, M., P{\'e}rez, J.~F., and Wang, W.}
\newblock Quality-of-service in cloud computing: modeling techniques and their
  applications.
\newblock {\em Journal of Internet Services and Applications 5}, 1 (2014), 11.

\bibitem{2012-app-VM-mapping-2012}
{\sc Ashraf, A., Byholm, B., Lehtinen, J., and Porres, I.}
\newblock Feedback control algorithms to deploy and scale multiple web
  applications per virtual machine.
\newblock In {\em Software Engineering and Advanced Applications (SEAA), 2012
  38th Euromicro Conference on\/} (Sept 2012), pp.~431--438.

\bibitem{EMA-CPU-memory-PD}
{\sc Ashraf, A., Byholm, B., and Porres, I.}
\newblock Cramp: Cost-efficient resource allocation for multiple web
  applications with proactive scaling.
\newblock In {\em Cloud Computing Technology and Science, 2012 IEEE 4th
  International Conference on\/} (2012), pp.~581--586.

\bibitem{2014-aco-app-to-VM-consolidation-2014}
{\sc Ashraf, A., and Porres, I.}
\newblock Using ant colony system to consolidate multiple web applications in a
  cloud environment.
\newblock In {\em Parallel, Distributed and Network-Based Processing (PDP),
  2014 22nd Euromicro International Conference on\/} (Feb 2014), pp.~482--489.

\bibitem{aslanpour2017auto}
{\sc Aslanpour, M.~S., Ghobaei-Arani, M., and Toosi, A.~N.}
\newblock Auto-scaling web applications in clouds: a cost-aware approach.
\newblock {\em Journal of Network and Computer Applications 95\/} (2017),
  26--41.

\bibitem{awada2017improving}
{\sc Awada, U., and Barker, A.}
\newblock Improving resource efficiency of container-instance clusters on
  clouds.
\newblock In {\em Cluster, Cloud and Grid Computing, IEEE 17th IEEE/ACM
  International Symposium on\/} (2017), IEEE Press, pp.~929--934.

\bibitem{baresi2016discrete}
{\sc Baresi, L., Guinea, S., Leva, A., and Quattrocchi, G.}
\newblock A discrete-time feedback controller for containerized cloud
  applications.
\newblock In {\em Foundations of Software Engineering, 24th ACM SIGSOFT
  International Symposium on\/} (2016), ACM, pp.~217--228.

\bibitem{epics}
{\sc Becker, T., Agne, A., Lewis, P., Bahsoon, R., Faniyi, F., Esterle, L.,
  Keller, A., Chandra, A., Jensenius, A., and Stilkerich, S.}
\newblock Epics: Engineering proprioception in computing systems.
\newblock In {\em Computational Science and Engineering (CSE), 2012 IEEE 15th
  International Conference on\/} (Dec 2012), pp.~353--360.

\bibitem{queue-prediction}
{\sc Bi, J., Zhu, Z., Tian, R., and Wang, Q.}
\newblock Dynamic provisioning modeling for virtualized multi-tier applications
  in cloud data center.
\newblock In {\em Cloud Computing (CLOUD), 2010 IEEE 3rd International
  Conference on\/} (July 2010), pp.~370--377.

\bibitem{change-point}
{\sc Bod\'{\i}k, P., Griffith, R., Sutton, C., Fox, A., Jordan, M., and
  Patterson, D.}
\newblock Statistical machine learning makes automatic control practical for
  internet datacenters.
\newblock In {\em Hot Topics in Cloud Computing, 2009 Conference on\/}
  (Berkeley, CA, USA, 2009), USENIX Association.

\bibitem{Compsac_2010_I_Brandic}
{\sc Brandic, I., Emeakaroha, V., Maurer, M., Dustdar, S., Acs, S., Kertesz,
  A., and Kecskemeti, G.}
\newblock Laysi: A layered approach for sla-violation propagation in
  self-manageable cloud infrastructures.
\newblock In {\em Computer Software and Applications Conference Workshops, IEEE
  34th Annual\/} (2010), pp.~365--370.

\bibitem{Brun:2009}
{\sc Brun, Y., Marzo~Serugendo, G., Gacek, C., Giese, H., Kienle, H., Litoiu,
  M., M\"{u}ller, H., Pezz\`{e}, M., and Shaw, M.}
\newblock Software engineering for self-adaptive systems.
\newblock Springer-Verlag, Berlin, Heidelberg, 2009, ch.~Engineering
  Self-Adaptive Systems Through Feedback Loops, pp.~48--70.

\bibitem{2013-JRAO-most-closest-work-2013}
{\sc Bu, X., Rao, J., and zhong Xu, C.}
\newblock Coordinated self-configuration of virtual machines and appliances
  using a model-free learning approach.
\newblock {\em IEEE Transactions on Parallel and Distributed Systems 24}, 4
  (April 2013), 681--690.

\bibitem{CloudSim}
{\sc Calheiros, R.~N., Ranjan, R., Beloglazov, A., De~Rose, C. A.~F., and
  Buyya, R.}
\newblock Cloudsim: A toolkit for modeling and simulation of cloud computing
  environments and evaluation of resource provisioning algorithms.
\newblock {\em Softw. Pract. Exper. 41}, 1 (Jan. 2011), 23--50.

\bibitem{tse1}
{\sc Calinescu, R., Grunske, L., Kwiatkowska, M., Mirandola, R., and
  Tamburrelli, G.}
\newblock Dynamic qos management and optimization in service-based systems.
\newblock {\em IEEE Transactions on Software Engineering 37}, 3 (May 2011),
  387--409.

\bibitem{2010-demand-pattern-match-2010}
{\sc Caron, E., Desprez, F., and Muresan, A.}
\newblock Forecasting for grid and cloud computing on-demand resources based on
  pattern matching.
\newblock In {\em Cloud Computing Technology and Science (CloudCom), 2010 IEEE
  Second International Conference on\/} (Nov 2010), pp.~456--463.

\bibitem{cloud_computing_2010_5_20_50060-ext}
{\sc Chazalet, A., Tran, F.~D., Deslaugiers, M., Lefebvre, A., Exertier, F.,
  and Legrand, J.}
\newblock Adding self-scaling capability to the cloud to meet service level
  agreements.
\newblock {\em International Journal on Advances in Intelligent Systems, 4}, 3
  (2011).

\bibitem{6008793}
{\sc Chen, T., and Bahsoon, R.}
\newblock Scalable service oriented replication in the cloud.
\newblock In {\em Cloud Computing (CLOUD), 2011 IEEE International Conference
  on\/} (July 2011), pp.~766--767.

\bibitem{Chen:2013:seams}
{\sc Chen, T., and Bahsoon, R.}
\newblock Self-adaptive and sensitivity-aware qos modeling for the cloud.
\newblock In {\em Software Engineering for Adaptive and Self-Managing Systems,
  8th International Symposium on\/} (2013), pp.~43--52.

\bibitem{Chen:2014:seams}
{\sc Chen, T., and Bahsoon, R.}
\newblock Symbiotic and sensitivity-aware architecture for globally-optimal
  benefit in self-adaptive cloud.
\newblock In {\em Software Engineering for Adaptive and Self-Managing Systems,
  9th International Symposium on\/} (2014), pp.~85--94.

\bibitem{Chen:2015:computer}
{\sc Chen, T., and Bahsoon, R.}
\newblock Towards a smarter cloud: Self-aware autoscaling of cloud
  configurations and resources.
\newblock {\em Computer, IEEE 48}, 9 (Sept 2015).

\bibitem{Chen:2015:tse-pending}
{\sc Chen, T., and Bahsoon, R.}
\newblock Self-adaptive and online qos modeling for cloud-based software
  services.
\newblock {\em IEEE Transactions on Software Engineering 43}, 5 (2017),
  453--475.

\bibitem{Chen:2015:tsc-pending}
{\sc Chen, T., and Bahsoon, R.}
\newblock Self-adaptive trade-off decision making for autoscaling cloud-based
  services.
\newblock {\em IEEE Transactions on Services Computing 10}, 4 (2017), 618--632.

\bibitem{Chen:2013:iccs}
{\sc Chen, T., Bahsoon, R., and Theodoropoulos, G.}
\newblock Dynamic qos optimization architecture for cloud-based dddas.
\newblock {\em Procedia Computer Science 18\/} (2013), 1881 -- 1890.
\newblock 2013 International Conference on Computational Science.

\bibitem{icpechen}
{\sc Chen, T., Bahsoon, R., Wang, S., and Yao, X.}
\newblock To adapt or not to adapt? technical debt and learning driven
  self-adaptation for managing runtime performance.
\newblock In {\em Performance Engineering, ACM/SPEC International Conference
  on\/} (April 2018).

\bibitem{Chen:2014:ucc}
{\sc Chen, T., Bahsoon, R., and Yao, X.}
\newblock Online qos modeling in the cloud: A hybrid and adaptive
  multi-learners approach.
\newblock In {\em Utility and Cloud Computing, 2014 IEEE/ACM 7th International
  Conference on\/} (2014), pp.~327--336.

\bibitem{Chen2016:book}
{\sc Chen, T., Faniyi, F., and Bahsoon, R.}
\newblock {\em Design Patterns and Primitives: Introduction of Components and
  Patterns for SACS}.
\newblock Springer International Publishing, Cham, 2016, pp.~53--78.

\bibitem{2014_epics_handbook}
{\sc Chen, T., Faniyi, F., Bahsoon, R., Lewis, P.~R., Yao, X., Minku, L.~L.,
  and Esterle, L.}
\newblock The handbook of engineering self-aware and self-expressive systems.
\newblock {\em arXiv preprint arXiv:1409.1793\/} (2014).

\bibitem{femosaa}
{\sc Chen, T., Li, K., Bahsoon, R., and Yao, X.}
\newblock Femosaa: Feature guided and knee driven multi-objective optimization
  for self-adaptive software.
\newblock {\em ACM Transactions on Software Engineering and Methodology\/}
  (2018).
\newblock in press.

\bibitem{roadmap}
{\sc Cheng, B., de~Lemos, R., Giese, H., Inverardi, P., Magee, J., Andersson,
  J., Becker, B., Bencomo, N., Brun, Y., Cukic, B., Di~Marzo~Serugendo, G.,
  Dustdar, S., Finkelstein, A., Gacek, C., Geihs, K., Grassi, V., Karsai, G.,
  Kienle, H., Kramer, J., Litoiu, M., Malek, S., Mirandola, R., Müller, H.,
  Park, S., Shaw, M., Tichy, M., Tivoli, M., Weyns, D., and Whittle, J.}
\newblock Software engineering for self-adaptive systems: A research roadmap.
\newblock In {\em Software Engineering for Self-Adaptive Systems}, B.~Cheng,
  R.~de~Lemos, H.~Giese, P.~Inverardi, and J.~Magee, Eds., vol.~5525 of {\em
  Lecture Notes in Computer Science}. Springer Berlin Heidelberg, 2009,
  pp.~1--26.

\bibitem{3-stages-game-theory}
{\sc Chi, R., Qian, Z., and Lu, S.}
\newblock A game theoretical method for auto-scaling of multi-tiers web
  applications in cloud.
\newblock In {\em Internetware, Fourth Asia-Pacific Symposium on\/} (2012),
  pp.~3:1--3:10.

\bibitem{tp38}
{\sc Chiang, R., and Huang, H.}
\newblock Tracon: Interference-aware scheduling for data-intensive applications
  in virtualized environments.
\newblock In {\em High Performance Computing, Networking, Storage and Analysis,
  2011 International Conference for\/} (2011), pp.~1--12.

\bibitem{2014-2-SVM-workload-type-2014}
{\sc Chiang, R.~C., Hwang, J., Huang, H.~H., and Wood, T.}
\newblock Matrix: Achieving predictable virtual machine performance in the
  clouds.
\newblock In {\em 11th International Conference on Autonomic Computing\/}
  (2014).

\bibitem{2013-self-organising-map-2013}
{\sc Chihi, H., Chainbi, W., and Ghedira, K.}
\newblock An energy-efficient self-provisioning approach for cloud resources
  management.
\newblock {\em SIGOPS Oper. Syst. Rev. 47}, 3 (Nov. 2013), 2--9.

\bibitem{linear-mapping-CP-bundles-2012}
{\sc Collazo-Mojica, X.~J., Sadjadi, S., Ejarque, J., and Badia, R.~M.}
\newblock Cloud application resource mapping and scaling based on monitoring of
  qos constraints.
\newblock Knowledge Systems Institute Graduate School, p.~88{\textendash}93.

\bibitem{2013-rule-based-multi-elasiticity-2013}
{\sc Copil, G., Moldovan, D., Truong, H.-L., and Dustdar, S.}
\newblock Multi-level elasticity control of cloud services.
\newblock In {\em Service-Oriented Computing}, S.~Basu, C.~Pautasso, L.~Zhang,
  and X.~Fu, Eds., vol.~8274 of {\em Lecture Notes in Computer Science}.
  Springer Berlin Heidelberg, 2013, pp.~429--436.

\bibitem{da2016autoelastic}
{\sc da~Rosa~Righi, R., Rodrigues, V.~F., da~Costa, C.~A., Galante, G.,
  De~Bona, L. C.~E., and Ferreto, T.}
\newblock Autoelastic: Automatic resource elasticity for high performance
  applications in the cloud.
\newblock {\em IEEE Transactions on Cloud Computing 4}, 1 (2016), 6--19.

\bibitem{2014-eplison-GA-weigh-h-scaling-2014}
{\sc El~Kateb, D., Fouquet, F., Nain, G., Meira, J.~A., Ackerman, M., and
  Le~Traon, Y.}
\newblock Generic cloud platform multi-objective optimization leveraging
  models@run.time.
\newblock In {\em Applied Computing, 29th Annual ACM Symposium on\/} (2014),
  pp.~343--350.

\bibitem{06032254}
{\sc Emeakaroha, V., Brandic, I., Maurer, M., and Breskovic, I.}
\newblock Sla-aware application deployment and resource allocation in clouds.
\newblock In {\em Computer Software and Applications Conference, IEEE 35th
  Annual\/} (2011), pp.~298--303.

\bibitem{HPCS_IWCMC_Vincent}
{\sc Emeakaroha, V., Brandic, I., Maurer, M., and Dustdar, S.}
\newblock Low level metrics to high level slas - lom2his framework: Bridging
  the gap between monitored metrics and sla parameters in cloud environments.
\newblock In {\em High Performance Computing and Simulation (HPCS), 2010
  International Conference on\/} (June 2010), pp.~48--54.

\bibitem{Emeakaroha_CloudComp2010}
{\sc Emeakaroha, V.~C., Calheiros, R.~N., Netto, M. A.~S., Brandic, I., and
  Rose, C. A. F.~D.}
\newblock Desvi: An architecture for detecting sla violations in cloud
  computing infrastructures.
\newblock In {\em Cloud computing, 2nd international ICST conference on\/}
  (2010).

\bibitem{farokhi2016hybrid}
{\sc Farokhi, S., Jamshidi, P., Lakew, E.~B., Brandic, I., and Elmroth, E.}
\newblock A hybrid cloud controller for vertical memory elasticity: A
  control-theoretic approach.
\newblock {\em Future Generation Computer Systems 65\/} (2016), 57--72.

\bibitem{2014-profiling-decision-tree-scaling-2014}
{\sc Fernandez, H., Pierre, G., and Kielmann, T.}
\newblock Autoscaling web applications in heterogeneous cloud infrastructures.
\newblock In {\em Cloud Engineering (IC2E), 2014 IEEE International Conference
  on\/} (March 2014), pp.~195--204.

\bibitem{05557978}
{\sc Ferretti, S., Ghini, V., Panzieri, F., Pellegrini, M., and Turrini, E.}
\newblock Qos-aware clouds.
\newblock In {\em Cloud Computing (CLOUD), 2010 IEEE 3rd International
  Conference on\/} (July 2010), pp.~321--328.

\bibitem{full-simulation-model}
{\sc Fittkau, F., Frey, S., and Hasselbring, W.}
\newblock Cdosim: Simulating cloud deployment options for software migration
  support.
\newblock In {\em Maintenance and Evolution of Service-Oriented and Cloud-Based
  Systems (MESOCA), 2012 IEEE 6th International Workshop on the\/} (Sept 2012),
  pp.~37--46.

\bibitem{GA-full-simulation}
{\sc Frey, S., Fittkau, F., and Hasselbring, W.}
\newblock Search-based genetic optimization for deployment and reconfiguration
  of software in the cloud.
\newblock In {\em Software Engineering, 2013 IEEE International Conference
  on\/} (2013), pp.~512--521.

\bibitem{2013-elasticity-primitives-2013}
{\sc Galante, G., and Bona, L.}
\newblock Constructing elastic scientific applications using elasticity
  primitives.
\newblock In {\em Computational Science and Its Applications}, vol.~7975 of
  {\em Lecture Notes in Computer Science}. 2013, pp.~281--294.

\bibitem{kriging-controller}
{\sc Gambi, A., Toffetti, G., Pautasso, C., and Pezze, M.}
\newblock Kriging controllers for cloud applications.
\newblock {\em Internet Computing, IEEE 17}, 4 (July 2013), 40--47.

\bibitem{2014-kalman+rule-based-2014}
{\sc Gandhi, A., Dube, P., Karve, A., Kochut, A., and Zhang, L.}
\newblock Adaptive, model-driven autoscaling for cloud applications.
\newblock In {\em 11th International Conference on Autonomic Computing\/}
  (2014).

\bibitem{gandhi2017model}
{\sc Gandhi, A., Dube, P., Karve, A., Kochut, A., and Zhang, L.}
\newblock Model-driven optimal resource scaling in cloud.
\newblock {\em Software \& Systems Modeling\/} (2017), 1--18.

\bibitem{kalman-clustering}
{\sc Ghanbari, H., Barna, C., Litoiu, M., Woodside, M., Zheng, T., Wong, J.,
  and Iszlai, G.}
\newblock Tracking adaptive performance models using dynamic clustering of user
  classes.
\newblock {\em SIGSOFT Softw. Eng. Notes 36}, 5 (Sept. 2011), 179--188.

\bibitem{rule-control-elasticity-cloud}
{\sc Ghanbari, H., Simmons, B., Litoiu, M., and Iszlai, G.}
\newblock Exploring alternative approaches to implement an elasticity policy.
\newblock In {\em Cloud Computing (CLOUD), 2011 IEEE International Conference
  on\/} (July 2011), pp.~716--723.

\bibitem{gill2017chopper}
{\sc Gill, S.~S., Chana, I., Singh, M., and Buyya, R.}
\newblock Chopper: an intelligent qos-aware autonomic resource management
  approach for cloud computing.
\newblock {\em Cluster Computing\/} (2017), 1--39.

\bibitem{signal-resource-trend-prediction}
{\sc Gong, Z., Gu, X., and Wilkes, J.}
\newblock Press: Predictive elastic resource scaling for cloud systems.
\newblock In {\em Network and Service Management (CNSM), 2010 International
  Conference on\/} (Oct 2010), pp.~9--16.

\bibitem{multitier-resalloc-Cloud11}
{\sc Goudarzi, H., and Pedram, M.}
\newblock Multi-dimensional sla-based resource allocation for multi-tier cloud
  computing systems.
\newblock In {\em Cloud Computing (CLOUD), 2011 IEEE International Conference
  on\/} (July 2011), pp.~324--331.

\bibitem{2013-software-CP-only-2013}
{\sc Guo, Y., Lama, P., Jiang, C., and Zhou, X.}
\newblock Automated and agile server parametertuning by coordinated learning
  and control.
\newblock {\em IEEE Transactions on Parallel and Distributed Systems 25}, 4
  (April 2014), 876--886.

\bibitem{scale-rule-based}
{\sc Han, R., Guo, L., Ghanem, M., and Guo, Y.}
\newblock Lightweight resource scaling for cloud applications.
\newblock In {\em Cluster, Cloud and Grid Computing (CCGrid), 2012 12th
  IEEE/ACM International Symposium on\/} (May 2012), pp.~644--651.

\bibitem{herbein2016resource}
{\sc Herbein, S., Dusia, A., Landwehr, A., McDaniel, S., Monsalve, J., Yang,
  Y., Seelam, S.~R., and Taufer, M.}
\newblock Resource management for running hpc applications in container clouds.
\newblock In {\em International Conference on High Performance Computing\/}
  (2016), Springer, pp.~261--278.

\bibitem{self-aware-ML-adaptive-control}
{\sc Hoffman, H.}
\newblock {\em Seec: A Framework for Self-aware Management of Goals and
  Constraints in Computing Systems}.
\newblock PhD thesis, Cambridge, MA, USA, 2013.

\bibitem{HuBrKo2011-SEAMS-ResAlloc}
{\sc Huber, N., Brosig, F., and Kounev, S.}
\newblock Model-based self-adaptive resource allocation in virtualized
  environments.
\newblock In {\em Software Engineering for Adaptive and Self-Managing Systems,
  6th International Symposium on\/} (2011), pp.~90--99.

\bibitem{ibm}
{\sc IBM}.
\newblock An architectural blueprint for autonomic computing.
\newblock {\em IBM Technical Report\/} (2003).

\bibitem{06119056}
{\sc Jiang, J., Lu, J., and Zhang, G.}
\newblock An innovative self-adaptive configuration optimization system in
  cloud computing.
\newblock In {\em Dependable, Autonomic and Secure Computing, 2011 IEEE Ninth
  International Conference on\/} (2011), pp.~621--627.

\bibitem{typical-navie-scaling-VM-number}
{\sc Jiang, J., Lu, J., Zhang, G., and Long, G.}
\newblock Optimal cloud resource auto-scaling for web applications.
\newblock In {\em Cluster, Cloud and Grid Computing (CCGrid), 2013 13th
  IEEE/ACM International Symposium on\/} (May 2013), pp.~58--65.

\bibitem{ensemble-prediction-VM-2014-full}
{\sc Jiang, Y., Perng, C.-S., Li, T., and Chang, R.}
\newblock Cloud analytics for capacity planning and instant vm provisioning.
\newblock {\em IEEE Transactions on Network and Service Management 10}, 3
  (September 2013), 312--325.

\bibitem{cache-static-ANN-bi-obj-2012}
{\sc Kabir, F., and Chiu, D.}
\newblock Reconciling cost and performance objectives for elastic web caches.
\newblock In {\em Cloud and Service Computing (CSC), 2012 International
  Conference on\/} (Nov 2012), pp.~88--95.

\bibitem{2014-kalman-pair-wised-coupling-on-tiers-2014}
{\sc Kalyvianaki, E., Charalambous, T., and Hand, S.}
\newblock Adaptive resource provisioning for virtualized servers using kalman
  filters.
\newblock {\em ACM Trans. Auton. Adapt. Syst. 9}, 2 (July 2014), 10:1--10:35.

\bibitem{DCSim}
{\sc Keller, G., Tighe, M., Lutfiyya, H., and Bauer, M.}
\newblock Dcsim: A data centre simulation tool.
\newblock In {\em Integrated Network Management (IM 2013), 2013 IFIP/IEEE
  International Symposium on\/} (May 2013), pp.~1090--1091.

\bibitem{PCA-model-scaling-2013}
{\sc Kim, S., Kim, J.-S., Hwang, S., and Kim, Y.}
\newblock An allocation and provisioning model of science cloud for high
  throughput computing applications.
\newblock In {\em Cloud and Autonomic Computing, 2013 ACM Conference on\/}
  (2013), pp.~1--8.

\bibitem{ISPASS07}
{\sc Koh, Y., Knauerhase, R., Brett, P., Bowman, M., Wen, Z., and Pu, C.}
\newblock An analysis of performance interference effects in virtual
  environments.
\newblock In {\em Performance Analysis of Systems Software, 2007. ISPASS 2007.
  IEEE International Symposium on\/} (April 2007), pp.~200--209.

\bibitem{JSS_Kousiouris}
{\sc Kousiouris, G., Cucinotta, T., and Varvarigou, T.}
\newblock The effects of scheduling, workload type and consolidation scenarios
  on virtual machine performance and their prediction through optimized
  artificial neural networks.
\newblock {\em J. Syst. Softw. 84}, 8 (Aug. 2011), 1270--1291.

\bibitem{MOCS2014-full}
{\sc Kousiouris, G., Menychtas, A., Kyriazis, D., Gogouvitis, S., and
  Varvarigou, T.}
\newblock Dynamic, behavioral-based estimation of resource provisioning based
  on high-level application terms in cloud platforms.
\newblock {\em Future Generation Computer Systems 32}, 0 (2014), 27 -- 40.

\bibitem{dynamic-model-comparison}
{\sc Kundu, S., Rangaswami, R., Gulati, A., Zhao, M., and Dutta, K.}
\newblock Modeling virtualized applications using machine learning techniques.
\newblock In {\em Virtual Execution Environments, 8th ACM SIGPLAN/SIGOPS
  Conference on\/} (2012), pp.~3--14.

\bibitem{lakew2017kpi}
{\sc Lakew, E.~B., Papadopoulos, A.~V., Maggio, M., Klein, C., and Elmroth, E.}
\newblock Kpi-agnostic control for fine-grained vertical elasticity.
\newblock In {\em Cluster, Cloud and Grid Computing, IEEE 17th IEEE/ACM
  International Symposium on\/} (2017), IEEE Press, pp.~589--598.

\bibitem{fuzzy-vm-interference}
{\sc Lama, P., Guo, Y., and Zhou, X.}
\newblock Autonomic performance and power control for co-located web
  applications on virtualized servers.
\newblock In {\em Quality of Service, 2013 IEEE/ACM 21st International
  Symposium on\/} (2013), pp.~1--10.

\bibitem{epics_survey}
{\sc Lewis, P., Chandra, A., Parsons, S., Robinson, E., Glette, K., Bahsoon,
  R., Torresen, J., and Yao, X.}
\newblock A survey of self-awareness and its application in computing systems.
\newblock In {\em Self-Adaptive and Self-Organizing Systems Workshops (SASOW),
  2011 Fifth IEEE Conference on\/} (Oct 2011), pp.~102--107.

\bibitem{7185305}
{\sc Lewis, P.~R., Chandra, A., Faniyi, F., Glette, K., Chen, T., Bahsoon, R.,
  Torresen, J., and Yao, X.}
\newblock Architectural aspects of self-aware and self-expressive computing
  systems: From psychology to engineering.
\newblock {\em Computer 48}, 8 (Aug 2015), 62--70.

\bibitem{wosp10sla}
{\sc Li, H., Casale, G., and Ellahi, T.}
\newblock Sla-driven planning and optimization of enterprise applications.
\newblock In {\em Performance Engineering, First Joint WOSP/SIPEW International
  Conference on\/} (2010), pp.~117--128.

\bibitem{li-opt-clouds-4}
{\sc Li, J., Chinneck, J., Woodside, M., Litoiu, M., and Iszlai, G.}
\newblock Performance model driven qos guarantees and optimization in clouds.
\newblock In {\em Software Engineering Challenges of Cloud Computing, ICSE
  Workshop on\/} (2009), pp.~15--22.

\bibitem{fine-grained-servce-LQM-MIP-power}
{\sc Li, J., Woodside, M., Chinneck, J., and Litoiu, M.}
\newblock Cloudopt: Multi-goal optimization of application deployments across a
  cloud.
\newblock In {\em Network and Service Management (CNSM), 2011 7th International
  Conference on\/} (Oct 2011), pp.~1--9.

\bibitem{li2015rest}
{\sc Li, L., Tang, T., and Chou, W.}
\newblock A rest service framework for fine-grained resource management in
  container-based cloud.
\newblock In {\em Cloud Computing (CLOUD), 2015 IEEE 8th International
  Conference on\/} (2015), IEEE, pp.~645--652.

\bibitem{acdc09}
{\sc Lim, H.~C., Babu, S., Chase, J.~S., and Parekh, S.~S.}
\newblock Automated control in cloud computing: Challenges and opportunities.
\newblock In {\em Automated Control for Datacenters and Clouds, 1st Workshop
  on\/} (2009), pp.~13--18.

\bibitem{determine-CP-compare-models}
{\sc Lloyd, W., Pallickara, S., David, O., Lyon, J., Arabi, M., and Rojas, K.}
\newblock Performance modeling to support multi-tier application deployment to
  infrastructure-as-a-service clouds.
\newblock In {\em Utility and Cloud Computing (UCC), 2012 IEEE Fifth
  International Conference on\/} (Nov 2012), pp.~73--80.

\bibitem{lorido2014review}
{\sc Lorido-Botran, T., Miguel-Alonso, J., and Lozano, J.~A.}
\newblock A review of auto-scaling techniques for elastic applications in cloud
  environments.
\newblock {\em Journal of Grid Computing 12}, 4 (2014), 559--592.

\bibitem{ma2016auto}
{\sc Ma, H., Wang, L., Tak, B.~C., Wang, L., and Tang, C.}
\newblock Auto-tuning performance of mpi parallel programs using resource
  management in container-based virtual cloud.
\newblock In {\em Cloud Computing (CLOUD), 2016 IEEE 9th International
  Conference on\/} (2016), IEEE, pp.~545--552.

\bibitem{2014-decision-tree-software-CP-2014}
{\sc Maji, A.~K., Mitra, S., Zhou, B., Bagchi, S., and Verma, A.}
\newblock Mitigating interference in cloud services by middleware
  reconfiguration.
\newblock In {\em Middleware, 15th International Conference on\/} (2014),
  pp.~277--288.

\bibitem{mann2015allocation}
{\sc Mann, Z.~{\'A}.}
\newblock Allocation of virtual machines in cloud data centers?a survey of
  problem models and optimization algorithms.
\newblock {\em ACM Computing Surveys (CSUR) 48}, 1 (2015), 11.

\bibitem{manvi2014resource}
{\sc Manvi, S.~S., and Shyam, G.~K.}
\newblock Resource management for infrastructure as a service (iaas) in cloud
  computing: A survey.
\newblock {\em Journal of Network and Computer Applications 41\/} (2014),
  424--440.

\bibitem{SEASS_2010_Michael_Maurer}
{\sc Maurer, M., Brandic, I., Emeakaroha, V., and Dustdar, S.}
\newblock Towards knowledge management in self-adaptable clouds.
\newblock In {\em Services (SERVICES-1), 2010 6th World Congress on\/} (July
  2010), pp.~527--534.

\bibitem{self-sla-and-resource}
{\sc Maurer, M., Brandic, I., and Sakellariou, R.}
\newblock Self-adaptive and resource-efficient sla enactment for cloud
  computing infrastructures.
\newblock In {\em Cloud Computing (CLOUD), 2012 IEEE 5th International
  Conference on\/} (June 2012), pp.~368--375.

\bibitem{megahed2017stochastic}
{\sc Megahed, A., Mohamed, M., and Tata, S.}
\newblock A stochastic optimization approach for cloud elasticity.
\newblock In {\em Cloud Computing (CLOUD), 2017 IEEE 10th International
  Conference on\/} (2017), IEEE, pp.~456--463.

\bibitem{2013-offline-profiling-2013}
{\sc Mian, R., Martin, P., Zulkernine, F., and Vazquez-Poletti, J.~L.}
\newblock Towards building performance models for data-intensive workloads in
  public clouds.
\newblock In {\em Performance Engineering, 4th ACM/SPEC International
  Conference on\/} (2013), pp.~259--270.

\bibitem{MIMO-fuzzy-hill-climbing-2013}
{\sc Minarolli, D., and Freisleben, B.}
\newblock Virtual machine resource allocation in cloud computing via
  multi-agent fuzzy control.
\newblock In {\em Cloud and Green Computing (CGC), 2013 Third International
  Conference on\/} (Sept 2013), pp.~188--194.

\bibitem{2014-navie-ANN-GA-2014}
{\sc Minarolli, D., and Freisleben, B.}
\newblock Distributed resource allocation to virtual machines via artificial
  neural networks.
\newblock In {\em Parallel, Distributed and Network-Based Processing, 22nd
  Euromicro International Conference on\/} (2014), pp.~490--499.

\bibitem{2014-linear-centralized-decision-making-2014}
{\sc Nikolov, V., Kachele, S., Hauck, F., and Rautenbach, D.}
\newblock Cloudfarm: An elastic cloud platform with flexible and adaptive
  resource management.
\newblock In {\em Utility and Cloud Computing (UCC), 2014 IEEE/ACM 7th
  International Conference on\/} (2014), pp.~547--553.

\bibitem{HPL-2008-123R1-mimo}
{\sc Padala, P., Hou, K.-Y., Shin, K.~G., Zhu, X., Uysal, M., Wang, Z.,
  Singhal, S., and Merchant, A.}
\newblock Automated control of multiple virtualized resources.
\newblock In {\em Computer Systems, 4th ACM European Conference on\/} (2009),
  pp.~13--26.

\bibitem{qu2016auto}
{\sc Qu, C., Calheiros, R.~N., and Buyya, R.}
\newblock Auto-scaling web applications in clouds: a taxonomy and survey.
\newblock {\em arXiv preprint arXiv:1609.09224\/} (2016).

\bibitem{qu2016reliable}
{\sc Qu, C., Calheiros, R.~N., and Buyya, R.}
\newblock A reliable and cost-efficient auto-scaling system for web
  applications using heterogeneous spot instances.
\newblock {\em Journal of Network and Computer Applications 65\/} (2016),
  167--180.

\bibitem{rameshan2016augmenting}
{\sc Rameshan, N., Liu, Y., Navarro, L., and Vlassov, V.}
\newblock Augmenting elasticity controllers for improved accuracy.
\newblock In {\em Autonomic Computing (ICAC), 2016 IEEE International
  Conference on\/} (2016), IEEE, pp.~117--126.

\bibitem{MASCOTS11}
{\sc Rao, J., Bu, X., Xu, C.-Z., and Wang, K.}
\newblock A distributed self-learning approach for elastic provisioning of
  virtualized cloud resources.
\newblock In {\em Modeling, Analysis Simulation of Computer and
  Telecommunication Systems (MASCOTS), 2011 IEEE 19th International Symposium
  on\/} (July 2011), pp.~45--54.

\bibitem{IWQOS11}
{\sc Rao, J., Wei, Y., Gong, J., and Xu, C.-Z.}
\newblock Dynaqos: Model-free self-tuning fuzzy control of virtualized
  resources for qos provisioning.
\newblock In {\em Quality of Service (IWQoS), 2011 IEEE 19th International
  Workshop on\/} (June 2011), pp.~1--9.

\bibitem{qcloud}
{\sc Ripal, N., Aman, K., and Alireza, G.}
\newblock Q-clouds: Managing performance interference effects for qos-aware
  clouds.
\newblock In {\em Computer Systems, 5th European Conference on\/} (2010),
  pp.~237--250.

\bibitem{landscape}
{\sc Salehie, M., and Tahvildari, L.}
\newblock Self-adaptive software: Landscape and research challenges.
\newblock {\em ACM Trans. Auton. Adapt. Syst. 4}, 2 (May 2009), 14:1--14:42.

\bibitem{ILP-cost-only-scaling-2013}
{\sc Sedaghat, M., Hernandez-Rodriguez, F., and Elmroth, E.}
\newblock A virtual machine re-packing approach to the horizontal vs. vertical
  elasticity trade-off for cloud autoscaling.
\newblock In {\em Cloud and Autonomic Computing, 2013 ACM Conference on\/}
  (2013), pp.~6:1--6:10.

\bibitem{seelam2015polyglot}
{\sc Seelam, S.~R., Dettori, P., Westerink, P., and Yang, B.~B.}
\newblock Polyglot application auto scaling service for platform as a service
  cloud.
\newblock In {\em Cloud Engineering (IC2E), 2015 IEEE International Conference
  on\/} (2015), IEEE, pp.~84--91.

\bibitem{shariffdeen2016workload}
{\sc Shariffdeen, R., Munasinghe, D., Bhathiya, H., Bandara, U., and Bandara,
  H.~D.}
\newblock Workload and resource aware proactive auto-scaler for paas cloud.
\newblock In {\em Cloud Computing, 2016 IEEE 9th International Conference on\/}
  (2016), pp.~11--18.

\bibitem{2011-cost-aware-v-h-scaling-2011}
{\sc Sharma, U., Shenoy, P., Sahu, S., and Shaikh, A.}
\newblock A cost-aware elasticity provisioning system for the cloud.
\newblock In {\em Distributed Computing Systems (ICDCS), 2011 31st
  International Conference on\/} (June 2011), pp.~559--570.

\bibitem{souza2015using}
{\sc Souza, A. A.~D., and Netto, M.~A.}
\newblock Using application data for sla-aware auto-scaling in cloud
  environments.
\newblock In {\em Modeling, Analysis and Simulation of Computer and
  Telecommunication Systems (MASCOTS), 2015 IEEE 23rd International Symposium
  on\/} (2015), IEEE, pp.~252--255.

\bibitem{sun2016roar}
{\sc Sun, Y., White, J., Eade, S., and Schmidt, D.~C.}
\newblock Roar: a qos-oriented modeling framework for automated cloud resource
  allocation and optimization.
\newblock {\em Journal of Systems and Software 116\/} (2016), 146--161.

\bibitem{sun2017automated}
{\sc Sun, Y., White, J., Li, B., Walker, M., and Turner, H.}
\newblock Automated qos-oriented cloud resource optimization using containers.
\newblock {\em Automated Software Engineering 24}, 1 (2017), 101--137.

\bibitem{van2015mnemos}
{\sc van Beek, V., Donkervliet, J., Hegeman, T., Hugtenburg, S., and Iosup, A.}
\newblock Mnemos: Self-expressive management of business-critical workloads in
  virtualized datacenters.

\bibitem{vecchiola2009aneka}
{\sc Vecchiola, C., Chu, X., and Buyya, R.}
\newblock Aneka: a software platform for .net-based cloud computing.
\newblock {\em High Speed and Large Scale Scientific Computing 18\/} (2009),
  267--295.

\bibitem{E3-R-extended}
{\sc Wada, H., Suzuki, J., Yamano, Y., and Oba, K.}
\newblock Evolutionary deployment optimization for service-oriented clouds.
\newblock {\em Softw. Pract. Exper. 41}, 5 (Apr. 2011), 469--493.

\bibitem{queue-VM-group}
{\sc Wang, C., Chen, J., Zhou, B.~B., and Zomaya, A.}
\newblock Just satisfactory resource provisioning for parallel applications in
  the cloud.
\newblock In {\em Services (SERVICES), 2012 IEEE Eighth World Congress on\/}
  (June 2012), pp.~285--292.

\bibitem{fuzzy-2-loop-control}
{\sc Wang, L., Xu, J., and Zhao, M.}
\newblock Application-aware cross-layer virtual machine resource management.
\newblock In {\em Autonomic Computing, ACM 9th International Conference on\/}
  (2012), pp.~13--22.

\bibitem{vm-fuzzy-MIMO}
{\sc Wang, L., Xu, J., and Zhao, M.}
\newblock Tracking adaptive performance models using dynamic clustering of user
  classes.
\newblock In {\em 7th International Workshop on Feedback Computing\/} (2012).

\bibitem{CloudAnalysis}
{\sc Wickremasinghe, B., Calheiros, R., and Buyya, R.}
\newblock Cloudanalyst: A cloudsim-based visual modeller for analysing cloud
  computing environments and applications.
\newblock In {\em Advanced Information Networking and Applications (AINA), 2010
  24th IEEE International Conference on\/} (April 2010), pp.~446--452.

\bibitem{Arc-cover-all-controller}
{\sc Wuhib, F., Stadler, R., and Lindgren, H.}
\newblock Dynamic resource allocation with management objectives:
  Implementation for an openstack cloud.
\newblock In {\em Network and service management (cnsm), 2012 8th international
  conference and 2012 workshop on systems virtualiztion management (svm)\/}
  (Oct 2012), pp.~309--315.

\bibitem{11icde_smartsla_full}
{\sc Xiong, P., Chi, Y., Zhu, S., Moon, H.~J., Pu, C., and Hacgumus, H.}
\newblock Smartsla: Cost-sensitive management of virtualized resources for
  cpu-bound database services.
\newblock {\em IEEE Transactions on Parallel and Distributed Systems 26}, 5
  (2015), 1441--1451.

\bibitem{2013-single-learner-filter-wrapper-LR-2013}
{\sc Xiong, P., Pu, C., Zhu, X., and Griffith, R.}
\newblock vperfguard: An automated model-driven framework for application
  performance diagnosis in consolidated cloud environments.
\newblock In {\em Performance Engineering, 4th ACM/SPEC International
  Conference on\/} (2013), pp.~271--282.

\bibitem{ICDCS2011}
{\sc Xiong, P., Wang, Z., Malkowski, S., Wang, Q., Jayasinghe, D., and Pu, C.}
\newblock Economical and robust provisioning of n-tier cloud workloads: A
  multi-level control approach.
\newblock In {\em Distributed Computing Systems (ICDCS), 2011 31st
  International Conference on\/} (June 2011), pp.~571--580.

\bibitem{MASCOTS11-bu-software-CP-full}
{\sc Xu, C.-Z., Rao, J., and Bu, X.}
\newblock Url: A unified reinforcement learning approach for autonomic cloud
  management.
\newblock {\em Journal of Parallel and Distributed Computing 72}, 2 (2012), 95
  -- 105.

\bibitem{2014-ARMA-single-agg-objective-2014}
{\sc Yang, J., Liu, C., Shang, Y., Cheng, B., Mao, Z., Liu, C., Niu, L., and
  Chen, J.}
\newblock A cost-aware auto-scaling approach using the workload prediction in
  service clouds.
\newblock {\em Information Systems Frontiers 16}, 1 (2014), 7--18.

\bibitem{parallel-RL-vertical-QoS-2013}
{\sc Yazdanov, L., and Fetzer, C.}
\newblock Vscaler: Autonomic virtual machine scaling.
\newblock In {\em Cloud Computing, 2013 IEEE Sixth International Conference
  on\/} (2013), pp.~212--219.

\bibitem{zhang2016container}
{\sc Zhang, H., Ma, H., Fu, G., Yang, X., Jiang, Z., and Gao, Y.}
\newblock Container based video surveillance cloud service with fine-grained
  resource provisioning.
\newblock In {\em Cloud Computing, IEEE 9th International Conference on\/}
  (2016), pp.~758--765.

\bibitem{MPC-price}
{\sc Zhang, Q., Zhu, Q., and Boutaba, R.}
\newblock Dynamic resource allocation for spot markets in cloud computing
  environments.
\newblock In {\em Utility and Cloud Computing (UCC), 2011 Fourth IEEE
  International Conference on\/} (Dec 2011), pp.~178--185.

\bibitem{software-RP-two-loops}
{\sc Zhang, Y., Huang, G., Liu, X., and Mei, H.}
\newblock Integrating resource consumption and allocation for infrastructure
  resources on-demand.
\newblock In {\em Cloud Computing (CLOUD), 2010 IEEE 3rd International
  Conference on\/} (July 2010), pp.~75--82.

\bibitem{kalman-AR}
{\sc Zheng, T., Litoiu, M., and Woodside, M.}
\newblock Integrated estimation and tracking of performance model parameters
  with autoregressive trends.
\newblock In {\em Performance Engineering, ACM/SPEC International Conference
  on\/} (2011), pp.~157--166.

\bibitem{TR-10-full-version}
{\sc Zhu, Q., and Agrawal, G.}
\newblock Resource provisioning with budget constraints for adaptive
  applications in cloud environments.
\newblock {\em IEEE Transactions on Services Computing 5}, 4 (Fourth 2012),
  497--511.

\bibitem{sla-provision}
{\sc Zhu, Z., Bi, J., Yuan, H., and Chen, Y.}
\newblock Sla based dynamic virtualized resources provisioning for shared cloud
  data centers.
\newblock In {\em Cloud Computing (CLOUD), 2011 IEEE International Conference
  on\/} (July 2011), pp.~630--637.

\end{thebibliography}
